\documentclass[iop,revtex4]{emulateapj}
\usepackage{amsmath}
\usepackage{natbib}
\usepackage{graphicx}
\usepackage{lscape}

\newcommand       \h          {\,{\rm H_2 O}}

\begin{document}

\title{Measurements of water surface snow lines in classical protoplanetary disks}

\author{Sandra M. Blevins\altaffilmark{1,2}, Klaus M. Pontoppidan\altaffilmark{1}, Andrea Banzatti\altaffilmark{1}, Ke Zhang\altaffilmark{3}, \\  
Joan R. Najita\altaffilmark{4}, John S. Carr\altaffilmark{5}, Colette Salyk\altaffilmark{6}, Geoffrey A. Blake\altaffilmark{7}}
\affil{\altaffilmark{1} Space Telescope Science Institute, 3700 San Martin Drive, Baltimore, MD 21218, USA}
\affil{\altaffilmark{2} The Catholic University of America, Department of Physics, 620 Michigan Avenue NE, Washington, DC 20064, USA}
\affil{\altaffilmark{3} University of Michigan, Department of Astronomy, 1085 S University Avenue, Ann Arbor, MI 48109, USA}
\affil{\altaffilmark{4} National Optical Astronomy Observatory, 950 N Cherry Avenue, Tucson, AZ 85719, USA}
\affil{\altaffilmark{5} Naval Research Laboratory, Code 7211, Washington, DC 20375, USA}
\affil{\altaffilmark{6} Vassar College, Department of Physics and Astronomy, 124 Raymond Avenue, Poughkeepsie, NY 12604, USA}
\affil{\altaffilmark{7} California Institute of Technology, Division of Geological \& Planetary Sciences, MC 150-21, Pasadena, CA 91125, USA}

\email{blevins@stsci.edu}

\begin{abstract}
We present deep {\it Herschel}-PACS spectroscopy of far-infrared water lines from a sample of four protoplanetary disks around solar-mass stars, selected to have strong water emission at mid-infrared wavelengths. By combining the new {\it Herschel} spectra with archival {\it Spitzer}-IRS spectroscopy, we retrieve a parameterized radial surface water vapor distribution from 0.1-100\,AU using two-dimensional dust and line radiative transfer modeling. The surface water distribution is modeled with a step model comprising of a constant inner and outer relative water abundance and a critical radius at which the surface water abundance is allowed to change. We find that the four disks have critical radii of $\sim 3-11$\,AU, at which the surface water abundance decreases by at least 5 orders of magnitude. The measured values for the critical radius are consistently smaller than the location of the surface snow line, as predicted by the observed spectral energy distribution. This suggests that the sharp drop-off of the surface water abundance is not solely due to the local gas-solid balance, but may also be driven by the de-activation of gas-phase chemical pathways to water below 300\,K. Assuming a canonical gas-to-dust ratio of 100, as well as coupled gas and dust temperatures $T_{\rm gas}=T_{\rm dust}$, the best-fit inner water abundances become implausibly high (0.01-1.0\,${\rm H_{2}}^{-1}$). Conversely, a model in which the gas and dust temperatures are decoupled leads to canonical inner disk water abundances of $\sim 10^{-4}\, \rm H_{2}^{-1}$, while retaining gas-to-dust ratios of 100. That is, the evidence for gas-dust decoupling in disk surfaces is stronger than for enhanced gas-to-dust ratios. 


\end{abstract}

\section{Introduction}

The composition of planets is closely linked to the gas/solid balance between molecular volatiles at the radius of their primary feeding zones. Volatiles in the form of ice are present beyond their respective sublimation radius, known as the {\it snow line}, and are available to catalyze planet formation \citep{Ida04b, Johansen07}, and contribute to the bulk mass of terrestrial planets, super-Earths, and giant planet cores. Volatiles in the gas-phase are located within the snow line and are only available to gravitationally formed giant planet atmospheres \citep{Oberg11}. The bulk composition of planets is therefore a tracer of their formation radius, even if they later experience significant orbital migration \citep{Madhusudhan14}. This has become particularly apparent as the combination of planet radii measurements from Kepler data and radial velocity mass-measurements have demonstrated that a wide range of planet densities exist for Earth-like planets and super-Earths orbiting within a fraction of an AU from their parent star \citep{Wolfgang12}. Indeed, the population is not consistent with a single planet mass-radius relation, suggesting that multiple modes of planet formation operate, each resulting in different planet bulk compositions; some super-Earths may be stripped-down Neptunes, some may be entirely rocky/metallic, and some may be water-worlds that migrated inwards from beyond the snow line \citep{Schlaufman09, Lopez12}. 

Many models for the formation of close-in super-Earths require assembly at large radii, beyond the snow line, followed by late-time gas-disk induced migration \citep{Alibert06,Mordasini12} because insufficient solid mass is available within 1\,AU to allow in-situ formation \citep{Hansen12}. This suggests that many super-Earths will be water-rich. The answer to migration versus in-situ formation of super-Earths lies, in part, in the measurements of bulk exoplanet compositions. However, to know whether, and to which degree, a volatile-rich super-Earth migrated, we also must know the location of the snow line at the time of planetesimal formation. 

In the Solar System, the present-day snow line is located near 2.5\,AU, roughly where water-rich C-type asteroids take over from water-poor S-type asteroids \citep{Bus02}. However, the record of the ancient solar snow line in asteroids is likely obscured due to dynamical mixing \citep{Walsh11}. Consequently, the radius of the water snow line has been estimated based on various theoretical models. The classical work by \cite{Hayashi81} assumes an optically thin disk and a present-day solar luminosity to recover a radius of 2.7\,AU. For optically thick protoplanetary disks, the location of the snow line is a more complex function of, at least, stellar evolutionary stage, optical depth of the disk and viscous dissipation. \cite{Lecar06} calculates the dependence of the water sublimation temperature on density in an optically thick disk and find a snow line location just within 2\,AU for accretion rates between $10^{-8}$ and $10^{-7}\,M_{\odot}\,\rm yr^{-1}$. \cite{Garaud07} and \cite{Kennedy08} use a similar approach to model the time evolution of the midplane snow line for optically thick disks affected by both direct stellar irradiation and accretion heating, and showed that at early stages the snow line begins at large radii and moves inwards as the system ages and both the stellar luminosity and accretion rates drop. For a Solar-mass star, the snow line is initially located as far out as 10\,AU but decreases to $\lesssim$1\,AU once the stellar accretion rate falls below $10^{-9}\,M_{\odot}\,\rm yr^{-1}$. 

Because a snow line inside of 1\,AU during the epoch of planetesimal formation in a young solar analog is seemingly in conflict with the low water abundance on the Earth, the dependence of the snow line on secondary parameters are currently being investigated. \cite{Oka11} calculated the effects of varying grain sizes and found that sufficiently large grains may push the snow line beyond 1\,AU during the entire evolution of a solar-type disk. \cite{Martin12} consider a more detailed model in which multiple snow lines may form simultaneously at different radii, especially at early times ($<1$\,Myr). In their model, this is due to the existence of the dead zone at intermediate disk radii in which the ionization fraction is low, slowing the accretion rate through the disk. Just outside of the dead zone the disk may become self-gravitating, leading to heating and sublimation of water ice at radii of 0.5-10\,AU. A key driver for this model is its ability to prevent the Earth from acquiring large amounts of water at its formation radius at 1 AU. 

\begin{figure}[ht!]
\centering
\includegraphics[width=9cm]{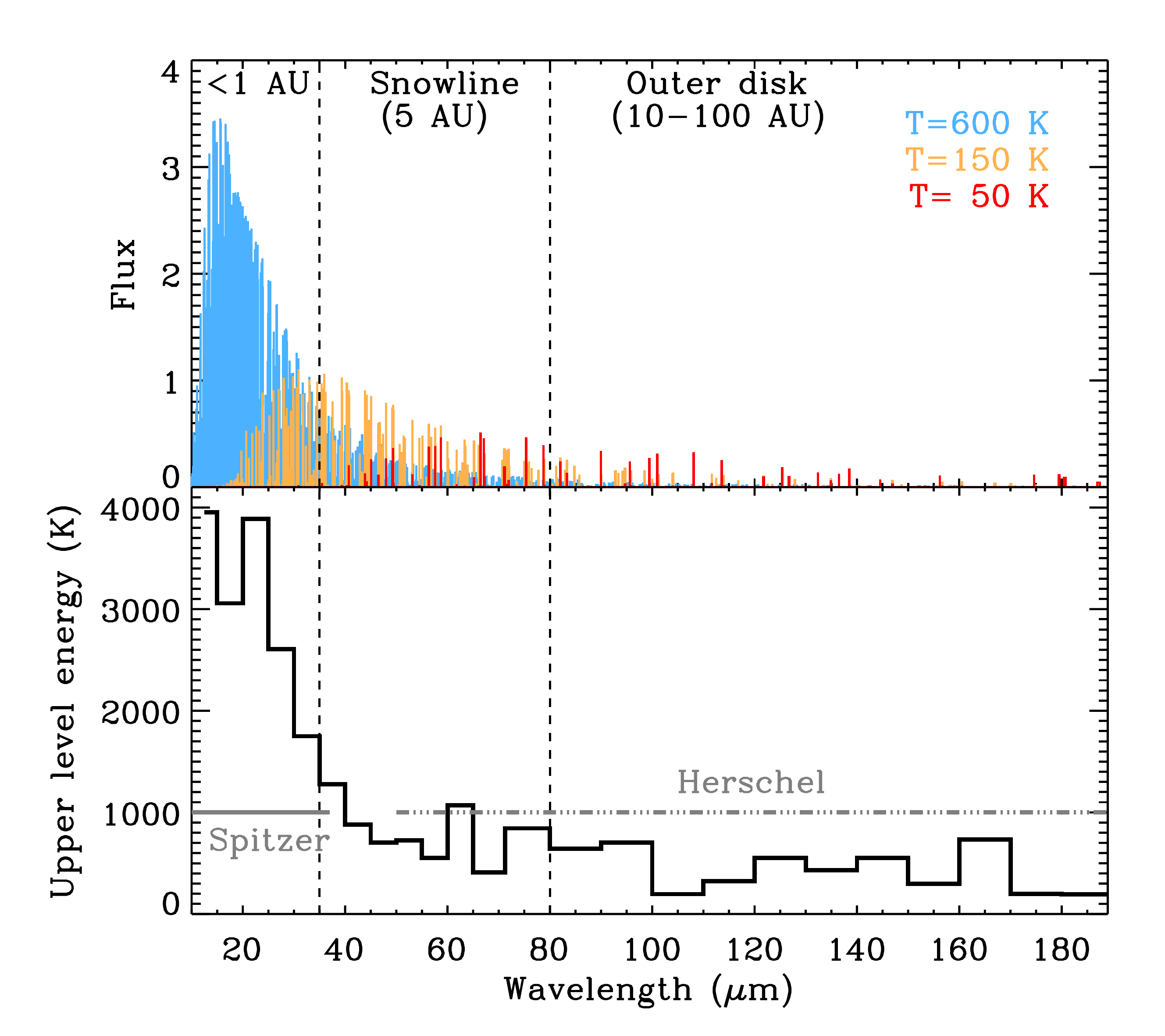} 
\caption{Top panel: Spectral line flux distribution of infrared water lines for three different single-temperature slab models. Bottom panel: The binned median $E_u$ of the strongest water lines as a function of wavelength. The $E_u$ distribution decreases with wavelength, reaching values most sensitive to the snowline location, defined as the peak $E_u$ for a water component with $T=150$\,K, around 60\,$\mu$m. At shorter wavelengths, the emission is dominated by hot water in the inner disk. At longer wavelengths the emission is dominated by cooler water vapor in the outer disk.}
\label{fig: spec_map_tech1}
\end{figure}

Measurements of the distribution of water vapor in present-day protoplanetary disks became possible, at least in principle, with the discovery of the mid-infrared molecular forest due to warm water vapor \citep{Carr08, Salyk08}, and far-infrared lines tracing cold water vapor \citep{Hogerheijde11}. The mid-infrared spectra of at least half of all young gas-rich disks around low-mass stars contain thousands of rotational water transitions with excitation temperatures $\sim 500-3000$\,K indicating the presence of copious amounts ($\gtrsim 10^{-4}\, \rm H_{2}^{-1}$) of warm water vapor emitting from within the inner few AU of the disk \citep{Pontoppidan10,Salyk11}. In strong contrast to the mid-infrared, water lines tracing cooler gas at 10s of AU are absent or weak in the far-infrared \citep{Meeus12,Riviere-Marichalar12}. Detections of cool water were made in very deep {\it Herschel Space Observatory} observations, but abundances are exceedingly low -- $10^{-10}-10^{-11}\, \rm H_{2}^{-1}$ \citep{Hogerheijde11}. Between the inner and outer disk, the water vapor abundance drops by about 6 orders of magnitude. \cite{Hogerheijde11} demonstrates that the low abundance in the outer disk is consistent with a photo-desorbed gaseous population derived from a massive reservoir of water ice. The inference is that water is present at all radii at high abundances, but that it freezes out, settles to the midplane beyond the snow line, and becomes available as planet-building material.   

In this paper, we present deep {\it Herschel Space Observatory} Photodetector Array Camera and Spectrograph ({\it Herschel}-PACS) spectroscopy of water lines, spanning a range of excitation temperatures, in four classical protoplanetary disks around young solar-mass stars. The disks are selected to have strong water vapor emission at mid-infrared wavelengths, as observed by the {\it Spitzer Space Telescope} InfraRed Spectrograph ({\it Spitzer}-IRS). We combine the new PACS spectra with archival {\it Spitzer} spectra to trace gas spanning a wide range of excitation temperatures ($\sim 100-5000$\,K) spanning the water freeze-out temperature of 120-170\,K \citep{Lecar06, Meijerink09}. By fitting a two-dimensional radiative transfer model to the broad-band spectral energy distributions and line fluxes, we measure the radial distribution of water vapor, including the location of the water snow line. This spectroscopic mapping technique was first deployed by \cite{Zhang13} in the case of the transition disk TW Hya, and here we apply it to classical, optically thick T Tauri disks. In Section \ref{sec: data} we describe the multi-wavelength spectroscopic observations from {\it Spitzer} and {\it Herschel} used to map the water vapor distribution in the four protoplanetary disks. Section \ref{sec: methods} describes the spectral modeling approach and how we retrieve the water vapor distribution. The results of the abundance retrieval are discussed in Section \ref{sec: results} and the implications for our understanding of disk volatiles are discussed in Section \ref{sec: discussion}.

\section{Observations} 
\label{sec: data}

\begin{figure*}[ht!]
\centering
\includegraphics[width=18.0cm]{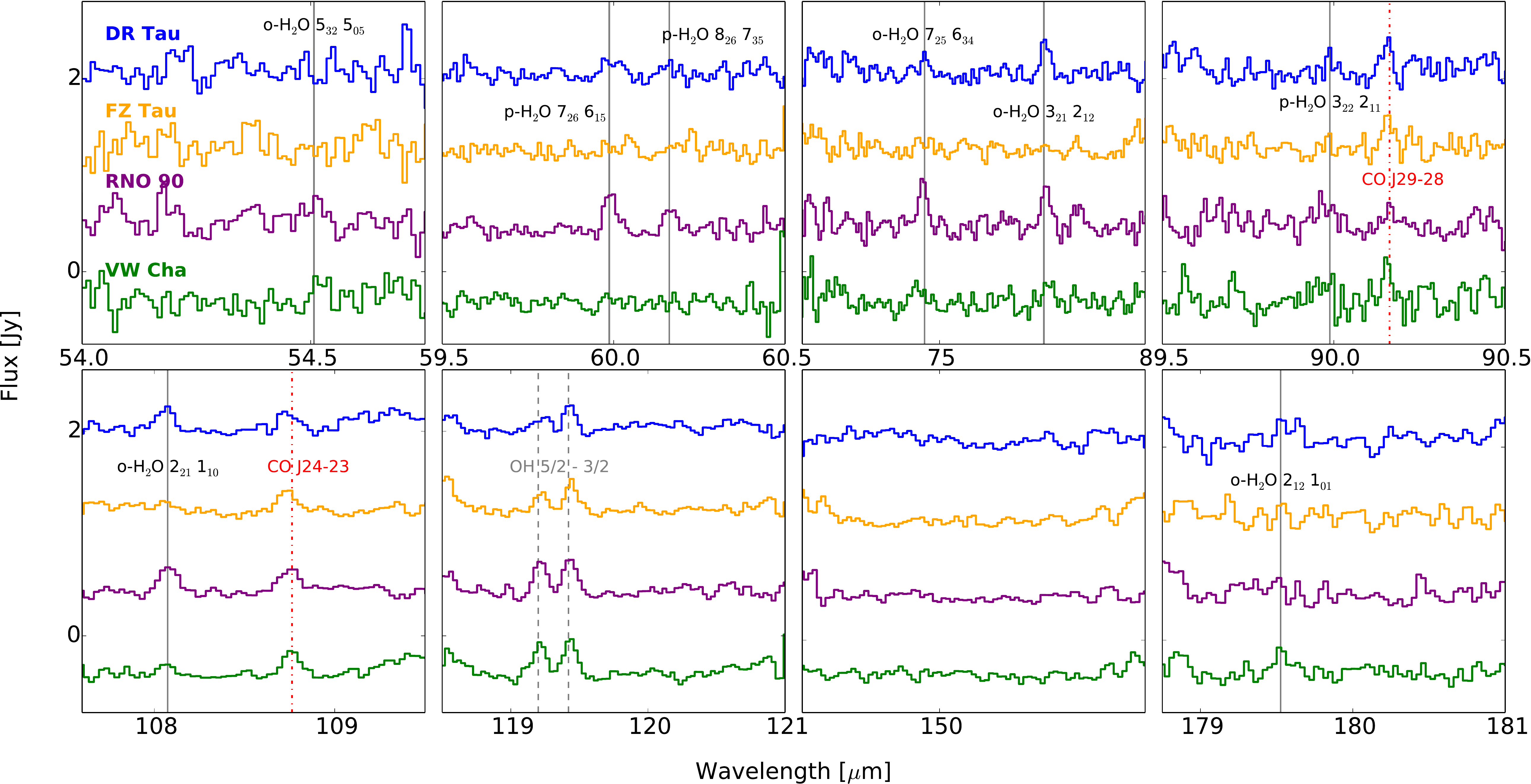}
\caption{New {\it Herschel}-PACS molecular spectra of DR Tau, FZ Tau, RNO 90, and VW Cha (from top to bottom). The vertical, solid lines mark water transitions included in the analysis. The dashed line marks the 5/2--3/2 OH doublet at 119.3 $\mu$m, and the dash-dotted line marks the CO J29-28 and CO J24-23 transitions at 90.2 and 108.8 $\mu$m, respectively; these lines are not analyzed in this paper.}
\label{fig: pacs_data}
\end{figure*}

Several thousand strong water lines from rotational and ro-vibrational transitions exist in the infrared ($\sim$ 1--200 $\mu$m). The upper level energies $E_u$ of the strongest transitions (those with large Einstein $A$ values) tend to decrease with increasing wavelength, from $\sim$10,000\,K at 1\,$\mu$m, $\sim$1,000\,K at 60\,$\mu$m to $\sim$100\,K at 200\,$\mu$m. In Figure \ref{fig: spec_map_tech1}, this is qualitatively illustrated using a single-temperature slab model. In a disk where the gas temperature decreases radially with increasing distance from the star, the high-energy water lines trace the inner, hotter disk, while the the low-energy lines are dominated by the large emitting areas of the cooler, outer disk. For instance, the upper levels feeding the mid-infrared transitions are populated when the gas temperature exceeds $\sim$300\,K and the density exceeds $\sim 10^8\,\rm cm^{-3}$ \citep{Ewine13}. Therefore, measuring line strengths spanning a factor $\sim$100 in $E_u$ and wavelength provides information over similar scales in disk radii, that is, 1--100\,AU. This allows a spectral mapping technique to retrieve the radial profile of the water abundance in protoplanetary disks \citep{Zhang13}.

\begin{deluxetable*}{lllcccccc}
\centering
\tablecolumns{9}
\tablewidth{0pt} 
\tablecaption{Log of water line spectroscopy}

\tablehead{
\colhead{Mode}           & \colhead{SH\tablenotemark{a}} & \colhead{LH\tablenotemark{a}} &\colhead{PACS\tablenotemark{b}} &\colhead{PACS\tablenotemark{b}} &\colhead{PACS\tablenotemark{b}} &\colhead{PACS\tablenotemark{b}} &\colhead{Spitzer AOR} &\colhead{Herschel OBSIDs} \\
\colhead{Range}          & \colhead{10--19}              & \colhead{20--37}        &\colhead{74.64--75.63}   &\colhead{107.7--109.1} &\colhead{118.8--120.5} &\colhead{178.9--181.1} &\colhead{}            &\colhead{} \\
\colhead{($\mu$m)}       & \colhead{$t_{\rm int}$}       & \colhead{$t_{\rm int}$} &\colhead{149.28--151.26} &\colhead{53.85--54.55} &\colhead{59.4--60.25}  &\colhead{89.45--90.55} &\colhead{} & \colhead{} \\
\\ [-1.5ex]
\hline  \\ [-1.5ex]
%\cline{1-9}
\colhead{  }           & \colhead{(s)} & \colhead{(s)} &\colhead{(Reps)}    &\colhead{(Reps)}  &\colhead{(Reps)}  &\colhead{(Reps)}  &\colhead{}            &\colhead{}
}
\startdata
DR Tau                   & $48\times 6.3$          & $40\times 14.7$    &3&5&4&3& r27067136  & 1342239700, 1342240158 \\
FZ Tau                   & $12\times 31.5$         & $12\times 61$      &3&5&4&3& r27058688  & 1342239722, 1342239723 \\
RNO 90                   & $36\times 6.3$          & $24\times 14.7$    &3&5&4&3& r27061760  & 1342241706, 1342250577 \\
VW Cha                   & $24\times 31.5$         & $24\times 61$ 	    &3&5&4&3& r27066112  & 1342238360, 1342238361 
\enddata
\tablecomments{
\tablenotetext{a}{SH and LH refer to the short-high and long-high {\it Spitzer}-IRS spectroscopic modes.}
\tablenotetext{b}{{\it Herschel}-PACS range scans. The spectral ranges are indicated (note that each scan results in two spectral orders). All PACS data were taken as part of program OT1\_kponto01\_1.}
}
\label{tab: obslog}
\end{deluxetable*}

Conversely, the spectral coverage typically provided by a single infrared spectrometer (near, mid- or far-infrared) can obtain only limited information about localized regions in disks. The wide spectral coverage needed to observe water in both the warm and the cold regions throughout a disk can be obtained by combining data from {\it Spitzer}-IRS (10--37\,$\mu$m) and from {\it Herschel}-PACS (55--200\,$\mu$m). \textit{Spitzer} spectra are sensitive to high abundances of water vapor at temperatures higher than several 100\,K, inwards of the snow lines, while \textit{Herschel} spectra are sensitive to the far lower water abundances and lower gas temperatures of 50-300\,K found across and beyond the snow line.

\subsection{Protoplanetary disk sample}
To map the radial distribution of water vapor, we selected a small number of protoplanetary disks known to have strong emission from water vapor at mid-infrared (10--37\,$\mu$m) wavelengths, as observed with {\it Spitzer} \citep{Salyk08, Carr11}, and conducted a deep spectroscopic survey with \textit{Herschel}-PACS of selected water lines between 54 and 180\,$\mu$m. The \textit{Herschel} lines extend the \textit{Spitzer} spectral mapping region from $\sim$1\,AU to $\gtrsim 50$\,AU, in particular enabling a measurement of the snow line in the line-emitting layers of the disks. To maximize the chances for detecting far-infrared water lines from disk regions with low abundances, the targeted sample was selected from disks observed to have the strongest water lines in the mid-infrared, as well as with the highest mid-infrared line-to-continuum ratios \citep{Pontoppidan10,Salyk11,Najita13}. Our disk sample is comprised of four classical (without known inner holes or gaps) protoplanetary disks surrounding the solar-mass pre-main sequence stars DR Tau, FZ Tau, RNO 90, and VW Cha with spectral types spanning M0 to G5. The water vapor observations are summarized in Table \ref{tab: obslog}. The properties of the young stars and their disks are summarized in Table \ref{tab:source_modelparams}.
 
\subsection{\textit{Herschel}-PACS spectroscopy}
Deep chopping/nodding spectral range scans ($R=940-5500$ or $\Delta v=55-320\,\rm km\,s^{-1}$) were obtained with the Photoconductor Array Camera and Spectrometer (PACS) \citep{Poglitsch10} onboard the \textit{Herschel Space Observatory} \citep{Pilbratt10} (see Figure \ref{fig: pacs_data}). The observations include four range scans for each disk, each providing two simultaneous spectral orders for a total of eight spectral segments per disk, five of which target rotational water transitions near 60, 75, 90, 108 and 179.5\,$\mu$m (see Table \ref{tab: obslog}). These spectral regions were selected to include intermediate to low-excitation water transitions with upper level energies spanning $114$\,K to $1414$\,K, to complement the higher energy (1000--4000\,K) water vapor lines detected in the IRS spectra (see Figure \ref{fig: spec_map_tech1}). Emission from two CO lines near 90.2\,$\mu$m \citep{Meeus12} and 108.8\,$\mu$m \citep{Karska13}, and the 5/2--3/2 OH doublet centered at 119.3\,$\mu$m were also observed, but not analyzed in this paper.  

We reduced the spectra using the \textit{Herschel Interactive Processing Environment} HIPE version 13.0.0, and extracted the spectra from the central spaxel from the rebinned data cubes. The extracted spectra were corrected to account for the flux that falls outside of the central spaxel using the HIPE aperture correction procedure, {\tt pointSourceLossCorrection}.  

\subsection{\textit{Herschel}-SPIRE spectro-photometry}
All four disks were also observed using the \textit{Herschel}-SPIRE Fourier Transform Spectrometer \citep{Griffin10}. The SPIRE spectra were obtained in order to provide far-infrared spectro-photometry for dust modeling, as well as to measure the warm CO content of the disks. The SPIRE spectra and CO lines were analyzed in detail by \cite{vanderWiel14}. 

\subsection{Archival Spitzer-IRS spectroscopy}
We use $R\sim 700$ ($\Delta v=400\,\rm km\,s^{-1}$) mid-infrared (10--37\,$\mu$m) spectra obtained with the InfraRed Spectrograph (IRS) onboard the \textit{Spitzer Space Telescope} and first presented by \cite{Pontoppidan10} (RNO 90, DR Tau and VW Cha) and \cite{Najita13} (FZ Tau). The \textit{Spitzer} spectra were reduced using the pipeline described in \cite{Pontoppidan10}, following the methods by \cite{Carr08}. Briefly, the reduction uses dedicated off-source sky integrations and an optimal selection of relative spectral response functions using all available calibration observations of standard stars. This method minimizes residual fringing and prevents the defringing procedure from removing line power from the dense molecular forest. It also enables an efficient removal of cosmic ray hits and bad pixels. 

\subsection{Extraction of line fluxes} 
\label{sec: line_fluxes}
Due to the relatively low spectral resolving power of {\it Spitzer}-IRS, the observed water emission lines are typically blended complexes of multiple transitions \citep{Pontoppidan10}. Further, there may be systematic errors that are difficult to quantify, especially near the edges of orders. Therefore, rather than fitting our model to the full spectral range, we opted to select 25 line complexes that have minimal systematic errors and contamination from emission by other species including, but not limited to, OH, C$_{2}$H$_{2}$, HCN, and CO$_{2}$ \citep{Banzatti12, Carr08, Carr11}. We experimented with fitting the full spectral range, and found that the results did not change significantly. In the \textit{Herschel} ranges, individual water lines are typically well-separated, with clean continua. 

Integrated fluxes of \textit{Spitzer} line complexes, as well as isolated \textit{Herschel} lines, were estimated by fitting a first-order polynomial to the local continuum, spanning a few resolution elements on each side of the lines, and integrating the flux density in the continuum-subtracted spectra using five-point Newton-Cotes integration. Figure \ref{fig:rno90_lines} summarizes the spectral regions used for the analysis. We estimated the formal statistical errors on the line fluxes by propagating the channel errors through the line integrals. In addition to the statistical, photon-noise dominated errors, $\sigma_{phot}$, the high signal-to-noise \textit{Spitzer} spectra often have significant, even dominant, systematic sources of error. These may include residual fringes and errors in the spectral response function. To estimate the systematic errors, we selected high signal-to-noise spectra from bright Herbig and transition disks from \cite{Pontoppidan10} without detected molecular line emission, and obtained a smoothed estimate of the continua, $F_C$, using a 40-pixel wide median filter. The effective contrast, that is, the standard deviation of the ratio of the spectrum, $F$, with the smoothed continuum, $C=\sigma(F/F_{\rm C})$, was estimated as a function of spectral channel using a moving box. The systematic error in a given spectrum was then estimated as:

\begin{equation}
\sigma_{\rm sys}(F) \sim F\times \sqrt{C^2-(\sigma_{\rm phot}(F)/F_{\rm C})^2}  
\end{equation}

Adding the systematic error to the formal statistical error in quadrature then provides the total pixel-to-pixel error of the \textit{Spitzer} spectra used in our model fits. We required a a signal-to-noise ratio on each extracted line of $\ge 5$ for a formal detection. The selected lines and their integrated fluxes are summarized in Tables \ref{tab:lineflux_spitzer} and \ref{tab:lineflux_herschel}.

\section{Spectral mapping procedure} 
\label{sec: methods}

Our method of using \textit{Spitzer} and \textit{Herschel} spectroscopy of water vapor in protoplanetary disks to construct a radial abundance structure was first described in \cite{Zhang13}. In the following, we describe a revised implementation of the technique as applied to our sample of protoplanetary disks. The temperature and density structure of each disk is retrieved by matching the available photometry and spectroscopy to a RADMC radiative transfer model of continuum dust emission (Section \ref{sec: dust_mod}). These disk structures are used as input for a line raytracer code, RADLite, which renders synthetic water spectra assuming a parameterized gas structure (Section \ref{sec: gas_mod}). Best-fit water models are identified using least-squares minimization of water line fluxes and confidence limits are estimated using constant $\chi^2$ boundaries following \cite{NR}.

\subsection{Dust temperature} 
\label{sec: dust_mod}

\begin{figure*}[ht!]
\centering
\includegraphics[width=4.65cm]{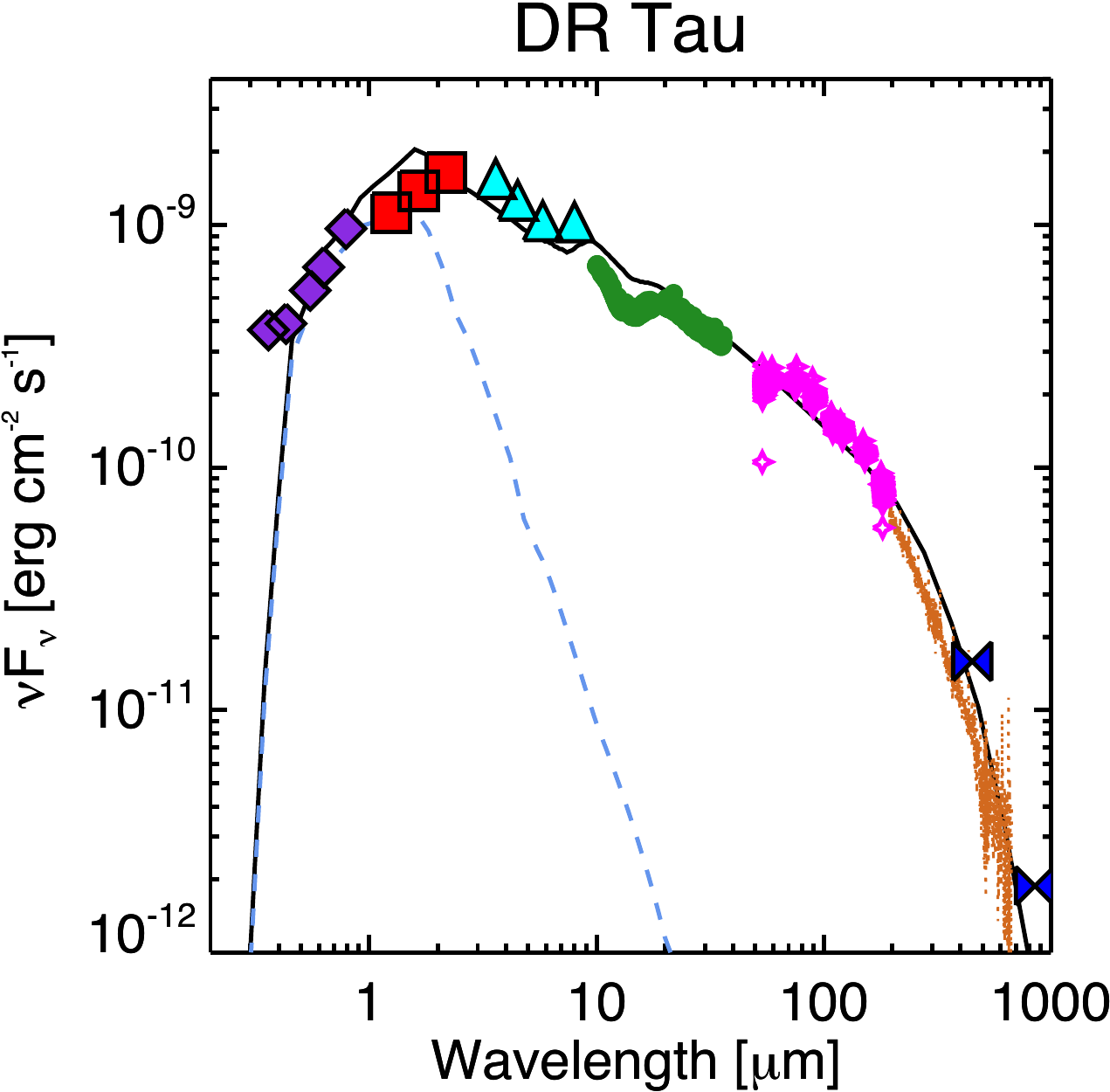}
\includegraphics[width=4.25cm]{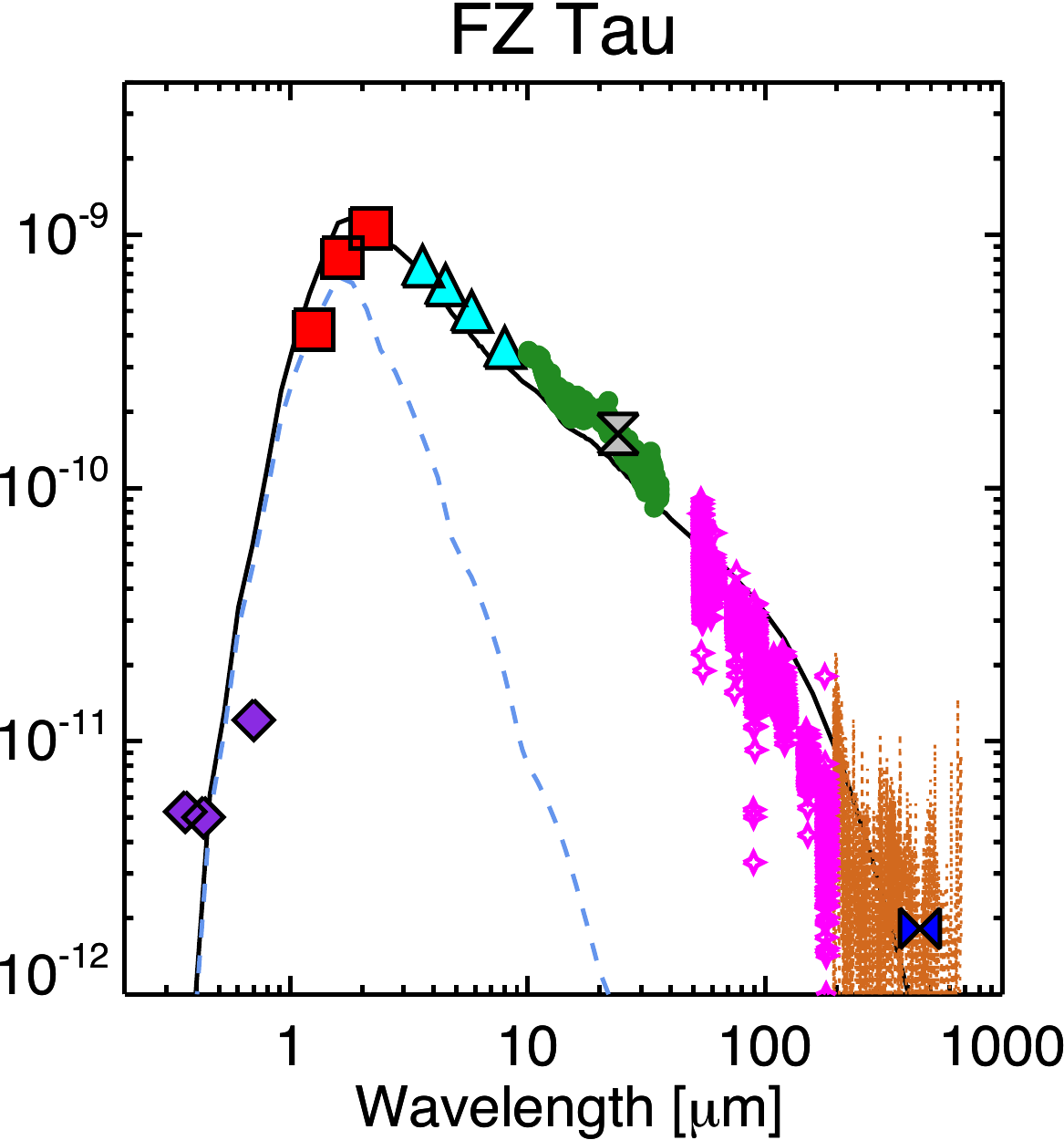}
\includegraphics[width=4.25cm]{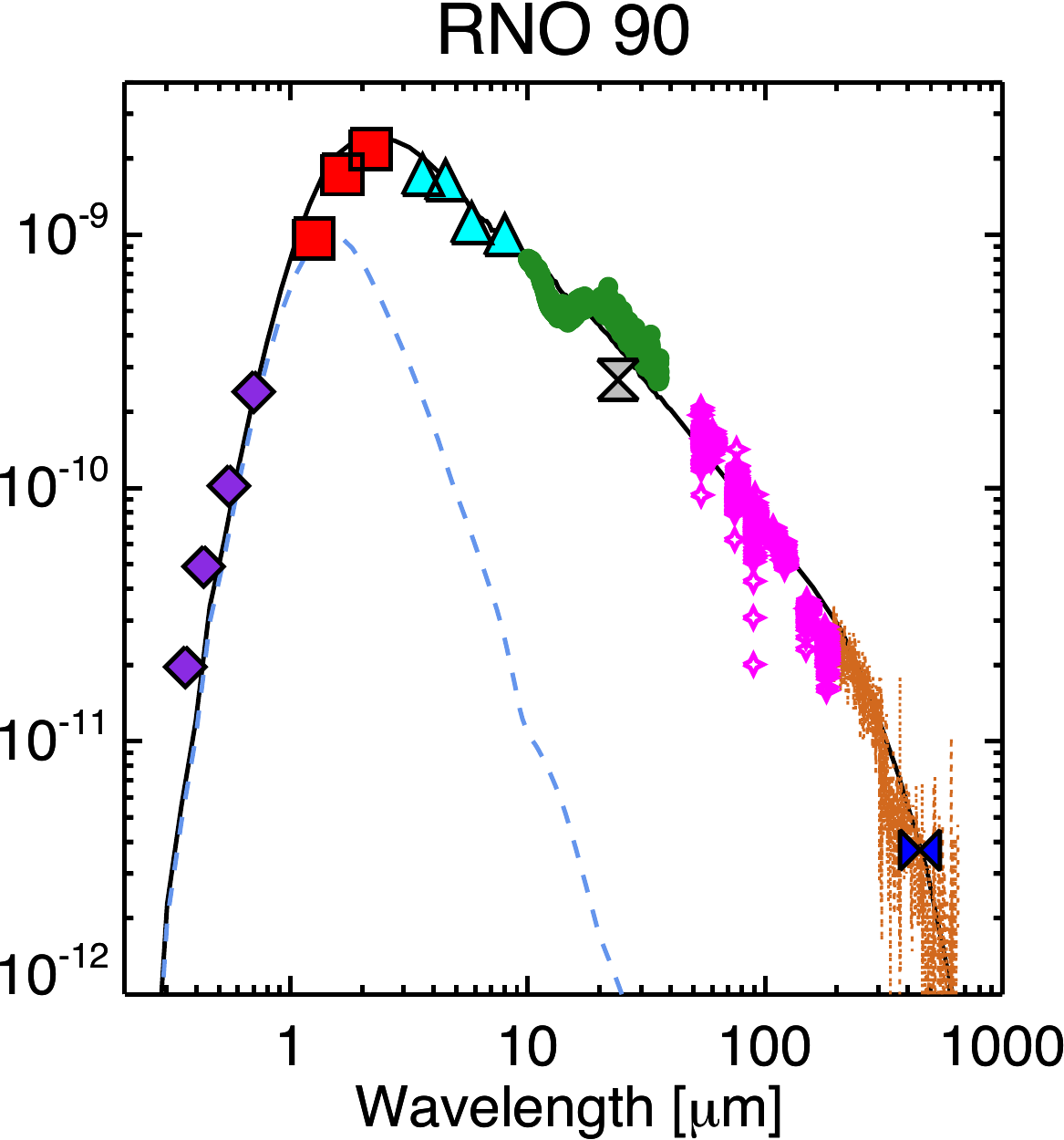}
\includegraphics[width=4.25cm]{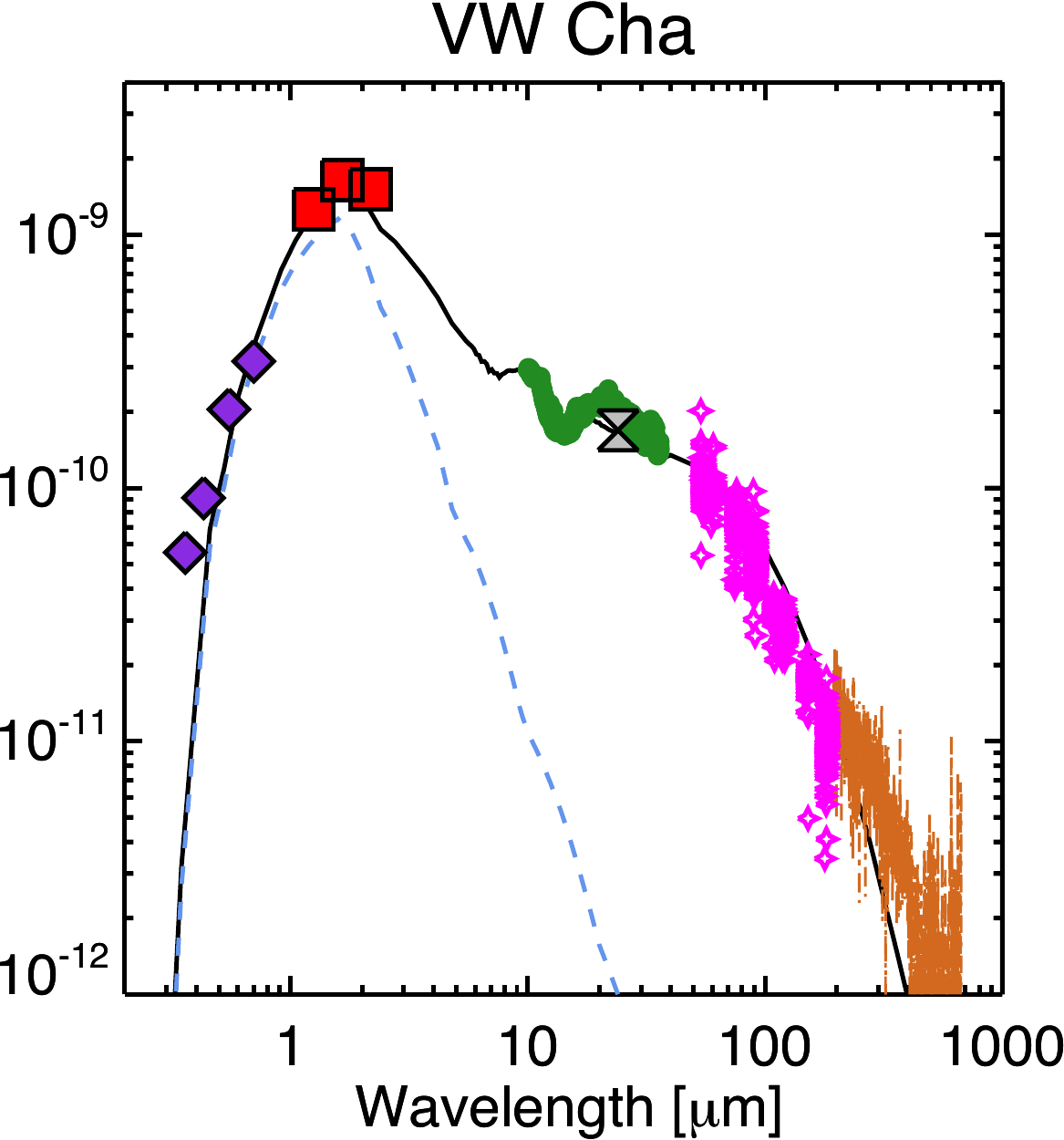}
\caption{Best-fit RADMC continuum models of the observed SEDs for each disk (solid curves). The dashed curves show the extincted stellar photospheres. Photometry references: UBVR from \cite{Herbig88, Kenyon95, Hogg00, Zacharias09} -- JHK from \cite{Cutri03} -- 3.6, 4.5, 5.8, 8.0, 24.0 $\mu$m from the Cores to Disks (c2d) \textit{Spitzer} Legacy catalog \citep{Evans03, Evans09} -- sub-mm from \cite{Andrews05}}
\label{fig:allseds}
\end{figure*}

To constrain the dust distribution of each disk, we combined broad-band photometry with the \textit{Spitzer}-IRS, \textit{Herschel}-PACS and SPIRE spectra to construct spectral energy distributions (SEDs; see Figure \ref{fig:allseds}). The online SED fitting tool by \cite{Robitaille06} was used for an initial exploration of the parameter space, while the fine-tuning of the model parameters to match the SEDs was done by hand to arrive at a plausible, but probably not unique, density structure. The stellar radius, disk mass, outer radius, flaring index, and outer vertical scale height, as defined in \cite{Dullemond01} were varied as free parameters, while the stellar effective temperature and inclination were kept fixed. The inner disk radius is assumed to be truncated by dust sublimation at 1500\,K. For comparison to our best-fit dust models, the outer disk scale heights relevant for isothermal disks in hydrostatic equilibrium were estimated using the model stellar radii, masses, temperatures, and disk flaring indices. The best-fit model scale height for VW Cha is close to equilibrium, but exceeds that of equilibrium for FZ Tau, RNO 90, and DR Tau by factors of 1.7, 1.4, and 2.0. In particular the mid-infrared continuum of DR Tau was difficult to fit within the formal hydrostatic limit. We note that DR Tau has the highest accretion rate of our sample, by about an order of magnitude ($\dot{M}$=1.5$\times$10$^{-7}$ $M_{\odot}$ yr$^{-1}$; \citealt{Salyk13}). High accretion rates lead to elevated midplane temperatures, in particular within $\sim$ 5 AU, and the passive RADMC model in general underestimates the midplane temperatures. This, in turn, may lead to the best-fit disk structure compensating by increasing the vertical scale heights to values that are formally too large compared to the disk temperature.

Given the observed SEDs, the dust temperature structure for each disk was calculated using the Monte Carlo code RADMC \citep{Dullemond04}, which solves the dust radiative transfer problem given an axisymmetric dust opacity and a density distribution. We use the dust opacity model from \cite{Meijerink09} (see their Figure 2 for a plot of the absorption and scattering coefficients). The dust is a mixture of 15\% amorphous carbon, by mass, from \cite{Zubko96} and 85\% amorphous silicate. It is assumed to have a size distribution parameterized using the \cite{Weingartner01} model, but with a maximum grain size of 40\,$\mu$m and a -2.5 power slope for the silicate grains. Some stellar and disk parameters were taken from the literature and fixed for individual systems, including the stellar mass, radius, effective temperature and inclination angle (see Table \ref{tab:source_modelparams}). The inclination angles for FZ Tau and VW Cha are not available and we adopt a value of 45$^{\circ}$ for both disks, which is consistent with inclinations constrained using velocity-resolved CO spectroscopy by \cite{Banzatti15}. The continuum models additionally include interstellar extinction using the \cite{Weingartner01} dust model with total-to-selective extinction of $R_V=5.5$.

\begin{figure}[ht!]
\includegraphics[width=8.cm]{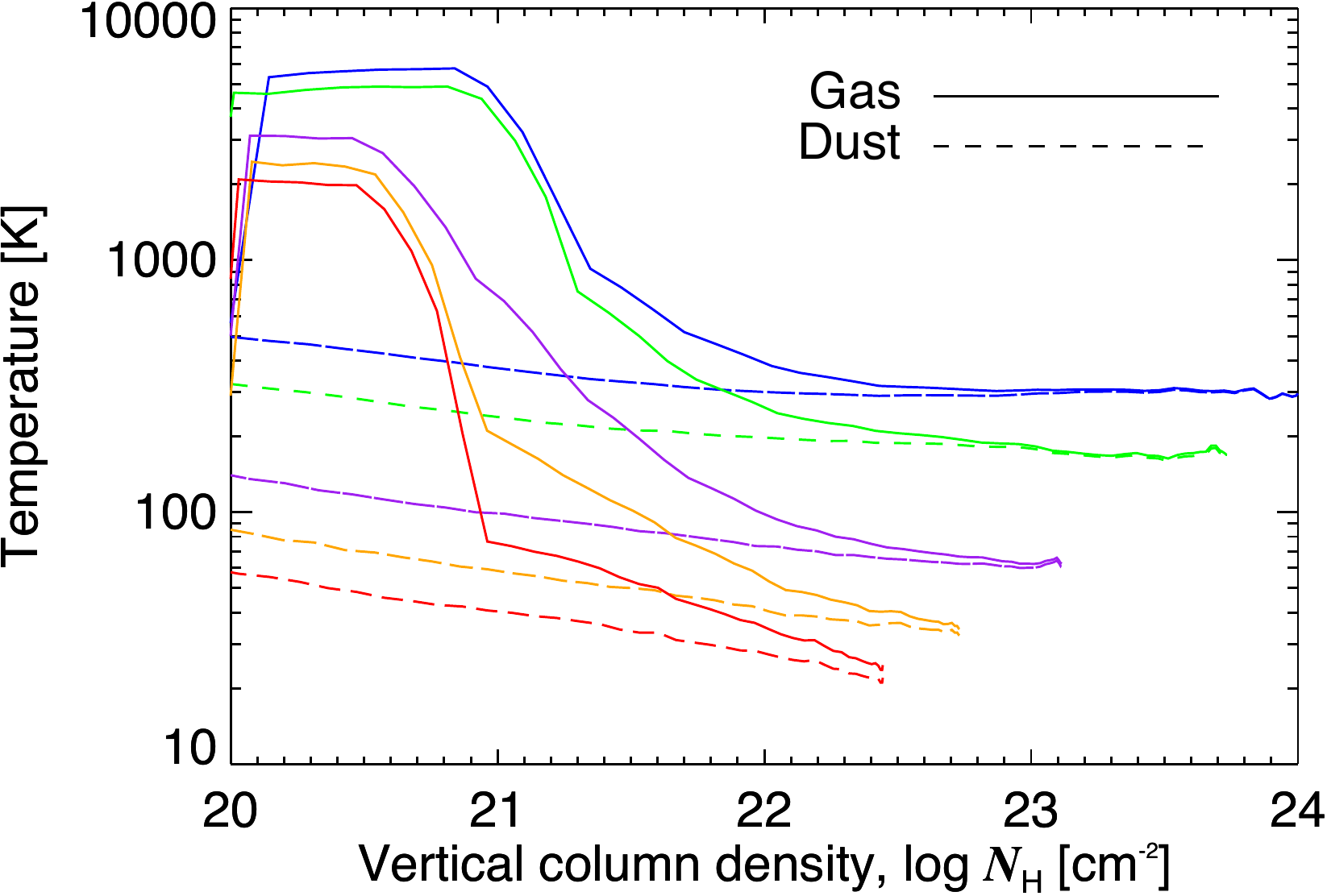}
\caption{Gas and dust temperatures as a function of vertical column density for the best-fitting model of RNO 90 for the Case II model. Temperatures are plotted for radii at 0.5, 1, 4, 10, and 20\,AU, ordered from top to bottom.}
\label{fig:najita}
\end{figure}

\begin{deluxetable*}{lccccccccccccr}[ht!]
\tablecolumns{14}
\tablewidth{0pt}
\tablecaption{Star and dust disk parameters}
\tablehead{
\colhead{Name}  &  \colhead{SpT} & \colhead{$M_*$} & \colhead{$R_*$} & \colhead{$T_{\rm eff}$} & \colhead{$L_{\rm tot}$\tablenotemark{a}} &\colhead{$d$} & \colhead{$M_{\rm disk}$\tablenotemark{e}}  & \colhead{$R_{\rm out}$\tablenotemark{e}} & \colhead{$H/R$\tablenotemark{b,e}} & \colhead{$\alpha$\tablenotemark{c,e}} & \colhead{$i$} & \colhead{$N_{\rm H}$\tablenotemark{d}} & \colhead{References}  \\
\colhead{}         & \colhead{}          & \colhead{($M_{\odot}$)} & \colhead{($R_{\odot}$)} & \colhead{(K)} & \colhead{($L_{\odot}$)} &\colhead{(pc)}  & \colhead{($M_{\odot}$)} & \colhead{(AU) } & \colhead{} & \colhead{} &\colhead{ (deg)}   & \colhead{(cm$^{-2}$)} & \colhead{} 
}
\startdata
DR Tau     & K7 & 0.8  & 2.1  & 4060  & 1.08  & 140  &  0.04  & 200   &   0.20  & 0.0357    & 9     &  $4.5\times10^{20}$       & 1,2,3,4,5,6,7,8   \\
FZ Tau     & M0 & 0.7  & 2.3  & 3918  & 1.12  & 140  & 0.001  & 120   &   0.15  & 0.0357   & 45   &  $5.5\times10^{21}$       & 5,7,9,10   \\
RNO 90     & G5 & 1.5  & 2.0  & 5770  & 4.06  & 125  &  0.005  & 200   &   0.10  & 0.00143   & 37   &  $6.8\times10^{21}$       & 4,5,7,11,12   \\
VW Cha     & K5 & 0.6  & 3.1  & 4350  & 3.08  & 178  & 0.001  & 200   &   0.15  & 0.107  & 45   &  $3.6\times10^{21}$       & 4,7,13,14,15   
\enddata
\tablecomments{
\tablenotetext{a}{The luminosity is the total energy rate input into the RADMC model. It includes both stellar and accretion luminosity, but is assumed to all be launched from the stellar surface.}
\tablenotetext{b}{Disk scale height at $R_{\rm out}$.}
\tablenotetext{c}{The power law index describing the vertical scale height $H/R\propto R^{\alpha}$ of the disk.}
\tablenotetext{d}{Interstellar extinction column densities, assuming $R_V=5.5$ and $A_V/N_{\rm H}=5.3\times10^{-22}\,\rm cm^2$ \citep{Weingartner01}.}
\tablenotetext{e}{Free disk parameters.}
\tablenotetext{}{{\bf References.} (1) \citealt{LRBrown13}; (2) \citealt{Pontoppidan11}; (3) \citealt{Salyk08}; (4) \citealt{Salyk11}; (5) \citealt{Ricci10tau}; (6) \citealt{Valenti93}; (7) \citealt{Pontoppidan10}; (8) \citealt{Hartmann98}; (9) \citealt{Herczeg09}; (10) \citealt{Rebull10}; (11) \citealt{Levreault88}; (12) \citealt{Pontoppidan14b}; (13) \citealt{Bast11}; (14) \citealt{Natta00}; (15) \citealt{Guenther07}.}
}
\label{tab:source_modelparams}
\end{deluxetable*}

\subsection{Gas temperature}
The kinetic temperature of gas in protoplanetary disk surfaces is expected to be higher than that of the dust down to vertical column densities of $\sim 10^{22}\,\rm cm^{-2}$, primarily due to collisional heating by free electrons produced by a combination of ultra-violet and X-ray irradiation \citep{Glassgold04,Kamp04,Jonkheid04}. In deeper layers, the electron density decreases rapidly as the gas and dust density increase, leading to strong thermal coupling between gas and dust. It is not a priori known how much of the observed water line flux is formed in disk layers with decoupled gas and dust temperatures. Simple slab models of the mid-infrared water emission derive column densities of $N_{\rm H_2O}\sim 10^{18}-10^{19}\,\rm cm^{-2}$ \citep{Salyk11,Carr11}, which, assuming canonical water abundances of $10^{-4}\,\rm H_2^{-1}$, correspond to total gas column densities of $10^{22}-10^{23}\,\rm cm^{-2}$, suggesting that some emitting water is in the decoupled layer, while some is coupled to the dust.
In recent detailed disk models, the gas temperature is calculated including full chemical networks, and assuming thermal balance between heating and cooling processes \citep{Woitke09, Willacy09, Glassgold09, Heinzeller11, Du14}. However, such calculations are CPU intensive and depend on parameters that are often poorly constrained, such as chemical rates, ultraviolet and X-ray radiation fields and dust properties. 

In order to create model grids large enough to fit the water line spectra in detail, we simplify the gas temperature calculation by computing two sets of model grids: One grid (Case I) in which the gas and dust are thermally coupled ($T_{\rm dust}=T_{\rm gas}$), and one (Case II) in which the gas temperature is scaled relative to the dust temperature using the thermo-chemical model of \cite{Najita11}. For Case II, the gas temperature scaling is computed using the ratio of gas and dust temperature as a function of disk radius and hydrogen column density $C_{\rm scale} = T_{\rm gas}/T_{\rm dust}(R,N_{\rm H})$ as given by the reference thermo-chemical model of \cite{Glassgold09,Najita11}. At low column densities ($N_{\rm H} \lesssim 10^{21}\,\rm cm^{-2}$), the water is efficiently destroyed and gas temperatures rapidly rise above $\sim 1000\,$K. In this regime, the water abundance is zero. The thermo-chemical model has star-disk parameters that are similar to those of our observed systems ($\sim 0.9\,L_{\odot}, 4000\,K$ central star surrounded by a $0.005\,M_{\odot}$ disk). Further, it includes careful treatment of the inner 10\,AU of the disk, where most of our line emission originates. An example of the resulting gas and dust temperatures for one of our disks is shown in Fig. \ref{fig:najita}.

\subsection{Line radiative transfer} 
\label{sec: gas_mod}

The continuum models determine the radiation field at each spatial location in the disk, as well as the local dust temperatures. These quantities, along with a water abundance structure, forms the input to RADLite, a line raytracer designed to render images and spectra of complex line emission from a RADMC model \citep{Pontoppidan09}. RADLite uses an adaptive ray grid scheme to optimally sample all size scales in a protoplanetary disk, enabling the calculation of large model grids including thousands of water lines tracing both the innermost disk as well as the outer disk. The molecular parameters for water are taken from the HITRAN 2008 database \citep{Rothman09}. We assume that the level populations are in local thermodynamic equilibrium (LTE), set by the local gas kinetic temperature. The local line broadening is set to $0.9 c_s$, where $c_s$ is the sound speed, consistent with recent measurements of strong turbulence in protoplanetary disk surfaces \citep{Hughes11}. A discussion of the potential effects of assuming LTE versus non-LTE conditions is presented in $\S$ 5.1.

\begin{figure*}[ht!]
\centering
\includegraphics[width=8.95cm]{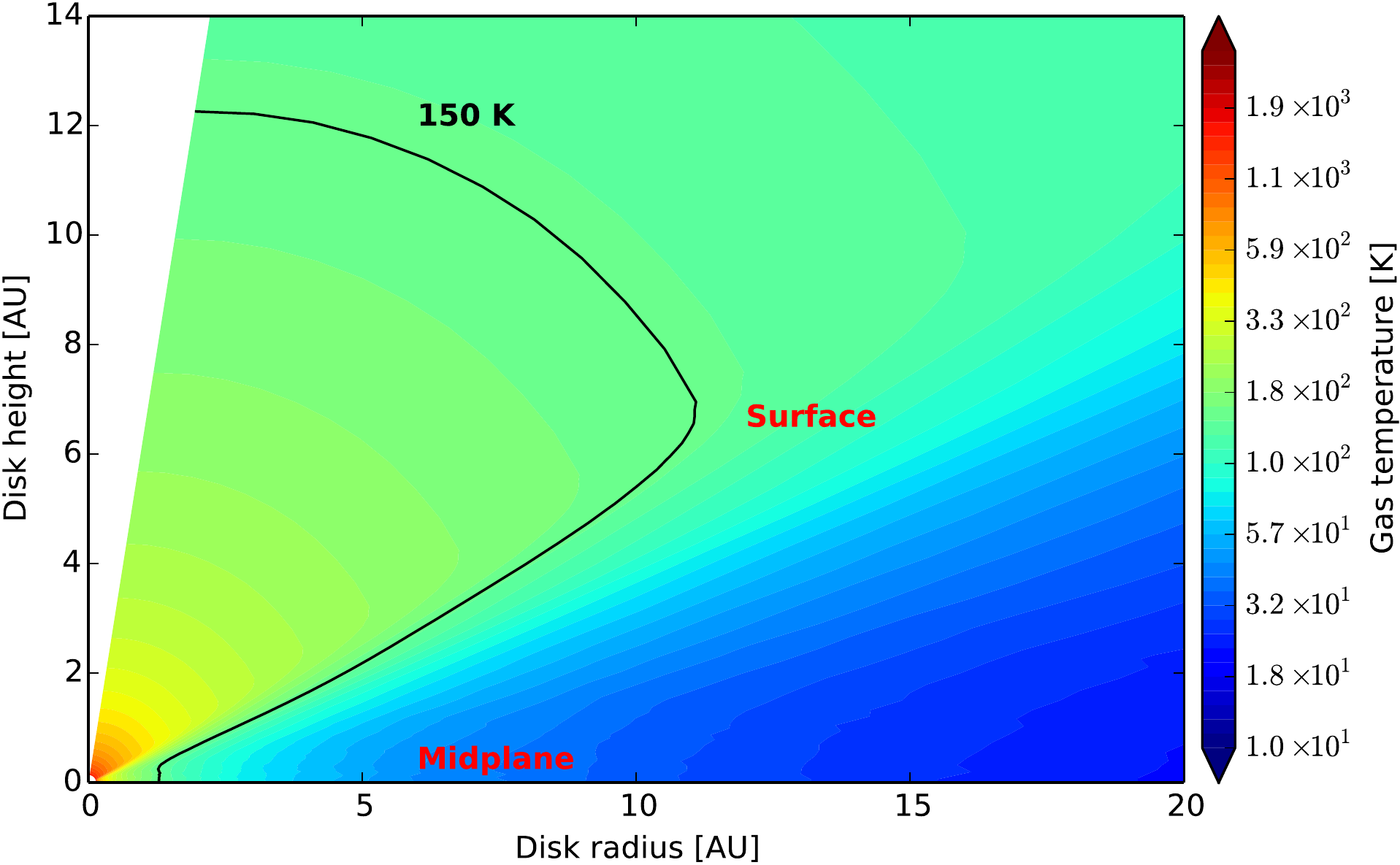}
\includegraphics[width=8.55cm]{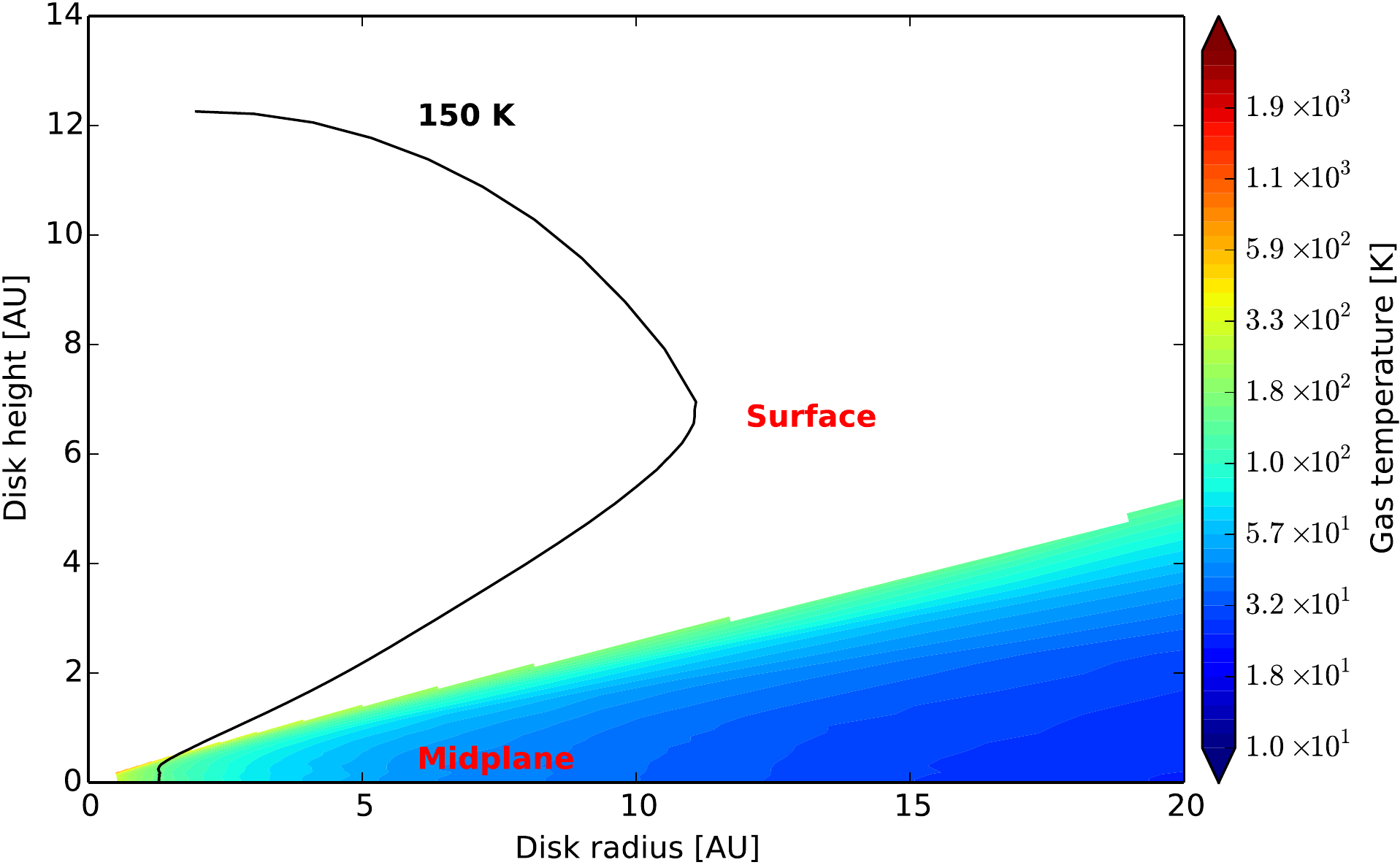}
\caption{Gas temperature structures for RNO 90. Left: Case I, with $T_{\rm gas}=T_{\rm dust}$. Right: Case II, scaling the gas temperature to that of \cite{Najita11}. The curve indicates the location of the 150\,K dust isotherm.}
\label{fig: gas_temperature}
\end{figure*}

\begin{figure*}[ht!]
\centering
\includegraphics[width=8.95cm]{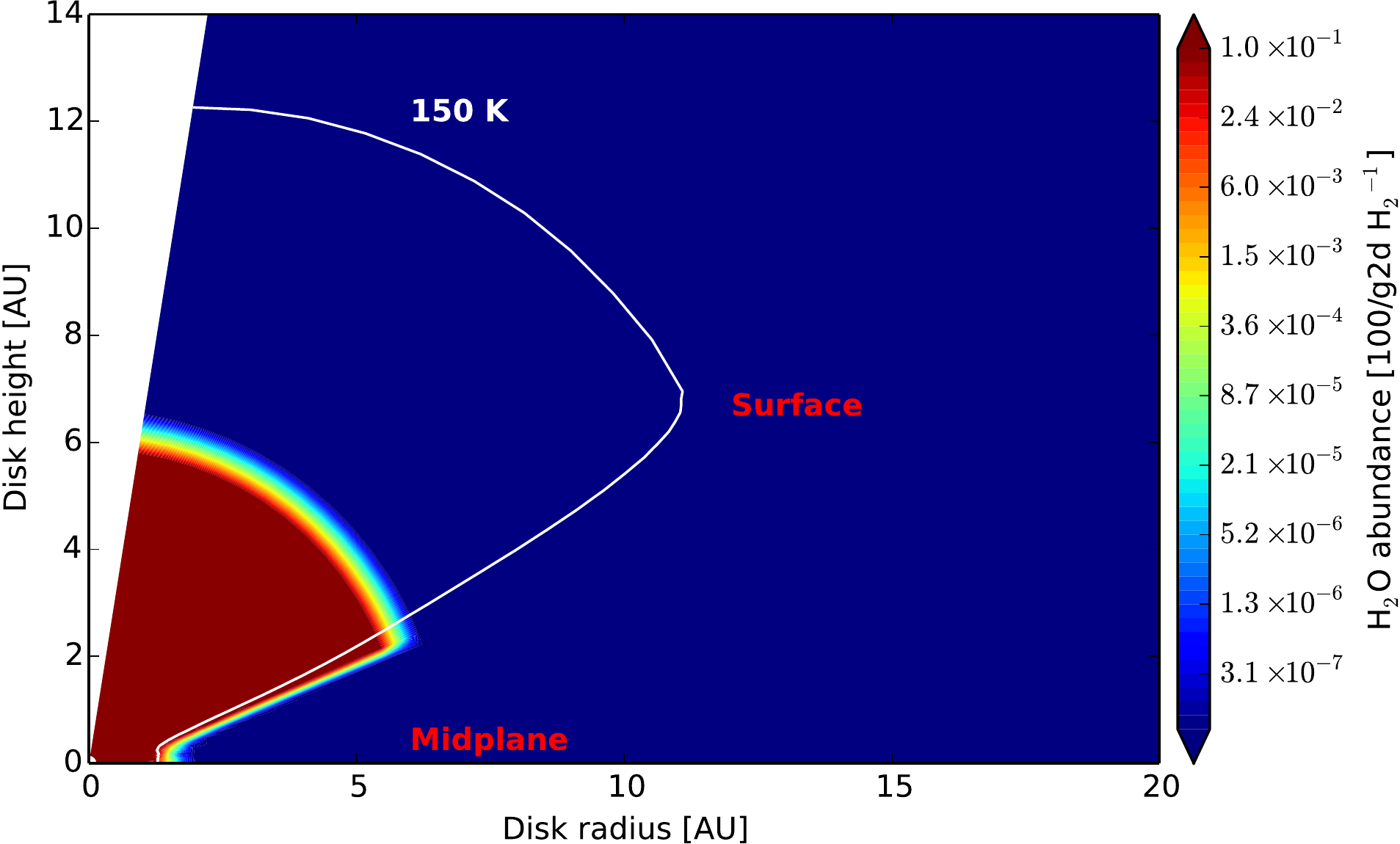}
\includegraphics[width=8.55cm]{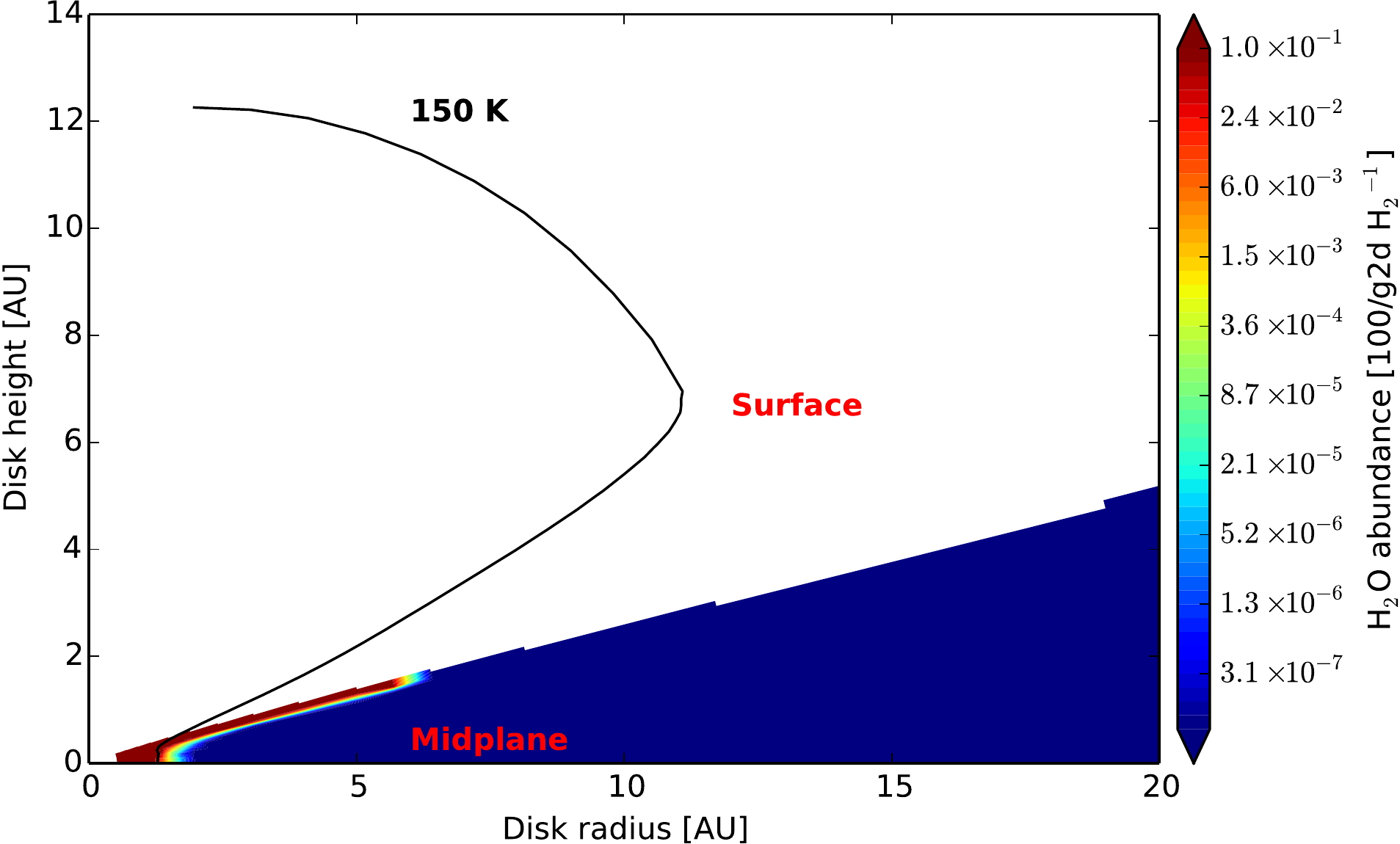}
\caption{Water abundance structures for RNO 90. Left: Case I, with temperature- and density-dependent water vapor abundance. The water abundance is set to a low value $X_{\rm min}$ at large radii, $R>R_{\rm crit}$. Right: The abundance structure of Case II. This is similar to Case I, with the addition that the water abundance is set to zero at low column densities, $N_{\rm H}<10^{21}\,\rm cm^{-2}$.}
\label{fig: gas_abundance}
\end{figure*}

\subsection{Water vapor abundance distribution} 
\label{sec: best_fits}

The objective of this study is to retrieve a parameterized water abundance structure for each disk, given all available water line fluxes. As a starting point, we use the density and temperature-dependent water freeze-out model from \cite{Meijerink09} and assume a constant ice+gas abundance $X_{\rm max}$. While the dust temperature is the parameter that, at the high gas densities in protoplanetary disks, primarily drives the gas/solid phase balance of water and other volatiles, there is also a weak density-dependence of the water vapor abundance. This basic model has a water vapor abundance that is high at temperatures $T\gtrsim 130-150\,$K and low elsewhere (see Figure \ref{fig: gas_abundance}). Since \textit{Herschel} observations have already demonstrated that even at low temperatures, the water vapor abundance is much higher than that expected for thermal desorption \citep{Oberg_h2o,Hogerheijde11}. Consequently, we additionally impose a minimum vapor abundance of $X_{\rm min}$ in the outer disk to account for the effects of non-thermal desorption mechanisms.

A fundamental property of the model is that, while the midplane water abundance drops by many orders of magnitude at the classical midplane snowline near $\sim 1\,$AU, the vapor abundance is high in the superheated disk surface out to $\sim 5-15\,$AU, depending on the stellar luminosity. For Case I, the radius at which optically thin dust cools below $\sim 150\,$K is the disk \textit{surface snow line} and is indicated by the 150\,K coupled gas and dust temperature isotherm. For Case II, the abundance model is necessarily more complex. The water vapor abundance is set to zero at low column densities, $N_{\rm H}\lesssim$10$^{21}$cm$^{-2}$, due to efficient photo-destruction, following the abundance structure of \cite{Najita11}. Conversely, a high abundance, even at low column densities, results in very strong water emission at temperatures in excess of 1000\,K, in clear contradiction with both data and the thermochemical models. However, we find that in this case the dust temperature is {\it below the freeze-out temperature at disk depths high enough to prevent photo-destruction of water beyond 1-2\,AU}. Since our aim is to retrieve the radii over which water is abundant, using the line data, we must allow for abundant water vapor in disk layers where the dust temperatures are as low as $T_{\rm dust}\sim60$\,K. This is illustrated in Figures \ref{fig: gas_temperature} and \ref{fig: gas_abundance}, which show the 150\,K dust isotherm as reference.  

We explore the possibility that the water vapor abundance drops at temperatures significantly higher than 150\,K. That is, we assume that the presence of cold dust always ensures that little water can remain in the gas phase while allowing for scenarios in which gas warmer than 150\,K may be dry. This is motivated by thermo-chemical models, which suggest that efficient gas-phase formation of water is triggered at relatively high temperatures \citep[200-300\,K,][]{Woitke09,Du14}, and the observational suggestion of \cite{Meijerink09} that the surface water abundance drops near 1\,AU, well within the surface snowline. Consequently, we define a critical radius $R_{\rm crit}<R_{\rm surface-snowline}$, beyond which the water abundance drops to $X_{\rm min}$ at all disk altitudes. If the water vapor abundance is driven by pure desorption from a constant reservoir of ice, $R_{\rm crit}$ would be expected to coincide with the surface snow line. Figure \ref{fig: gas_abundance} shows the resulting parameterized abundance structure for a case where $R_{\rm crit}$ is well inside the surface snow line. For both Case I and II, the free parameters being constrained are the inner ($X_{\rm max}$) and outer ($X_{\rm min}$) water abundances, and the critical radius ($R_{\rm crit}$).

While the amount of hydrogen present in the disks is not measured, the SED fit provides an estimate of the amount of dust in the upper disk layers. Assuming a gas-to-dust mass ratio leads to an estimate of the water abundance relative to hydrogen. However, since the gas-to-dust ratio is unknown, we report the water abundances relative to a canonical gas-to-dust ratio of 100. 

\subsection{Effects of gas-dust decoupling (Case II)}
\begin{figure}[ht!]
\centering
\includegraphics[width=8.5cm]{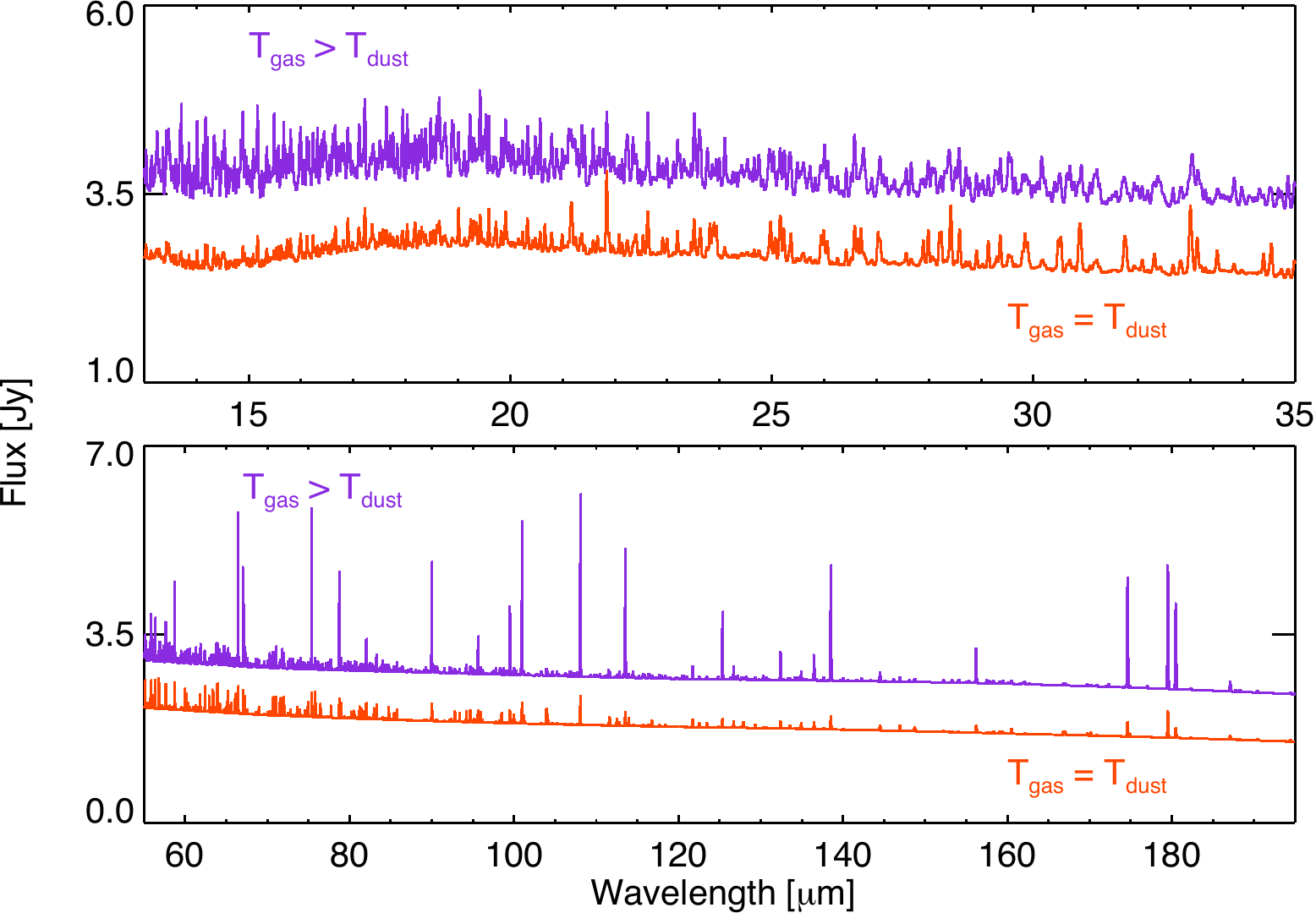}
\caption{Comparison of the RNO 90 Case I and Case II models for the same disk parameters: models generated with $T_{\rm gas}=T_{\rm dust}$, and with $T_{\rm gas}>T_{\rm dust}$ decoupling (offset for clarity) in the {\it Spitzer}-IRS (top) and {\it Herschel}-PACS (bottom) spectral ranges. The models shown have a critical radius of $R_{\rm crit}=7\,$AU, inner water abundance of $X_{\rm max}= 10^{-1}(\rm 100/g2d)\,\rm H_{2}^{-1}$, and outer water abundance of $X_{\rm min}=10^{-7} (\rm 100/g2d)\,\rm H_{2}^{-1}$.}
\label{fig: gd-decoup_line_comp}
\end{figure}

For Case II models the gas and dust temperatures are decoupled in warm molecular disk layers. In general we find, given identical water abundance parameters and the same critical radius, that water line strengths are strongly increased across both the {\it Spitzer}-IRS and {\it Herschel}-PACS spectral range. This is illustrated in Figure \ref{fig: gd-decoup_line_comp}, and suggests that the coupled case requires higher water abundances to reproduce the data.   

\begin{figure*}[ht!]
\centering
\includegraphics[width=4.45cm]{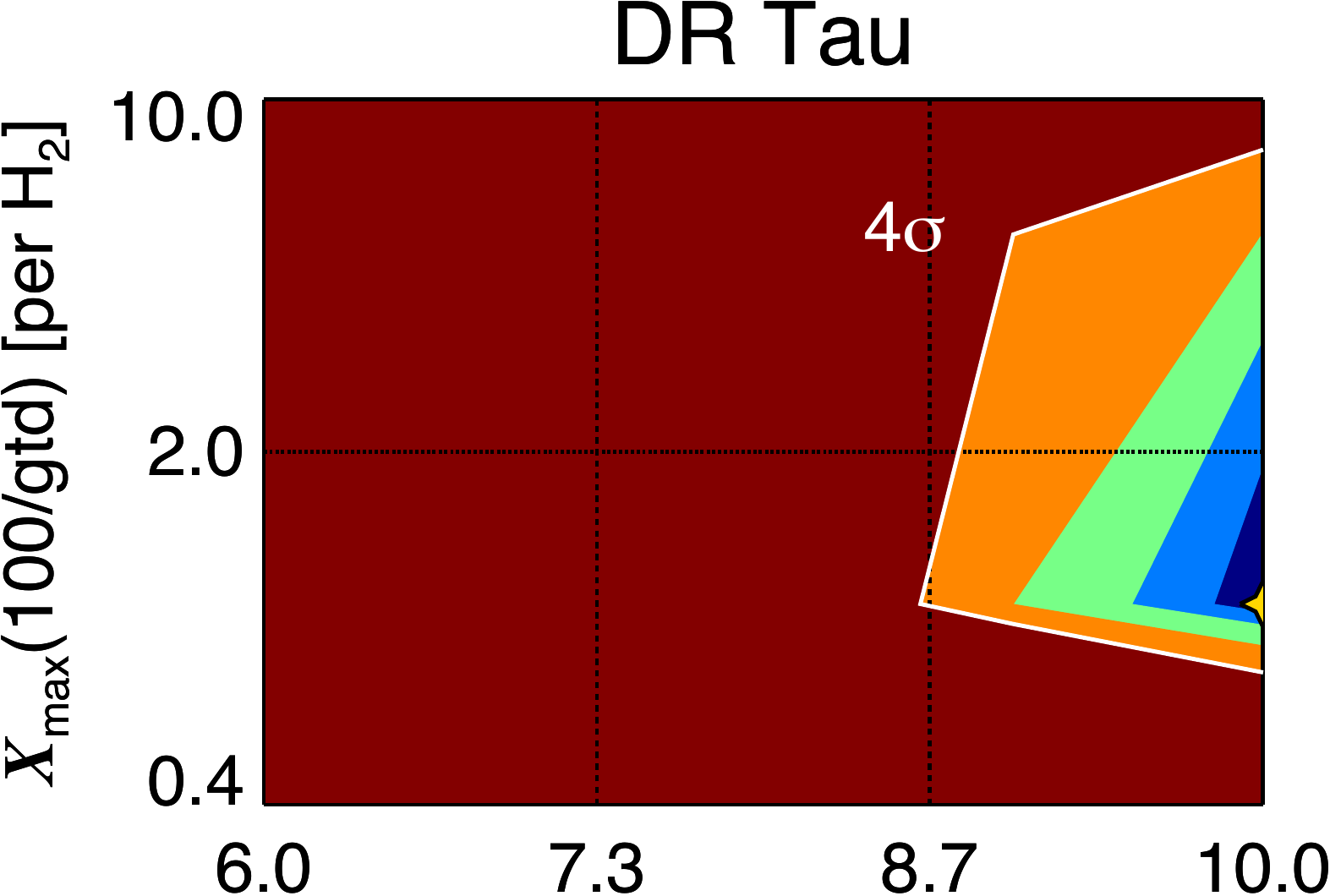}
\includegraphics[width=4.35cm]{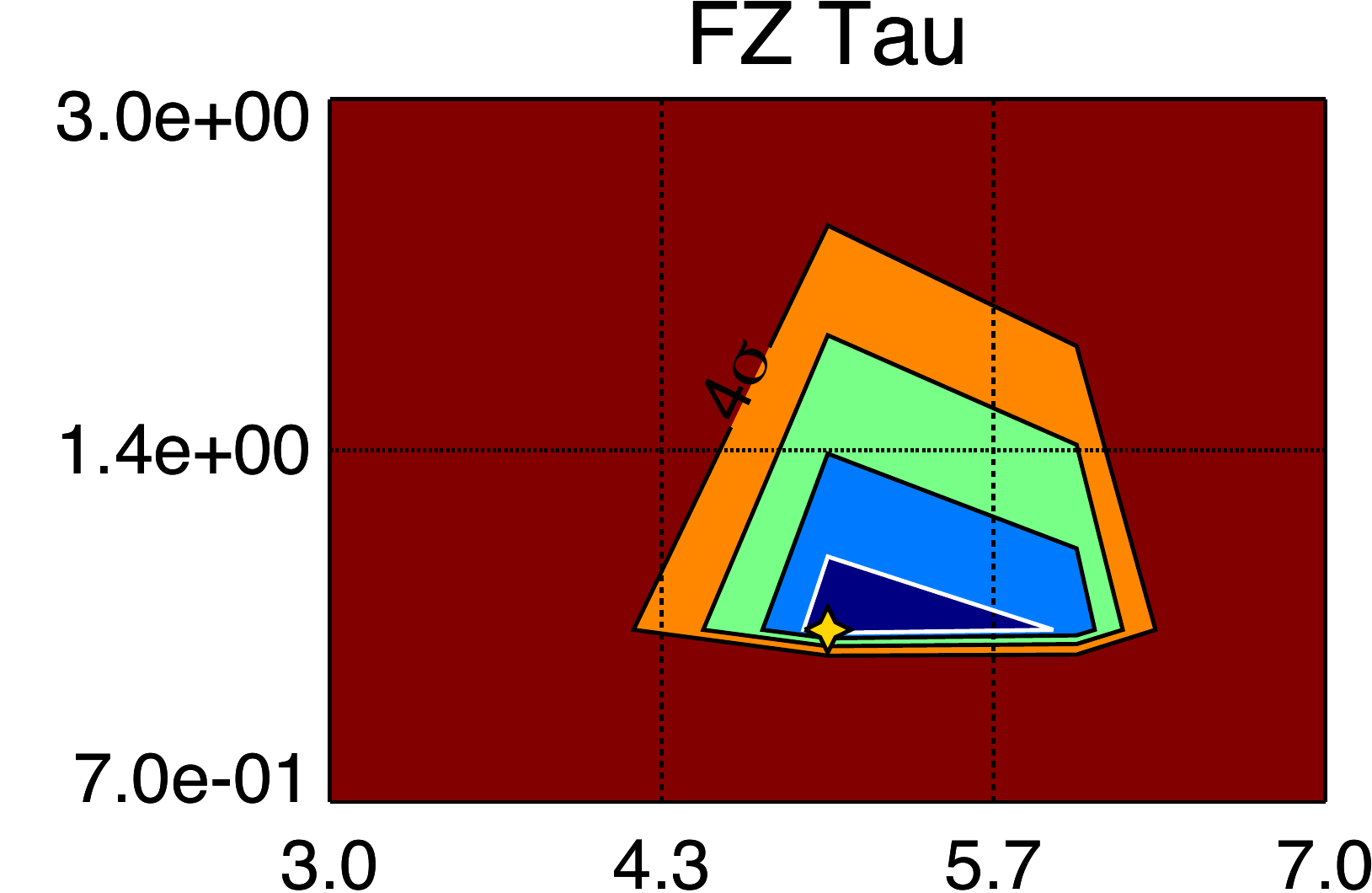}
\includegraphics[width=4.35cm]{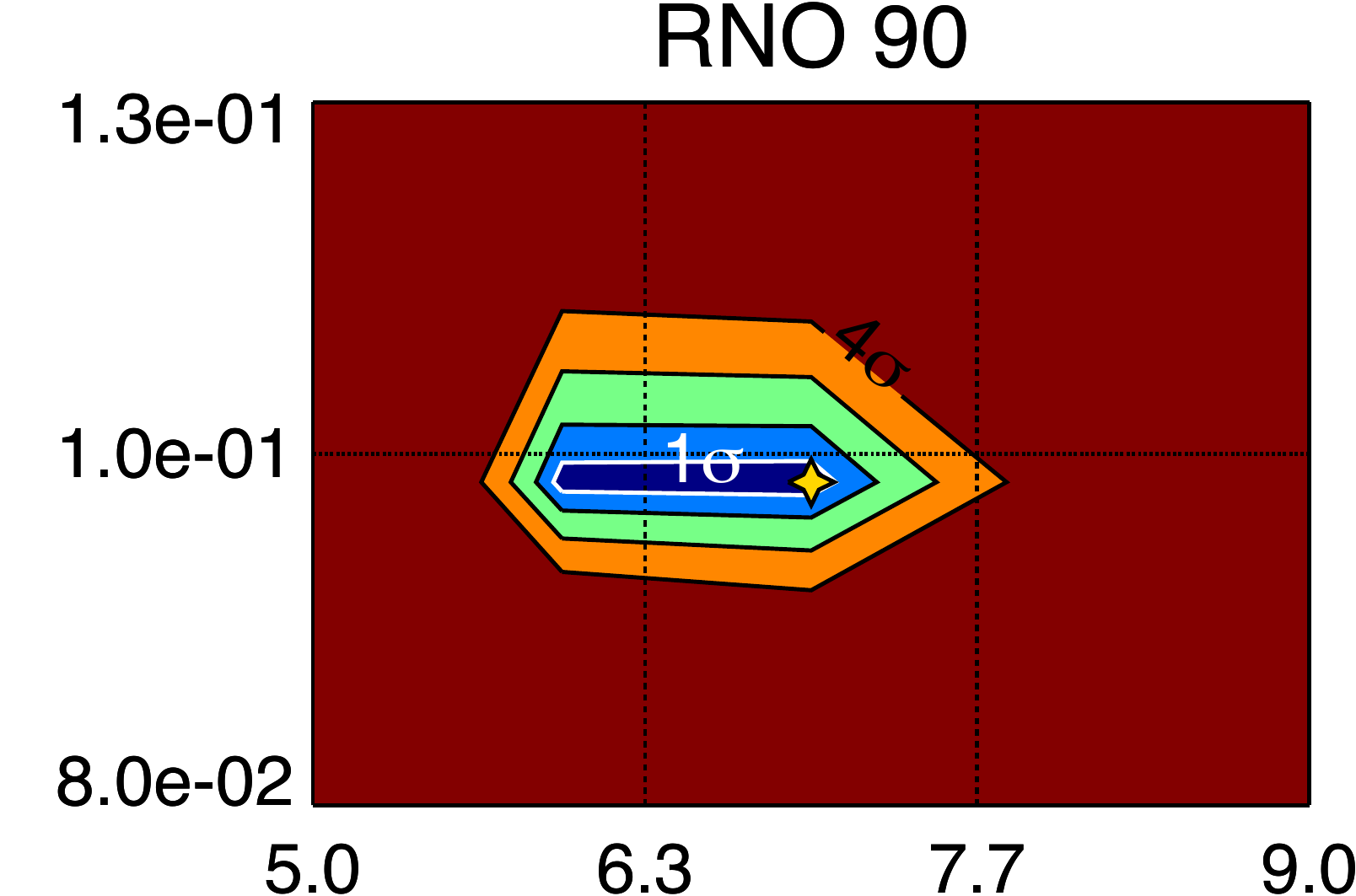}
\includegraphics[width=4.35cm]{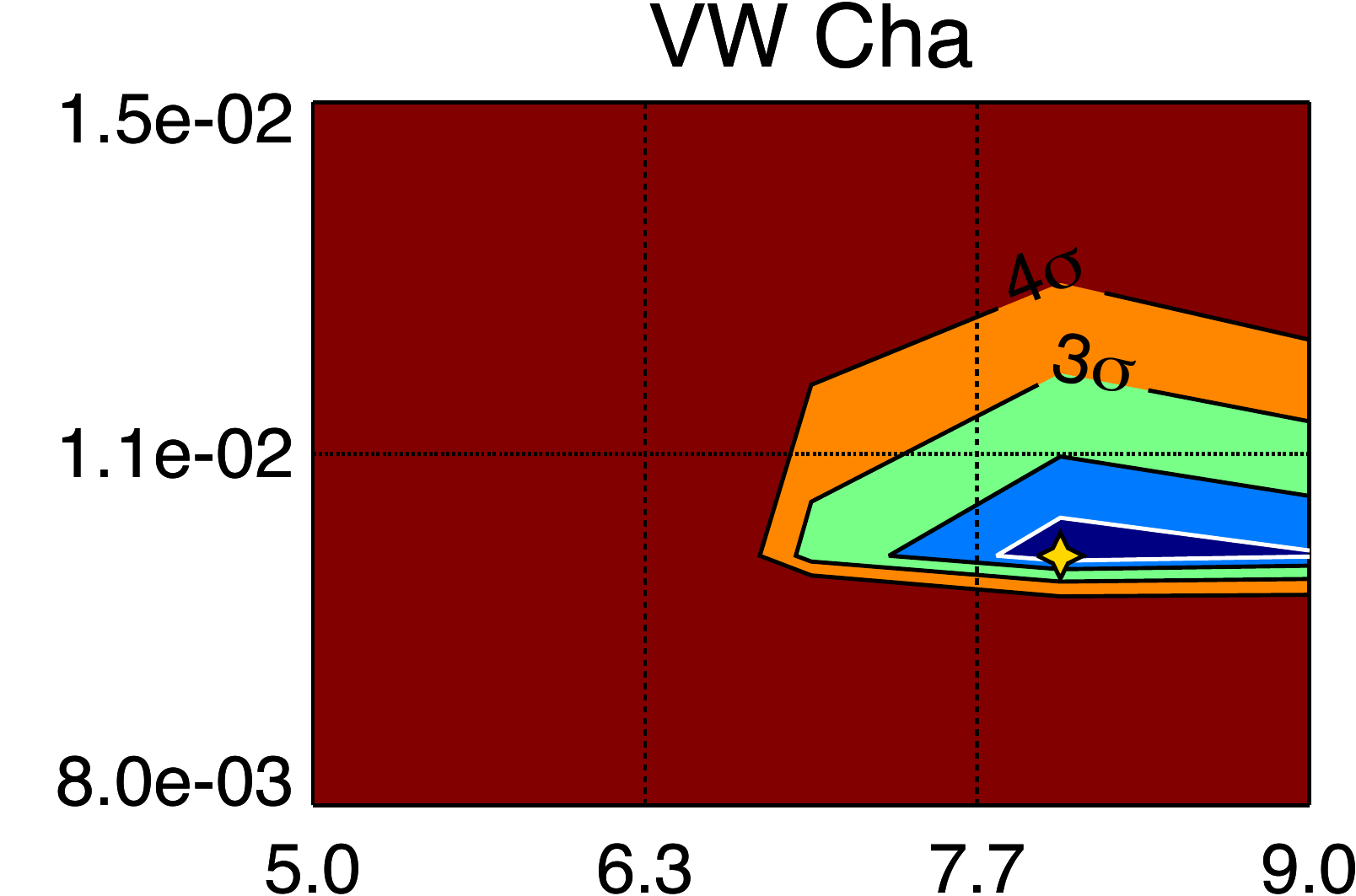}

\includegraphics[width=4.45cm]{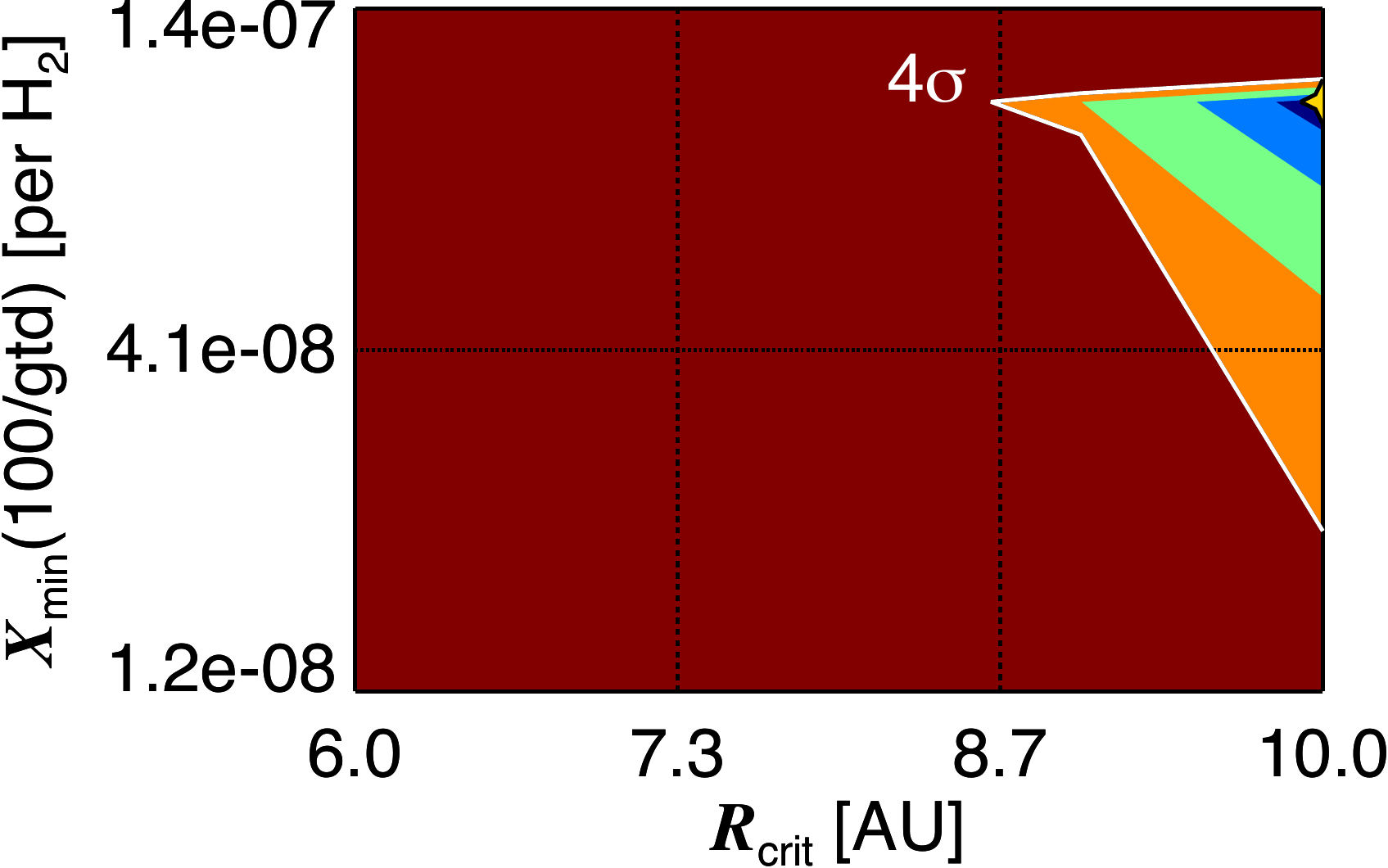}
\includegraphics[width=4.35cm]{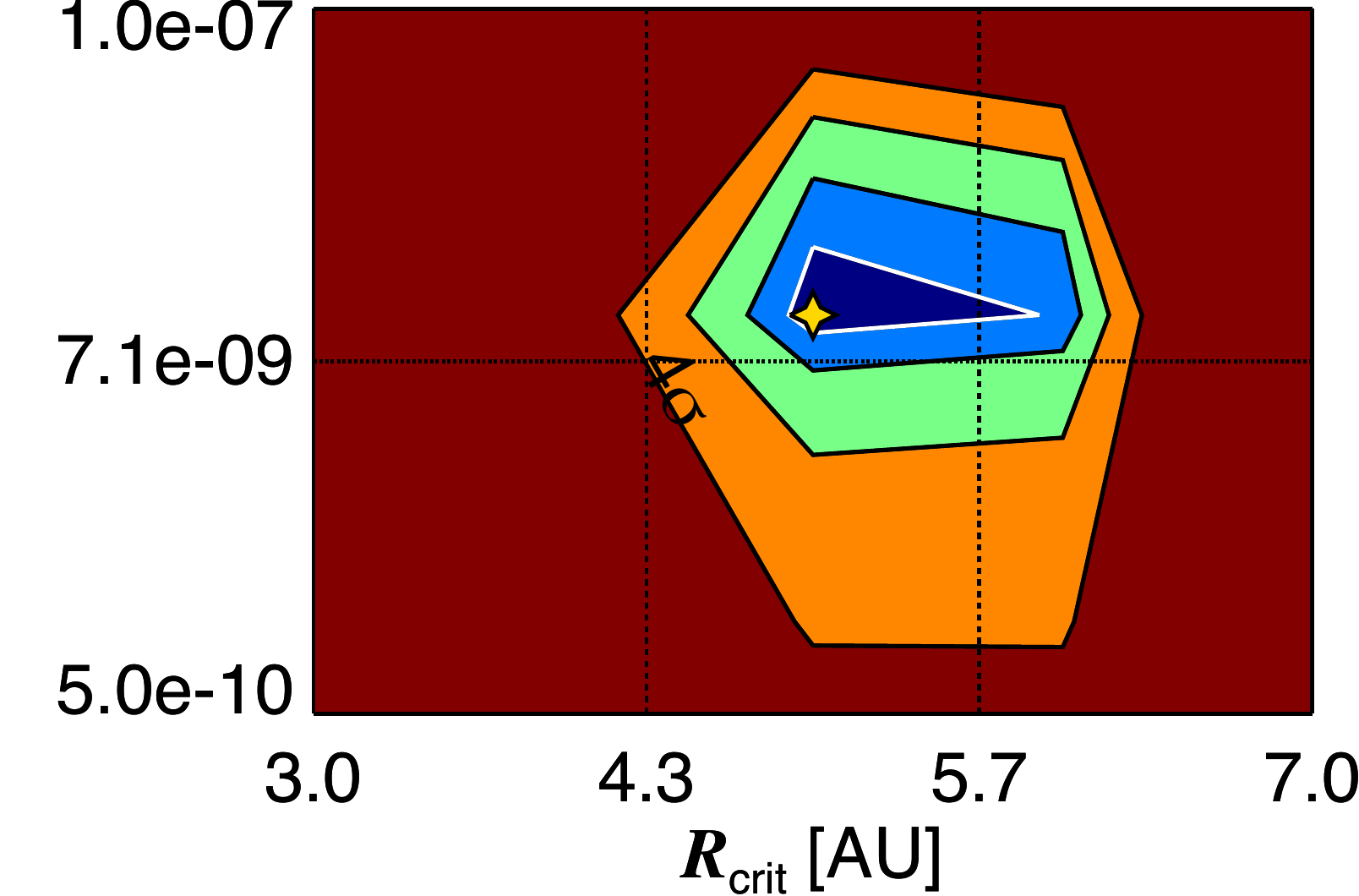}
\includegraphics[width=4.35cm]{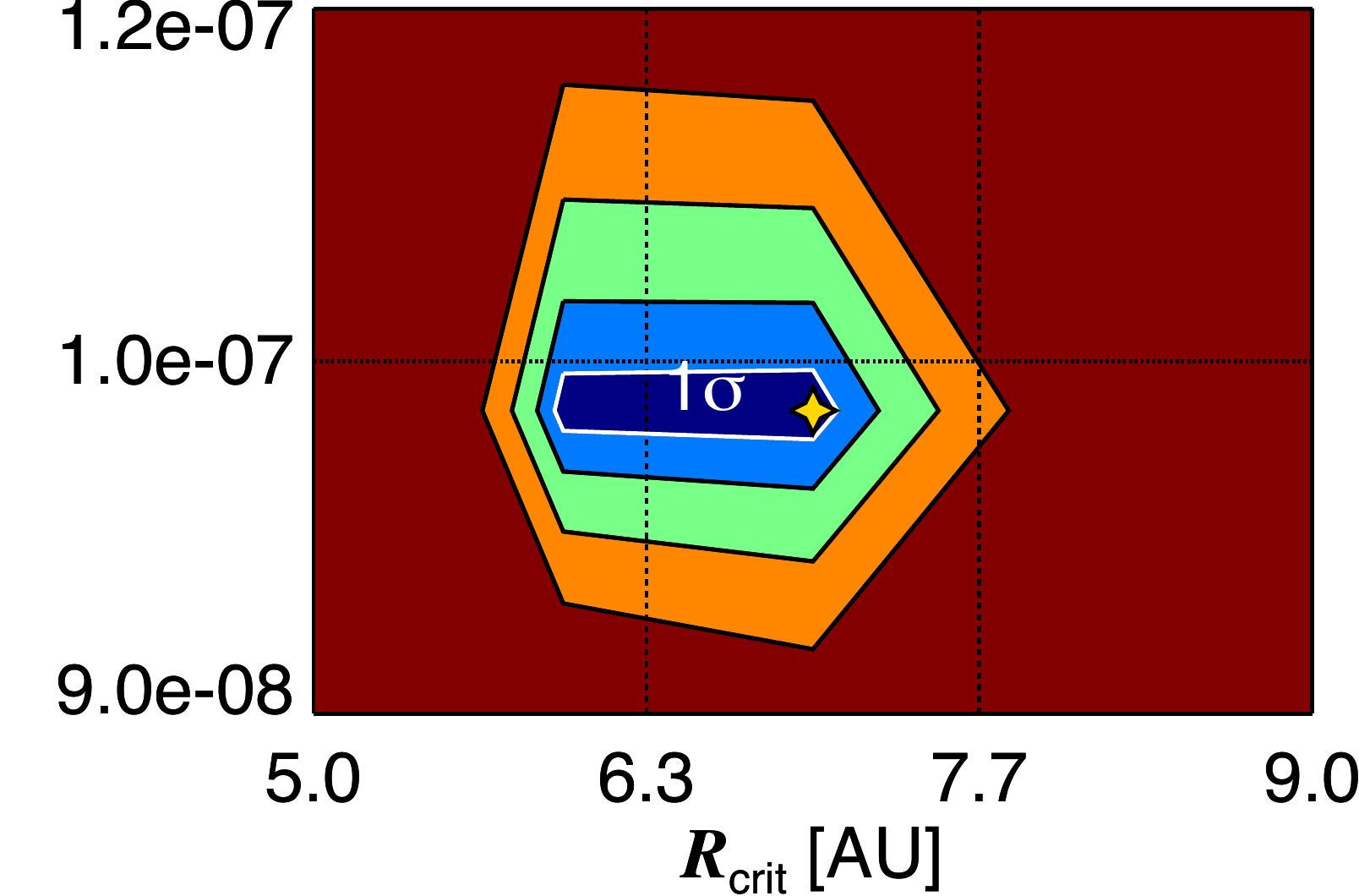}
\includegraphics[width=4.35cm]{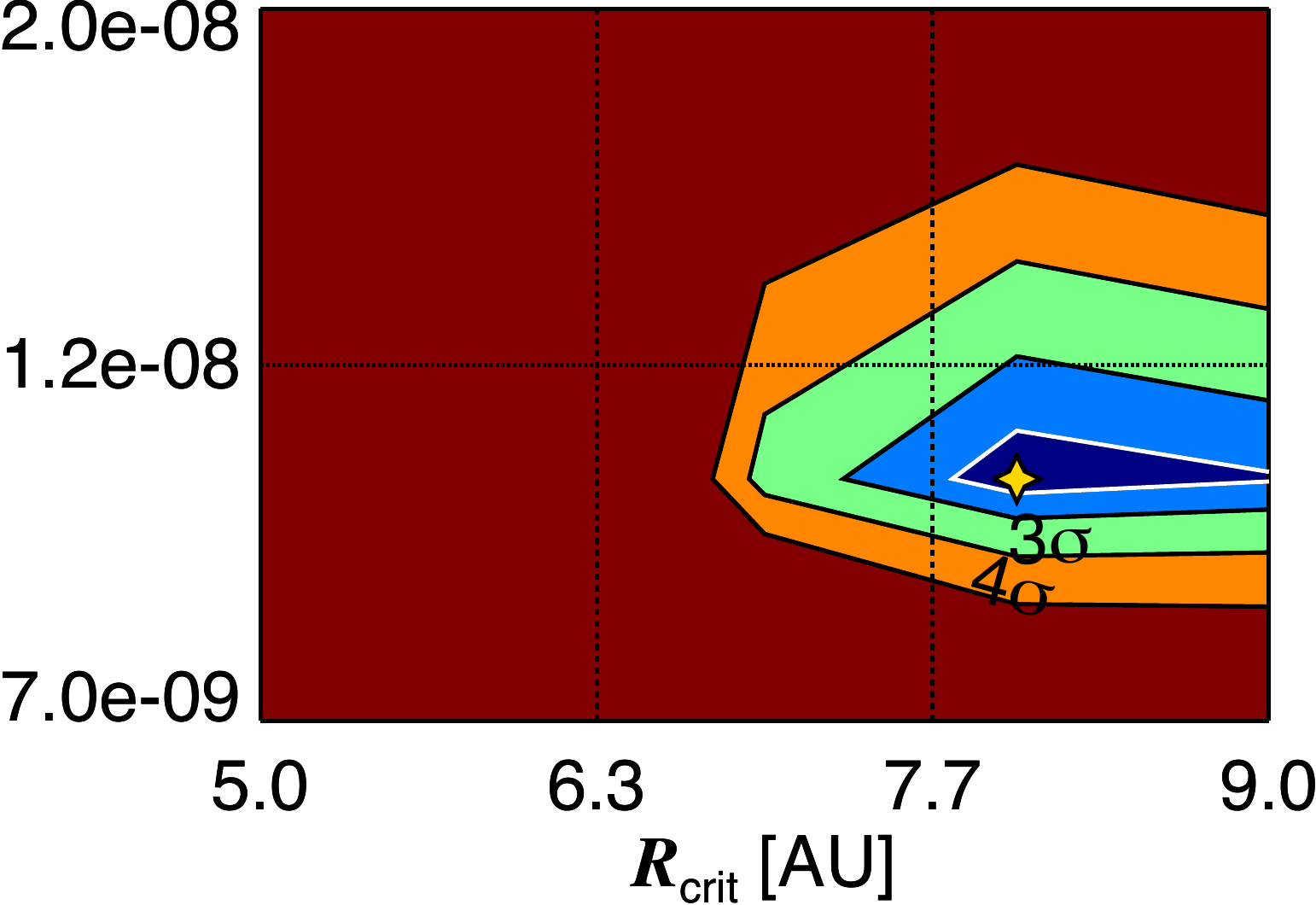}
\caption{Marginalized (projected) $\chi^{2}$ contours for each model grid generated with coupled gas and dust temperatures (Case I). All water abundance units are in (100/gtd) H$_{2}^{-1}$ (see Section \ref{sec: res_abund}).}
\label{fig: redchi_2D}
\end{figure*}

\begin{figure*}[ht!]
\centering
\includegraphics[width=4.55cm]{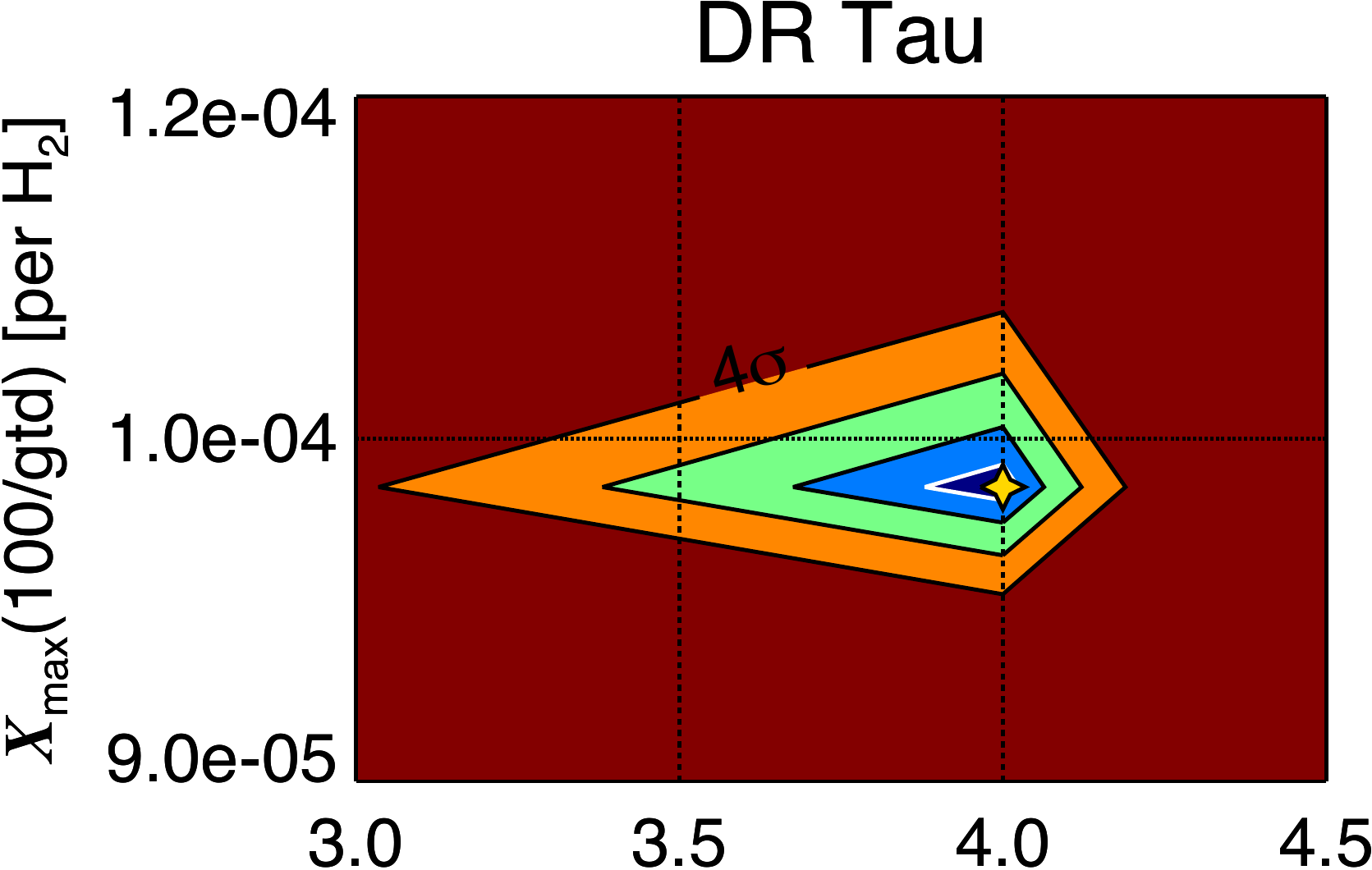}
\includegraphics[width=4.40cm]{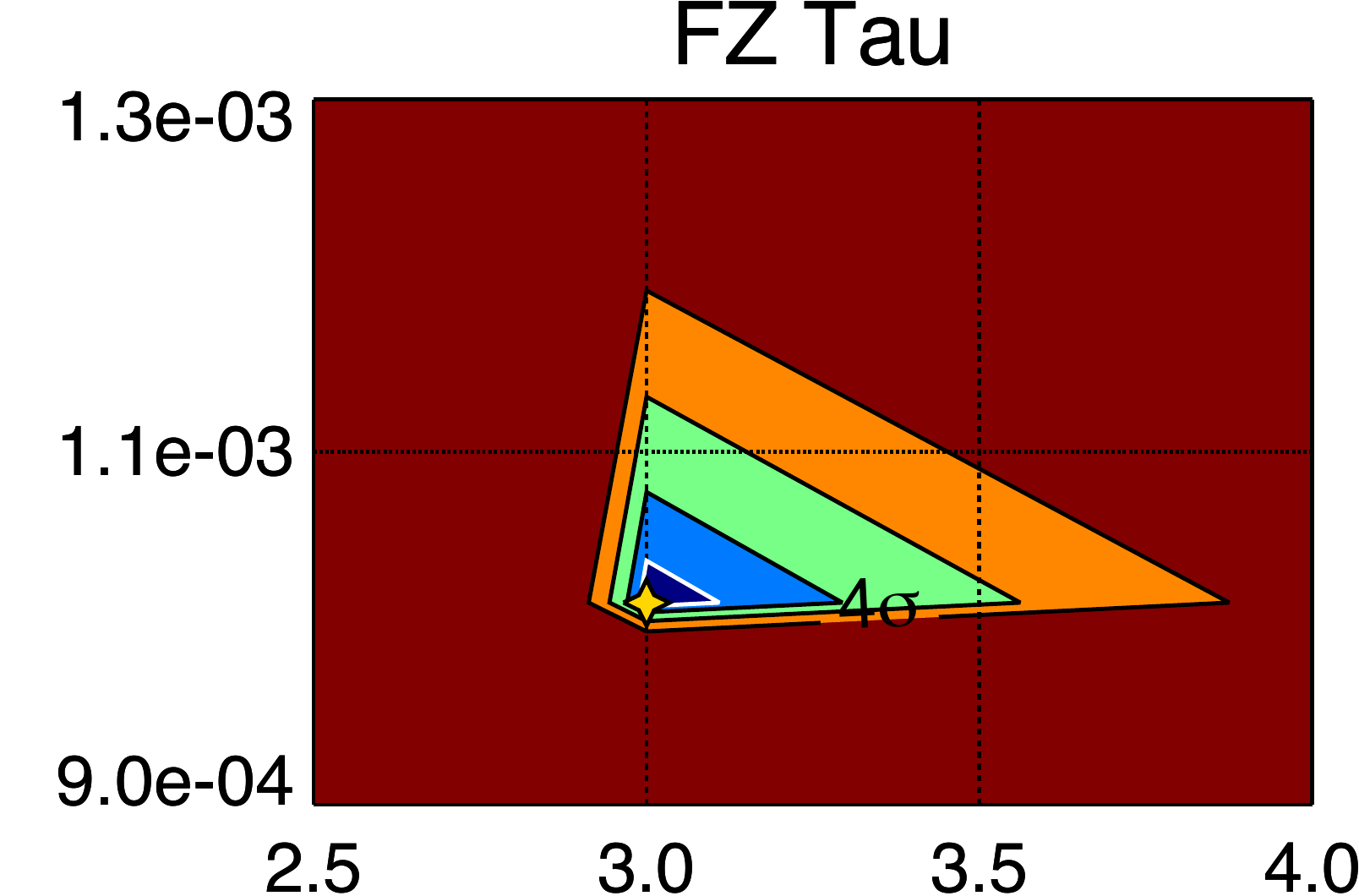}
\includegraphics[width=4.40cm]{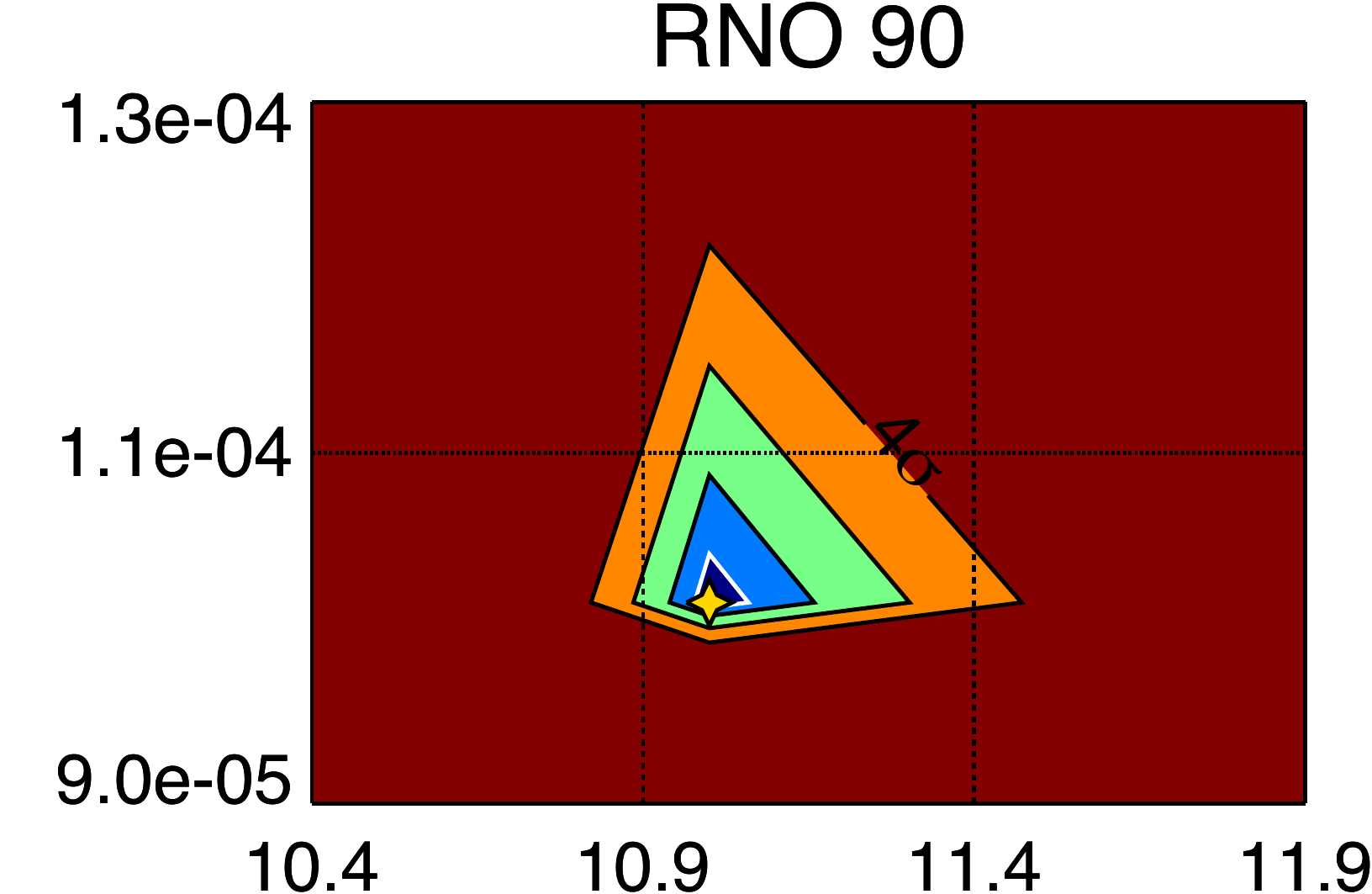}
\includegraphics[width=4.30cm]{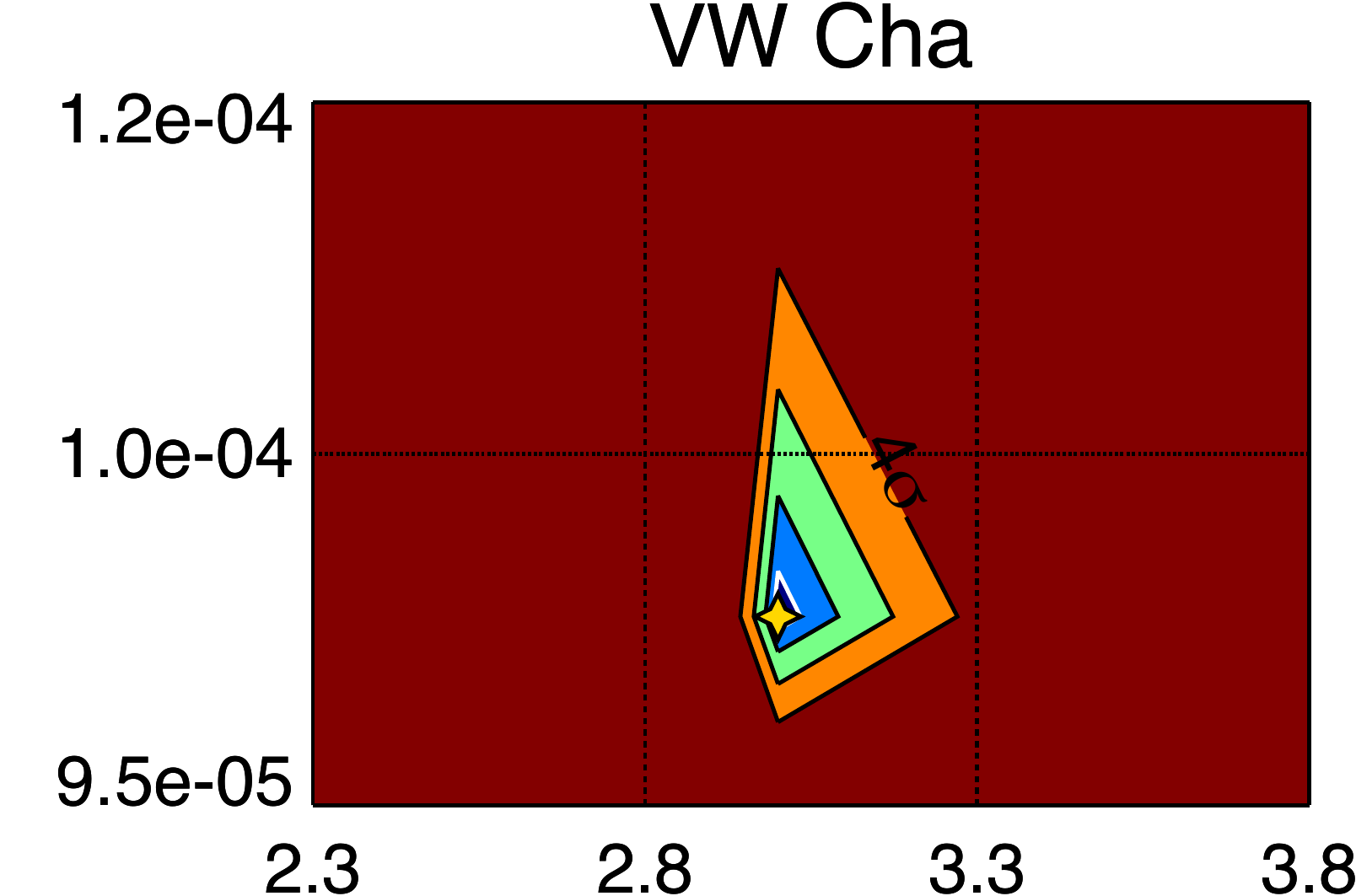}

\includegraphics[width=4.5cm]{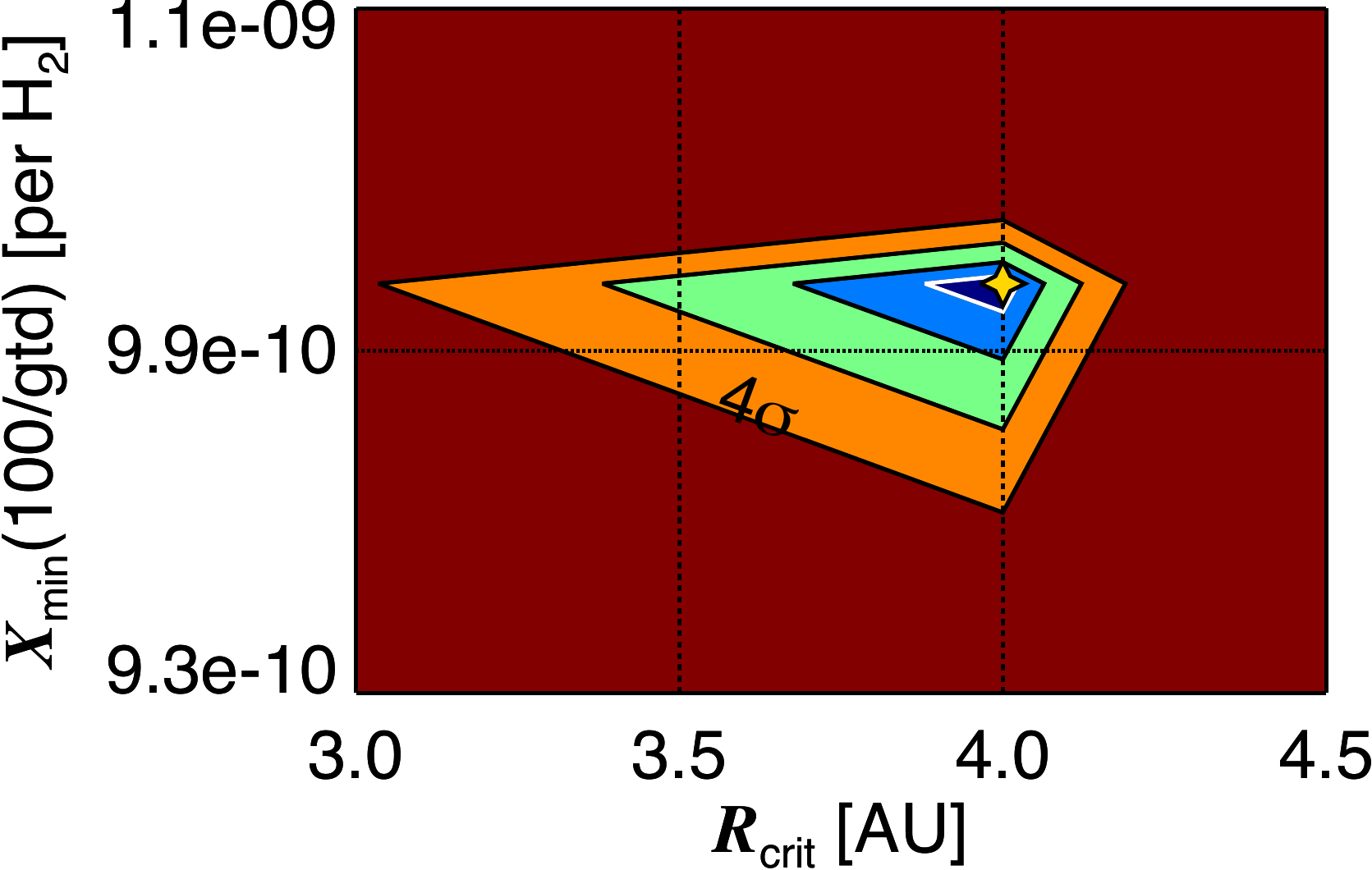}
\includegraphics[width=4.35cm]{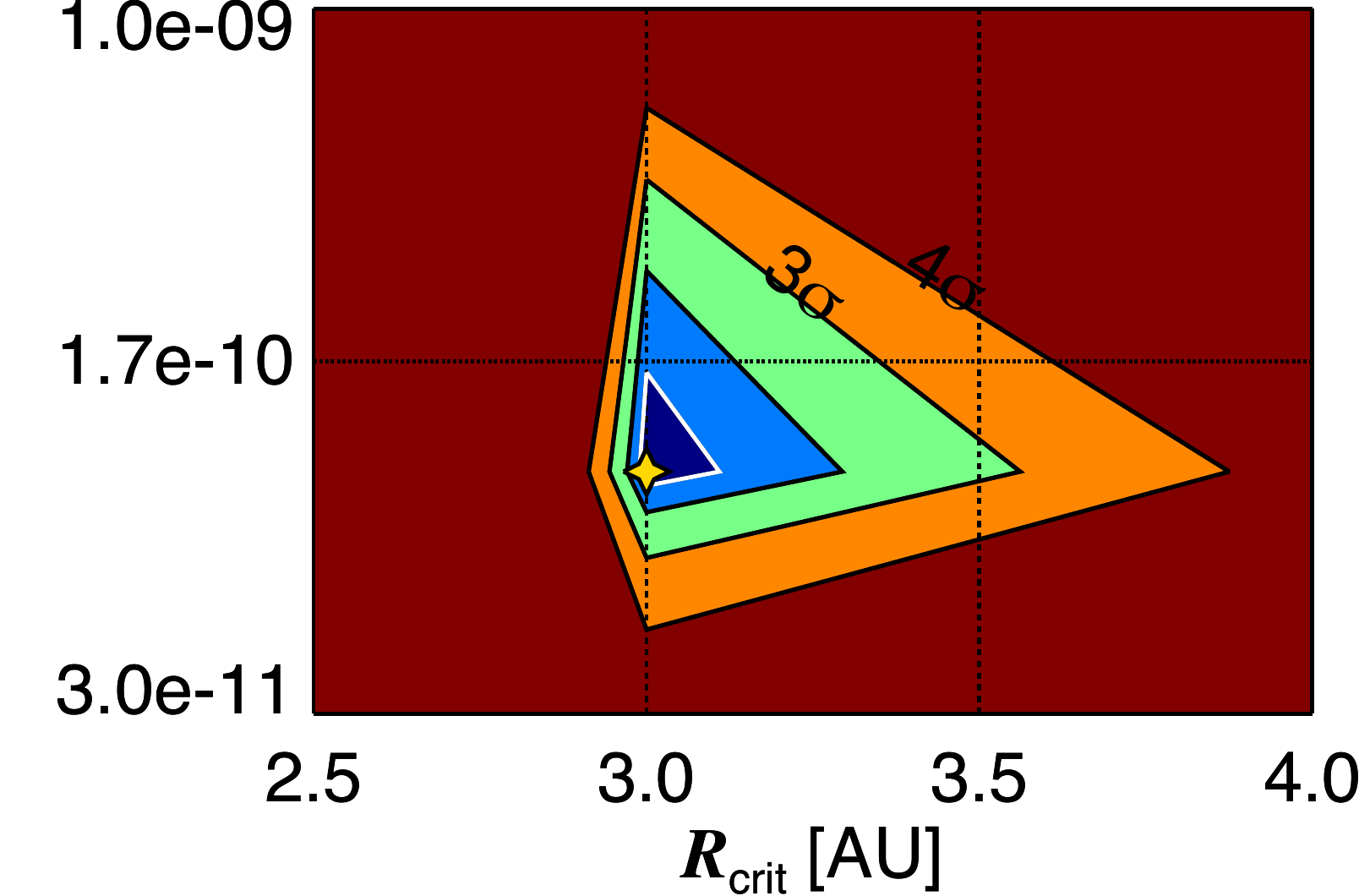}
\includegraphics[width=4.35cm]{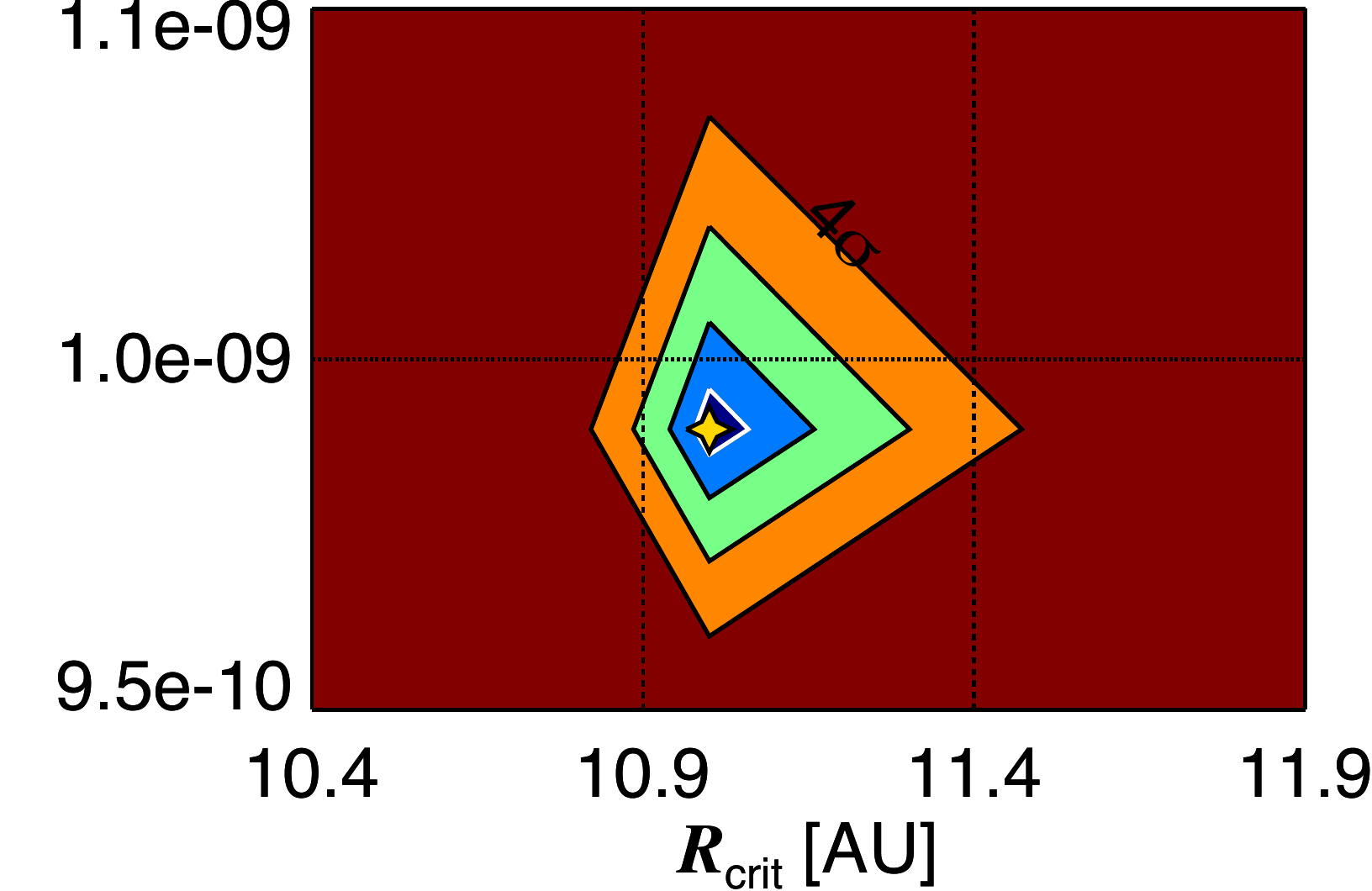}
\includegraphics[width=4.35cm]{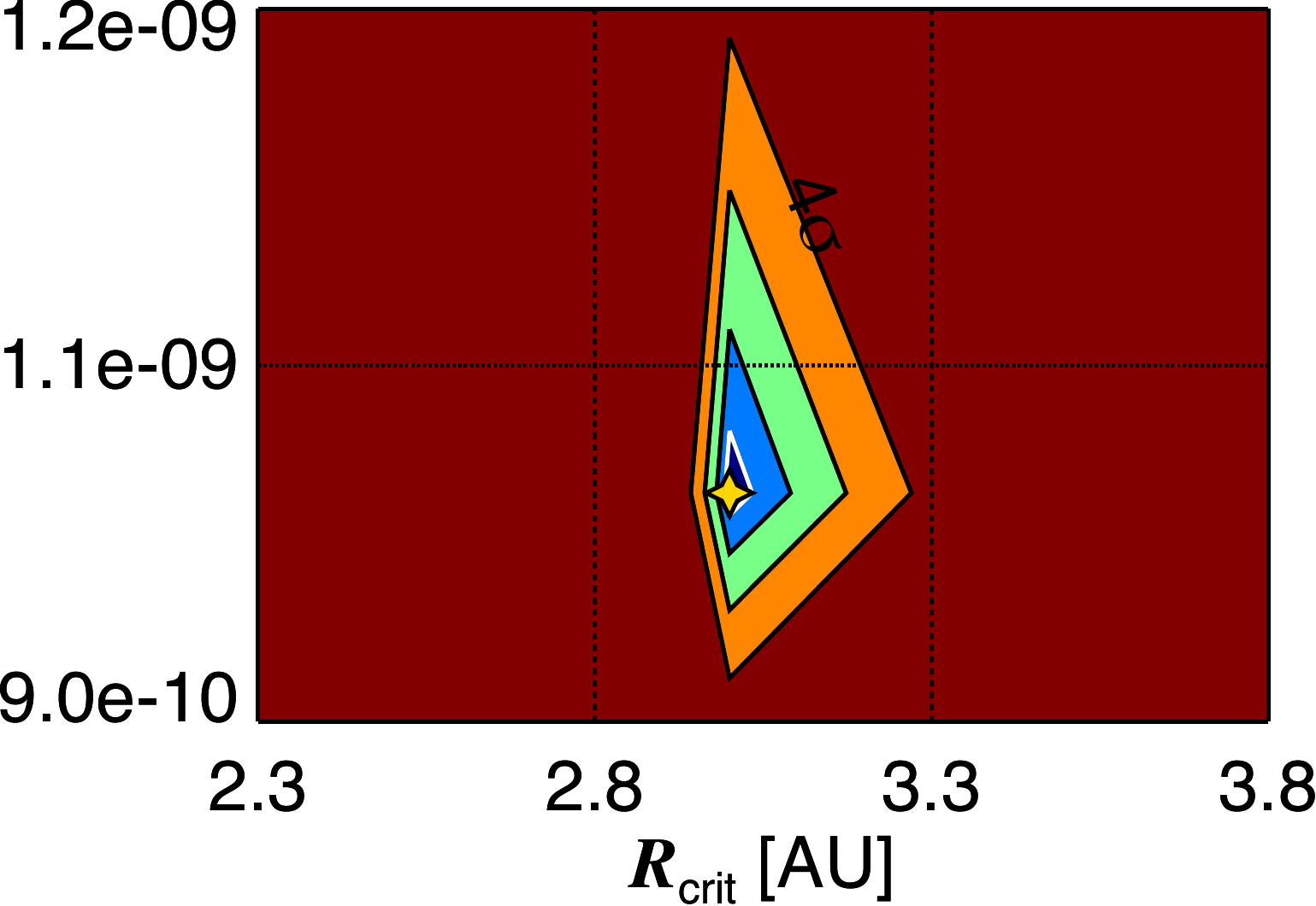}
\caption{Marginalized (projected) $\chi^{2}$ contours for each model grid generated with gas/dust temperature decoupling (Case II). All water abundance units are in (100/gtd) H$_{2}^{-1}$ (see Section \ref{sec: res_abund}).}
\label{fig: gd_redchi_2D}
\end{figure*}

\subsection{Model grid}
Grids of model \textit{Spitzer}-IRS and \textit{Herschel}-PACS spectra for Case I and II were created by varying the three free parameters. $R_{\rm crit}$ was sampled in 1 AU steps, while $X_{\rm max}$ and $X_{\rm min}$ were sampled in steps of 1 dex. Each model spectrum was convolved to the local spectral resolving power of the appropriate observation, and the resulting line complexes were integrated using the same procedure as for the observed spectra. The goodness-of-fit of the model relative to the data was evaluated using a least-squares measure. Figures \ref{fig: redchi_2D} and \ref{fig: gd_redchi_2D} show the $\chi^2$ surface projected onto two-dimensional marginalized space for Cases I and II, respectively. To determine the best fit parameters without being restricted to one of the grid values, we fit a polynomial to the $\chi^2$ curves in one-dimensional parameter space (Figure \ref{fig: redchi_1D_caseI} and \ref{fig: redchi_1D_caseII}). The 68\% confidence intervals are estimated using a $\Delta\chi^2$ step of 3.5, as appropriate for a three-dimensional parameter space. This leads to reduced $\chi^2$ values in the range 20--32 (see Table \ref{tab:results}). It is likely that a least-squares approach falls short for accounting for the full complexity of comparing two-dimensional dust-line radiative transfer models to high signal-to-noise infrared spectroscopy, but developing a more sophisticated method, such as a Bayesian Monte Carlo approach, is currently unfeasible, given the computational intensity of the models.

\begin{figure*}[ht!]
\centering
\includegraphics[width=16cm]{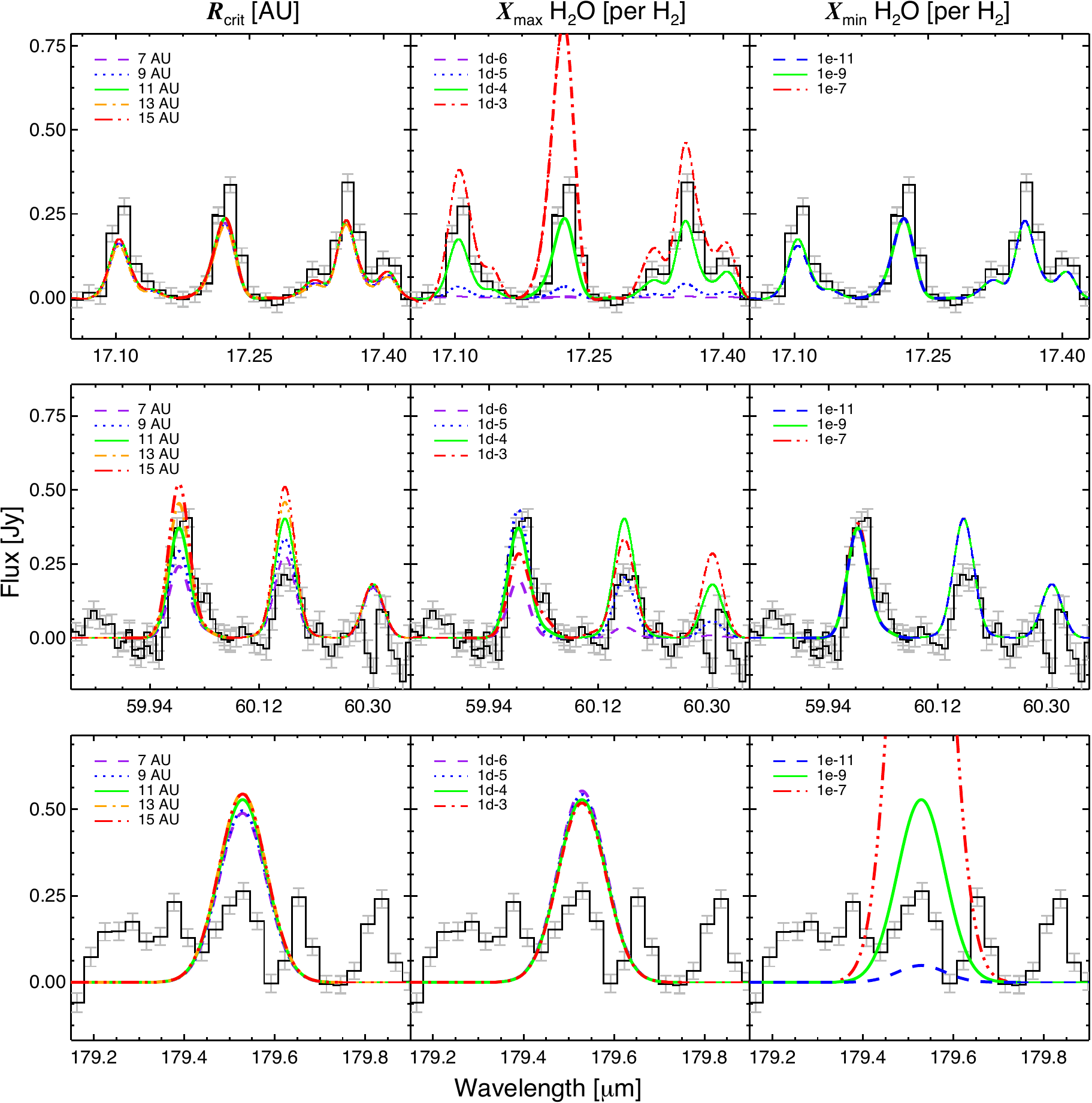}
\caption{Sensitivity of different water lines to variations in the water model parameters. The examples shown are for Case II modeling of RNO 90, with high-energy water transitions near 17.2\,$\mu$m ($E_{u} \sim 3000$\,K), intermediate-energy transitions near 60\,$\mu$m ($E_{u}$ $\sim 1000$\,K) and a low-energy transition near 179.5\,$\mu$m ($E_{u}\sim 100$\,K). Left column: Models for different values of $R_{\rm crit}$, with $X_{\rm in}$ and $X_{\rm out}$ are fixed at  $10^{-4}$ and $10^{-9} (\rm 100/g2d)\,\rm{\rm H}_{2}^{-1}$, respectively. Middle column: Models for different values of $X_{\rm in}$, with $R_{\rm crit}$ and $X_{\rm out}$ fixed at 11\,AU and $10^{-9}(\rm 100/g2d)\,\rm H_{2}^{-1}$, respectively. Right column: Models for different values of $X_{\rm out}$, with $R_{\rm crit}$ and $X_{\rm in}$ fixed at 11\,AU and $10^{-4} (\rm 100/g2d)\,\rm H_{2}^{-1}$, respectively.}
\label{fig:rno90_grid}
\end{figure*}

Figure \ref{fig:rno90_grid} illustrates the sensitivity of three representative water lines to different abundance parameters. Indeed, to constrain the full radial abundance structure, lines spanning the full range of upper level energies are needed. The higher energy lines near 17\,$\mu$m are sensitive to variations in the water abundance, insensitive to variations in the critical radius, and unaffected by variations in outer disk water abundance. The lower energy line near 179.5\,$\mu$m is sensitive to the outer disk water abundance but insensitive to the other two parameters. Intermediate energy lines near 60\,$\mu$m are sensitive to variations in the critical radius and inner water vapor abundances.

\section{Results}
\label{sec: results}
\subsection{Emitting regions of mid- and far-infrared water lines}

Emission lines from protoplanetary disks generally do not trace the cold midplane of the disk, but are formed in a layer closer to the surface due to the temperature inversion of the super-heated surface layer, line and dust opacity, or a combination of all these effects \citep{vanZadelhoff01, Glassgold04, Gorti04}. Since we search for a sharp decrease in water vapor abundance at a critical radius, it is important to consider which disk height corresponds to this critical radius. 

\begin{figure*}[ht!]
\centering
\includegraphics[width=14cm]{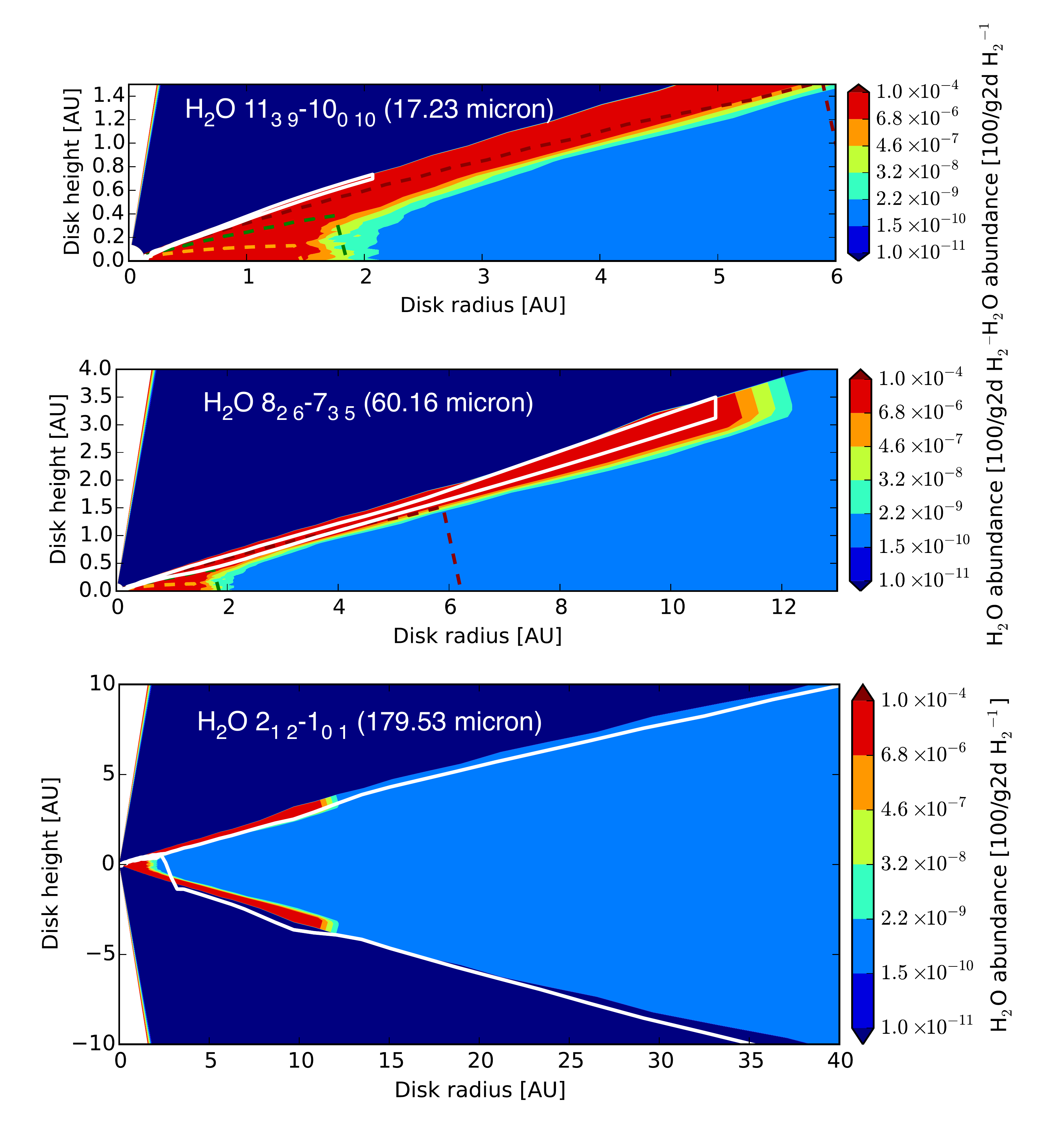}
\caption{The line emitting regions for the three representative water lines superimposed on the fractional water abundance for the best-fitting Case II RNO 90 model. The solid curve indicates the disk region responsible for between 10 and 90\% of the observed intensity in the line center. The dashed curves indicate the surfaces of water column densities of $10^{18}$, $10^{19}$ and $10^{20}\,\rm cm^{-2}$, respectively. The observer is viewing the disk from the top of the figure. }
\label{fig: line_surface}
\end{figure*}

The disk region traced by rotational water lines from warm gas is illustrated using three representative lines near 17, 60, and 179 $\mu$m (shown in Figure \ref{fig:rno90_grid}), spanning upper level energies from 2500 K to 114 K. The dominance of the surface layers in the line emission is illustrated in Figure \ref{fig: line_surface}. Here, the vertical layers responsible for emitting between 10 and 90\% of the specific intensity in the line centers are shown for three representative water lines. The line emitting areas are superimposed on contours showing the local water abundance. Also shown are the vertically integrated water column densities of $10^{18}$, $10^{19}$ and $10^{20}\,\rm cm^{-2}$ as functions of radius. These curves allow for direct comparison to the zero-dimensional slab models, which are commonly used in the literature for characterization of molecular emission. It is seen that the mid- to far-infrared water line emission indeed originates in the disk surface, and that they are not sensitive to the location of the midplane snow line. Beyond a few AU, the 179.5 micron line emitting region spans a much wider region because the disk dust is optically thin at these wavelengths. That is, line emission from both the near and far side of the disk contributes to the total line intensity.  

\begin{figure}[ht!]
\centering
\includegraphics[width=8.5cm]{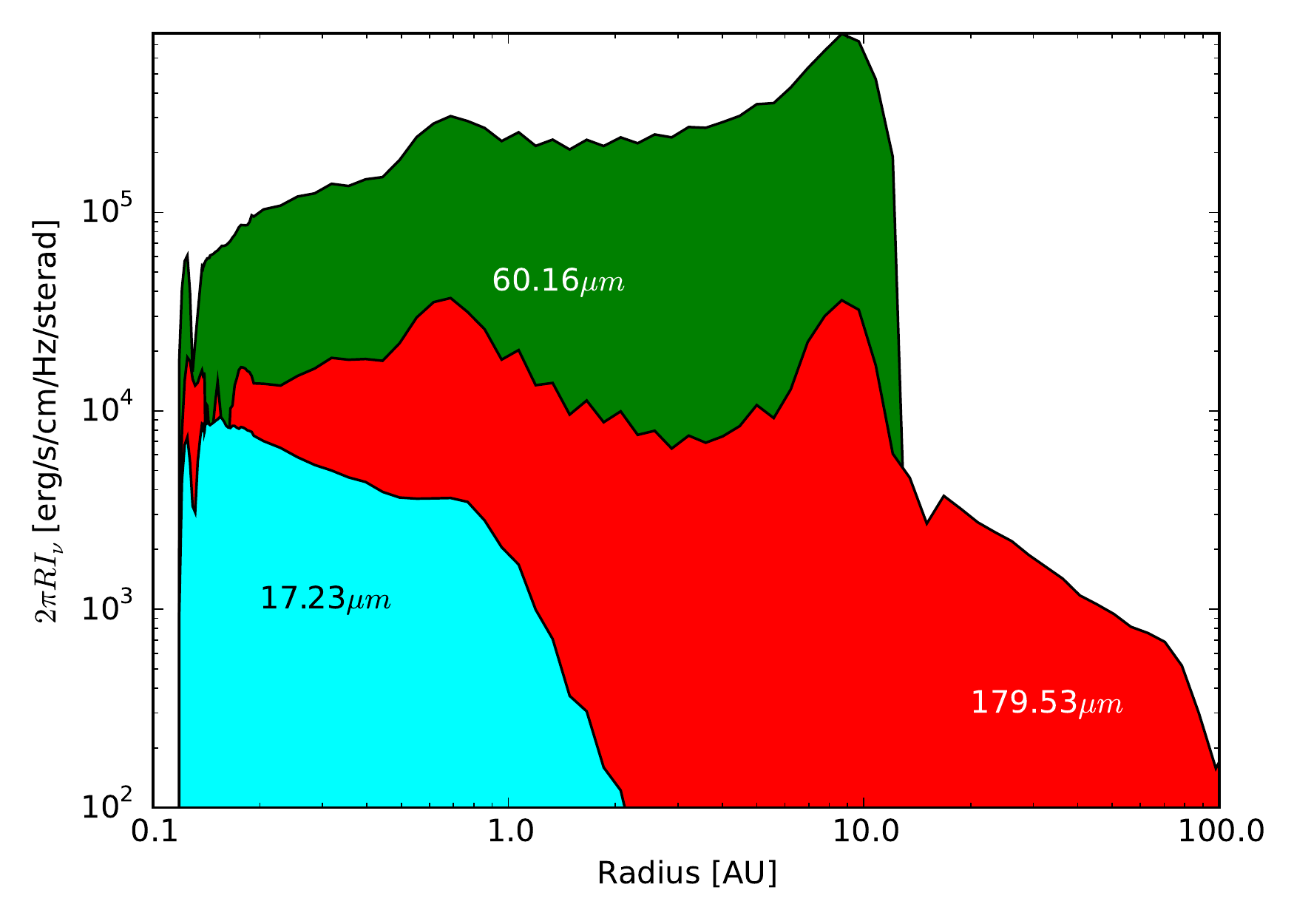}
\includegraphics[width=8.5cm]{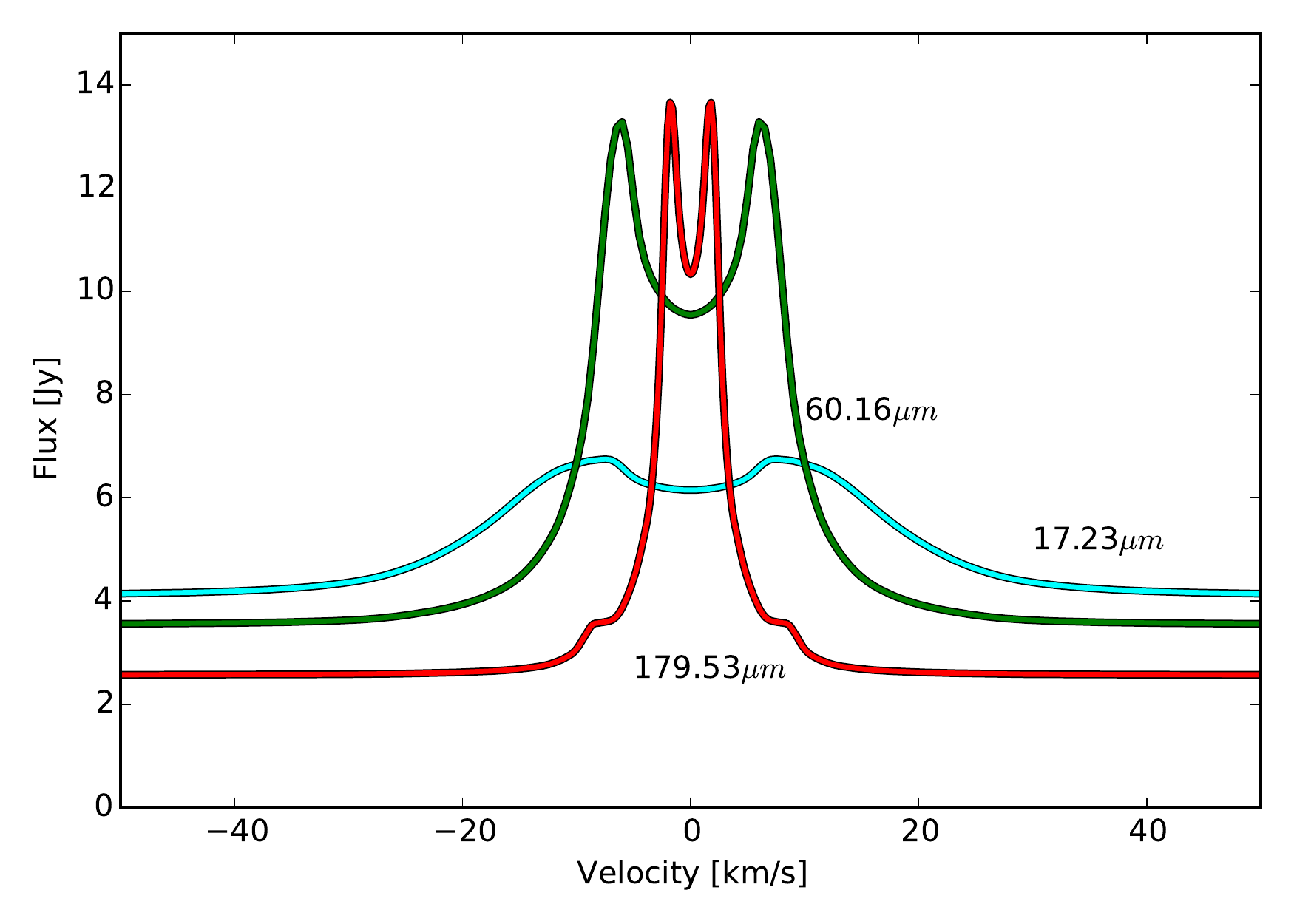}
\caption{Top: Radial specific intensity distribution for the best-fitting Case II RNO 90 model of the three reference water lines. The 17.23\,$\mu$m line is characterized by a sharp peak at the disk inner rim, and a range of strong emission out to $\sim$1\,AU, as is fairly typical for the \textit{Spitzer} range. The 60.16\,$\mu$m line traces 0.5 out to 7\,AU, where it is sharply truncated by the surface snow line. This sensitivity of the the short-wave PACS water lines to the surface distribution is also indicated by Figure \ref{fig:rno90_grid}. Bottom: The corresponding predicted line profiles to compare with the unresolved observed lines for this study.}
\label{fig: radial_intensity}
\end{figure}

In Figure \ref{fig: radial_intensity} the radial distribution of the specific line intensity is shown for the three representative water lines. The water line at 17.23\,$\mu$m, with an upper level energy of $\sim$2500\,K, traces radii from the inner rim of the disk out to 1\,AU, which is typical for lines in the \textit{Spitzer} range. The 60\,$\mu$m line with an upper level energy of $\sim$1500\,K traces a separate region of the disk, and increases in relative contribution until it is truncated by the critical radius. Since these two lines roughly bracket the range of upper level energies of the \textit{Spitzer} range, the RADLite models appear consistent with previously published single-temperature slab model fits to the \textit{Spitzer} range. For instance, \cite{Salyk11} report water emitting radii of 0.5-2.0\,AU, while \cite{Carr11} report radii of 0.8-1.5 for disks around typical solar-mass stars. The low-temperature lines, which the 179.5\,$\mu$m line represents, primarily trace the disk beyond the snow line where abundances are low. Indeed, although the outer disk water abundance is 5--6 orders of magnitude lower than the inner abundance, there is a significant contribution from the outer disk to the 179.5\,$\mu$m line. The general non-detection of this line implies very low outer disk water abundances \citep{Meeus12, Riviere-Marichalar12}. 

\subsection{Relative water abundance} 
\label{sec: res_abund}

Table \ref{tab:results} shows the best-fitting water abundance parameters for the Case I and Case II grids. For the model grid with coupled gas and dust temperatures, $T_{\rm dust}=T_{\rm gas}$ (Case I), we find that the best-fit inner disk water abundances are unrealistically high ($\ge 0.1\,\times (\rm 100/g2d)\,\rm H_2^{-1}$; see Figure \ref{fig: gd-decoup_line_comp}). Since these values are much higher than those of elemental oxygen, the Case I scenario would require that either the assumed gas-to-dust ratio, or another implicit assumption of the model, is incorrect. Increasing the gas-to-dust ratio to $10^4-10^5$ would be sufficient to bring the implied water vapor abundance down to the canonical value of $\sim 10^{-4}\,\rm H_2^{-1}$. However, since the dust mass is tied to data via the SED fit, increasing the gas-to-dust ratio uniformly at all radii would also increase the disk mass to implausibly high values of $\sim 1\,M_{\odot}$. Therefore, if Case I were true, significant increases in gas-to-dust ratio would apply only to the surface layers of the disk traced by the water line emission. 

Alternatively, allowing the gas temperature to increase, following a thermo-chemical calculation (Case II), leads to very different water abundances. In general, the Case II best-fit models have canonical inner disk water abundances, even for a gas-to-dust ratio of 100 ($\sim 10^{-4} (\rm 100/g2d)\,\rm H_2^{-1}$; see Table \ref{tab:results}).  The retrieved inner disk water abundances for the two cases suggest that the gas-to-dust ratio is either very high, or that the gas temperature is effectively decoupled. If both effects are significant, it would suggest that the water abundance is lower than canonical. 

The outer fractional water vapor abundances are representative of the cooler \textit{Herschel} water tracing the outer disk surface beyond the critical radius. Similar to the inner disk abundances, the outer disk abundances are lower for the Case II grid than the Case I grid, by 1-2 orders of magnitude. The Case II abundances of $10^{-9}-10^{-10}$ H$_2^{-1}$ are roughly consistent with previous estimates of outer disk water abundances \citep{Hogerheijde11}.

\subsection{Location of the surface snow line} 
\label{sec: res_snowline}

The model grids generally identify a best-fitting critical radius, $R_{\rm crit}$, at which the surface abundance of water vapor decreases by orders of magnitude. $R_{\rm crit}$ is located in the range 3-10\,AU, although there is a tendency for the Case II best-fit $R_{\rm crit}$ to be smaller than the Case I radii. This is plausible, since Case II is able to produce more warm gas for a given $R_{\rm crit}$, driving the parameter to smaller values. However, this effect is much less apparent than the strong decrease in fractional water abundance at all radii for Case II. Since the Case II models are more representative of the current theoretical understanding of the disk thermo-chemistry, and since there are many independent lines of evidence suggesting gas-dust decoupling in protoplanetary disk surface, we consider the Case II grid the best estimator of the critical radius.

\subsection{Comparison to theoretical midplane and surface snow lines}
\begin{figure}[ht!]
\includegraphics[width=8.5cm]{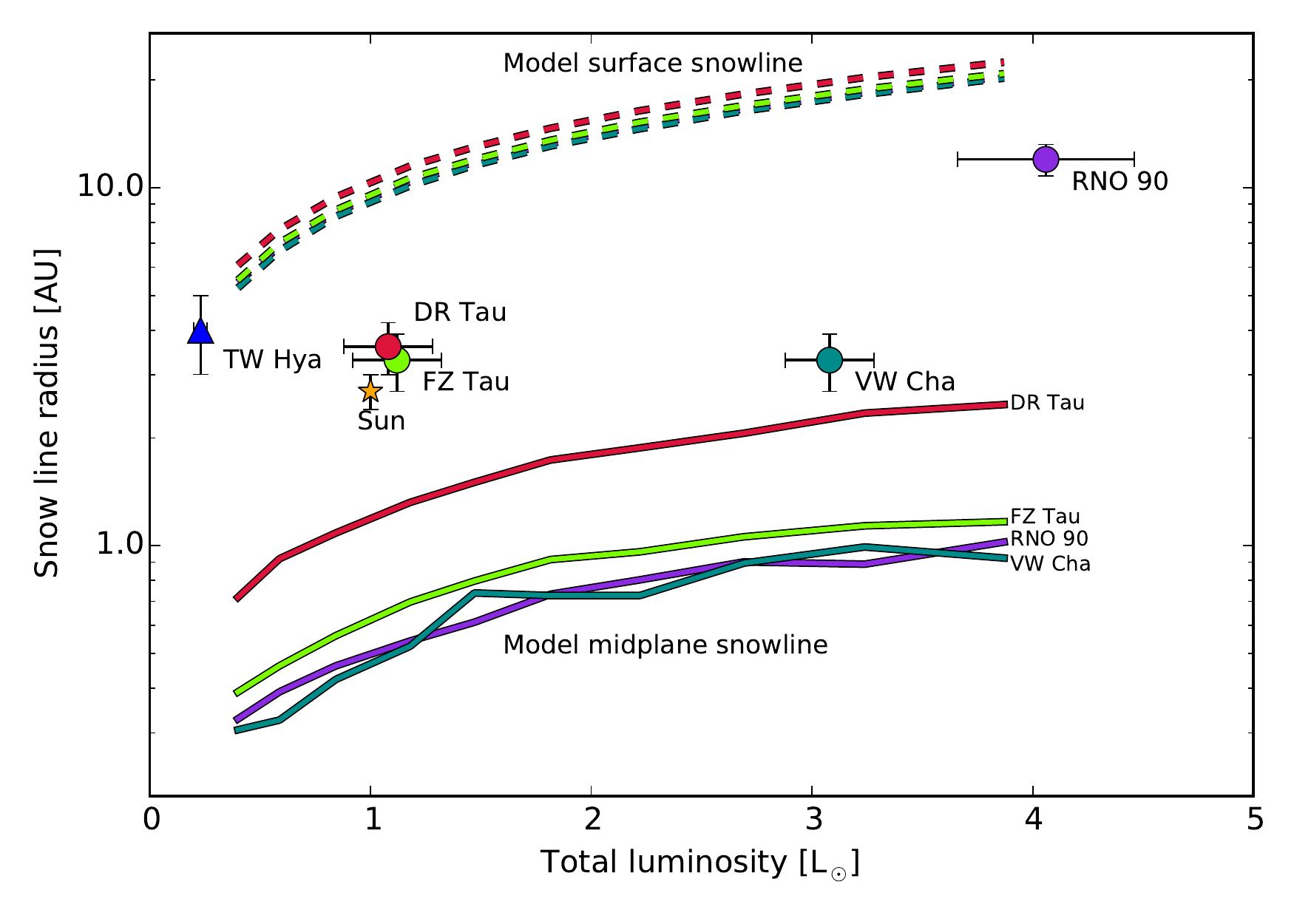}
\caption{The relation between the total luminosity and the theoretical locations of the midplane and surface snow lines, compared to our measured critical radii for the Case II model grid ($T_{\rm gas} > T_{\rm dust}$). We also include the location of the TW Hya snowline as measured by \cite{Zhang13} and the present-day canonical solar system snow line. The solid and dashed curves show the midplane and surface snow lines, respectively, for each of the four model disks as a function of the luminosity of the central star.}
\label{fig: snowline_vs_luminosity}
\end{figure}

In Figure \ref{fig: snowline_vs_luminosity}, we compare the Case II critical radius measurements to model expectations as a function of the total system luminosity. We also compare to the midplane and surface snow lines for a fiducial passive (no accretion heating) disk for both modeling cases. The passive disk snow line relation is derived by varying the luminosity of the central star of the RADMC model for each of the four disks, while keeping all other parameters constant. It can be seen that the surface snow lines are not strongly dependent on the disk model structure. This is expected, since the surface snow line is defined as the optically thin limit, and therefore approximates the isotherm of a spherically symmetric model. The midplane snowline does depend on the structure of the disk, with systematically larger snowline radii found for disks with larger scale heights, since more puffy disks intercept a larger fraction of the stellar light, which in turn leads to increasing heating of the disk at each radius.

The measured $R_{\rm crit}$ are significantly larger, by a factor $\sim 4$, than those expected for {\bf any of} the passive disk midplane snow line, and about two times larger than the snowline of a young accreting disk. The measured $R_{\rm crit}$ are also larger than the present-day solar snowline at $\sim 2.7$\,AU \citep{Bus02}. However, they are significantly smaller than the surface snow line as defined by the optically thin limit. This indicates the presence of warm, but dry, surface gas and dust at radii between 4 and 15\,AU.

\begin{deluxetable}{lcccc}[ht!]
\tablecolumns{5}
\tablewidth{0pt} 
\tablecaption{Best-fitting water distribution parameters}
\tablehead{
\colhead{Name} & \colhead{$R_{\rm crit}$} & \colhead{$X(\h)_{\rm max}$} & \colhead{$X(\h)_{\rm min}$} & \colhead{$\dfrac{\chi^2_{\rm min}}{\nu}$} \\
\colhead{} & \colhead{(AU)} & \colhead{(100/gtd H$_2^{-1}$)} & \colhead{(100/gtd H$_2^{-1}$)} & \colhead{}
}
\startdata

\cutinhead{Case I ($T_{\rm gas} = T_{\rm dust}$)}

DR Tau  & $>7$                 & $1.0 $                           & $1.0\times 10^{-7}$ & 28.5 \\
FZ Tau   & $5.3\pm 1.0$   & $1.0 $                           & $1.0\times 10^{-8}$ & 29.7 \\
RNO 90 & $6.5\pm 0.4$   & $1.0\times 10^{-1}$   & $1.0\times 10^{-7}$ &  32.2 \\
VW Cha & $7.9\pm 1.2$   & $1.0\times 10^{-2}$   & $1.0\times 10^{-8}$ & 24.8 \\

\cutinhead{Case II ($T_{\rm gas} > T_{\rm dust}$)}

DR Tau & $3.6\pm 0.2$      & $1.0\times 10^{-4}$ & $1.0\times 10^{-9}$ & 19.8  \\
FZ Tau   & $3.3\pm 0.2$     & $1.0\times 10^{-3}$ & $1.0\times 10^{-10}$ & 29.6 \\
RNO 90 & $11.1\pm 0.2$   & $1.0\times 10^{-4}$ & $1.0\times 10^{-9}$ & 27.6 \\
VW Cha & $3.3\pm 0.2$     & $1.0\times 10^{-4}$ & $1.0\times 10^{-9}$ & 25.8
\tablecomments{
This table lists the results obtained from modeling with (bottom) and without (top) gas-dust temperature decoupling.
}
\enddata
\label{tab:results}
\end{deluxetable}

\section{Discussion}
\label{sec: discussion}

\subsection{Potential effects of model assumptions}
Generating large grids of model water spectra is highly CPU intensive, and several major assumptions were made to keep the problem tractable. While we allow for gas-dust decoupling in the Case II grid, level populations are set to those of LTE conditions. It is known that LTE conditions likely do not hold in general \citep{Meijerink09}. Yet, it is also known that LTE populations lead to excellent fits of synthetic to observed spectra \citep{Carr08}. 

Part of the explanation for this may be that the effects of gas-dust temperature decoupling and non-LTE excitation counteract each other in the disk photosphere, as shown by \cite{Meijerink09}; high gas temperatures increase upper level populations, but the high critical densities of the water transitions conversely lead to sub-thermal populations at higher disk altitudes where the gas-dust decoupling is the strongest. A significant difference in our revised model, based on the \cite{Najita11} thermochemical gas temperatures, is that the water is effectively destroyed at high altitudes where the gas-dust decoupling is the strongest, and where densities are too low to populate the upper levels. Our LTE model spectra are therefore close to the non-LTE case of \cite{Meijerink09}. Infrared pumping tends to lead to excitation temperatures that mimic that of the local continuum color temperature, which in turn is driven by the dust temperature, rather than that of the gas. Indeed, \cite{Meijerink09} found that, while there are indications of gas-dust decoupling and non-LTE effects in the \textit{Spitzer} spectra, assuming LTE and coupled gas-dust keep line fluxes within a factor of two. While it is important to consider non-LTE effects, if such calculations can be made tractable for grid-based model fits, we do not expect them to affect the general conclusions of this paper. 

We found that, in the Case II models, the local dust temperature and the location of abundant water vapor are not mutually consistent; the dust temperature is too low to allow abundant water vapor at disk radii and heights where the model fit requires it to be. Further exploration of thermochemical models and increased gas-to-dust ratio may reconcile this discrepancy. 

We also neglect effects of accretion heating in the midplane by using a passive disk model. This assumption may be justified by comparing to the standard 1D+1D models of accreting protoplanetary disks by \cite{dAlessio05}. 
Using these models, \cite{Dullemond07} compares the regimes in which the disk midplane/surface temperature structures are dominated by irradiation or accretion heating for a star-disk system similar to those of our survey (see their Figure 3). Inspection of the figure indicates that for the highest accretion rate in our sample, of just over $10^{-7}\rm\,M_{\odot}\,yr^{-1}$ for DR Tau, the disk surface is irradiation-dominated beyond 2 AU. For our other three disks, the disk surface is irradiation-dominated beyond 1 AU. The D'Alessio model disk surface temperatures refer to the dust temperature. In our gas/dust decoupled case, we would expect that accretion heating plays an even smaller relative role in the gas temperature since additional external heating sources are included \citep{Najita11} compared to the D'Alessio models.

\subsection{High gas-to-dust ratios or decoupled gas temperatures?}

The best-fitting water vapor distribution models suggest that either the gas-to-dust ratio in the disk surface, at least inside the surface snow line, must be very high, or most of the water emission is formed in a layer in the disk where the gas temperature is higher than that of the dust. Both of these scenarios are theoretically plausible. High gas-to-dust ratios in the inner disk surface are predicted by many models of grain growth and settling \citep{Miyake95,Tanaka05,Fromang06,Ciesla07} and have been supported by other observations. Using near-infrared CO absorption, \cite{Rettig06} and \cite{Horne12} found gas-to-dust enhancements over ISM values of up to a factor 10 (or 50 in the extreme case of AA Tau). \cite{Furlan05} found indirectly from the distribution of mid-infrared colors of protoplanetary disks in Taurus that a high degree of dust setting is common, corresponding to gas-to-dust enhancement factors of 2-3 orders of magnitude. Citing similar reasons, high gas-to-dust ratios of up to 10,000 were also used in previous two-dimensional non-LTE models of {\it Spitzer} water emission from disks \citep{Meijerink09}.

Conversely, there is theoretical consensus that gas-dust temperature decoupling is generally significant in disk surfaces at column densities below $\sim 10^{22}\,\rm cm^{-2}$ \citep{Henning13}. If thermo-chemical gas-dust decoupling is taken into account, our best-fit Case II models indicate that there is no strong evidence for gas-to-dust mass ratios significantly in excess of 100. Other analyses of protoplanetary disk observations indicate canonical or sub-canonical gas-to-dust ratios. In transition disks, recent ALMA observations of CO coupled with detailed modeling have suggested low gas-to-dust ratios of $\sim$10 \citep{Bruderer14}. Thermo-chemical modeling of far-infrared atomic and molecular disk tracers, as observed with {\it Herschel}, have also indicated low gas-to-dust ratios in some cases \citep{Meeus10,Keane14}. However, it is not clear to which degree these findings are related to an advanced evolutionary stage of the disks, and the classical disks included in this paper may not be directly comparable. 

Finally, it should be noted that, since the sample of water-rich disks have been selected in part based on their high line-to-continuum ratios at {\it Spitzer} wavelengths, it is possible that the uniformly high gas-to-dust ratios of our sample is a selection effect. Indeed, at the nominal water abundance and gas-to-dust ratio of $10^{-4}\,\rm H^{-2}$ and 100, respectively, the water emission would have been below the detection limit of {\it Spitzer} throughout the Case I model grid (see Figure \ref{fig: gd-decoup_line_comp}). 

\subsection{Size of the warm molecular layer: Freeze-out or gas-phase formation of water?}
In Figure \ref{fig: snowline_vs_luminosity}, it is seen that the critical radii fall at smaller radii than the formal surface snow line, as defined by the optically thin dust limit. That is, there is a dry surface region at radii where the dust temperature is above the freeze-out temperature. There are several potential explanations for this result. Thermo-chemical models predict that the water abundance, in surface regions where chemical and photo-destruction time scales are short, is dominated by gas-phase pathways to water with significant activation barriers \citep{Woitke09, Glassgold09, Bethell09}. Thus, where the gas temperature drops below $\sim$200\,K, the water vapor abundance is predicted to decrease dramatically due to this chemical effect rather than freeze-out. Our observations are broadly consistent with this scenario. \cite{Meijerink09} also found that the surface snow line was found at small radii in at least one case, albeit based on {\it Spitzer} data only. They suggested an alternate scenario in which strong depletion of surface water could occur if water vapor is turbulently mixed to deeper, cooler layers in the disk, where it can efficiently freeze out onto dust grains. If the icy grains settle to the mid-plane the water is effectively depleted from the surface via this so-called ``vertical cold finger'' effect. In this case, it is predicted that the surface critical radius should roughly equal the radius of the midplane snow line at $\sim 1$\,AU. Figure \ref{fig: snowline_vs_luminosity} indicates that the critical radius is well beyond the midplane snow line, and does therefore not directly support the action of an efficient vertical cold finger effect, which depletes surface oxygen over cold midplane regions. 

\subsection{Future observations}
Further observations of water in disks are needed to confirm and refine the observed distribution of water in protoplanetary disks. We have demonstrated that, for constraining the water surface snow line, the most relevant transitions tracing 100-200\,K gas are found at 40-100\,$\mu$m. However, the \textit{Herschel} mission has ended, and there is currently no comparable observational capability in the far-infrared. SOFIA represents one option, but is about an order of magnitude less sensitive than {\it Herschel} and cannot observe low-excitation water, making it difficult to observe the disks around solar-type stars with strong water emission. Disk around brighter, more massive stars do not show strong water vapor emission \citep{Pontoppidan10, Fedele11}. ALMA may be able to observe cool water vapor using the 183\,GHz water line using the new Band 5 receivers currently being constructed. In the more distant future, SPICA may provide the next set of far-infrared water vapor spectroscopy in disks, exploring lower stellar masses, and disks with fainter lines to place the strong water emitters presented in this paper into a larger, perhaps more representative, context. 

A complementary option for constraining water in protoplanetary disk surfaces is to image the 3\,$\mu$m water ice feature in scattered light, as demonstrated by \cite{Honda09}. This is potentially a  powerful diagnostic, and may be able to confirm that the disk surface is dry (no vapor or ice) between $R_{\rm crit}$ and the surface snow line.

\acknowledgements
Support for SMB was provided by the STScI Director's Discretionary Fund (DDRF). KMP and AB acknowledge financial support by a NASA Origins of the Solar System grant No. OSS 11-OSS11-0120, a NASA Planetary Geology and Geophysics Program under grant NAG 5-10201.
This work is based in part on observations made with the \textit{Spitzer Space Telescope}, which is operated by the Jet Propulsion Laboratory, California Institute of Technology under a contract with NASA. This work is based in part on observations made with the \textit{Herschel Space Observatory}, a European Space Agency Cornerstone Mission with significant participation by NASA. Support for this work was provided by NASA through an award issued by JPL/Caltech.

\bibliographystyle{apj}
\bibliography{snowline_main}

\clearpage

\appendix
\label{appendix}

\section{Gallery of individual lines and line complexes in RNO 90} 

\begin{figure*}[ht!]
\centering
\includegraphics[width=18cm]{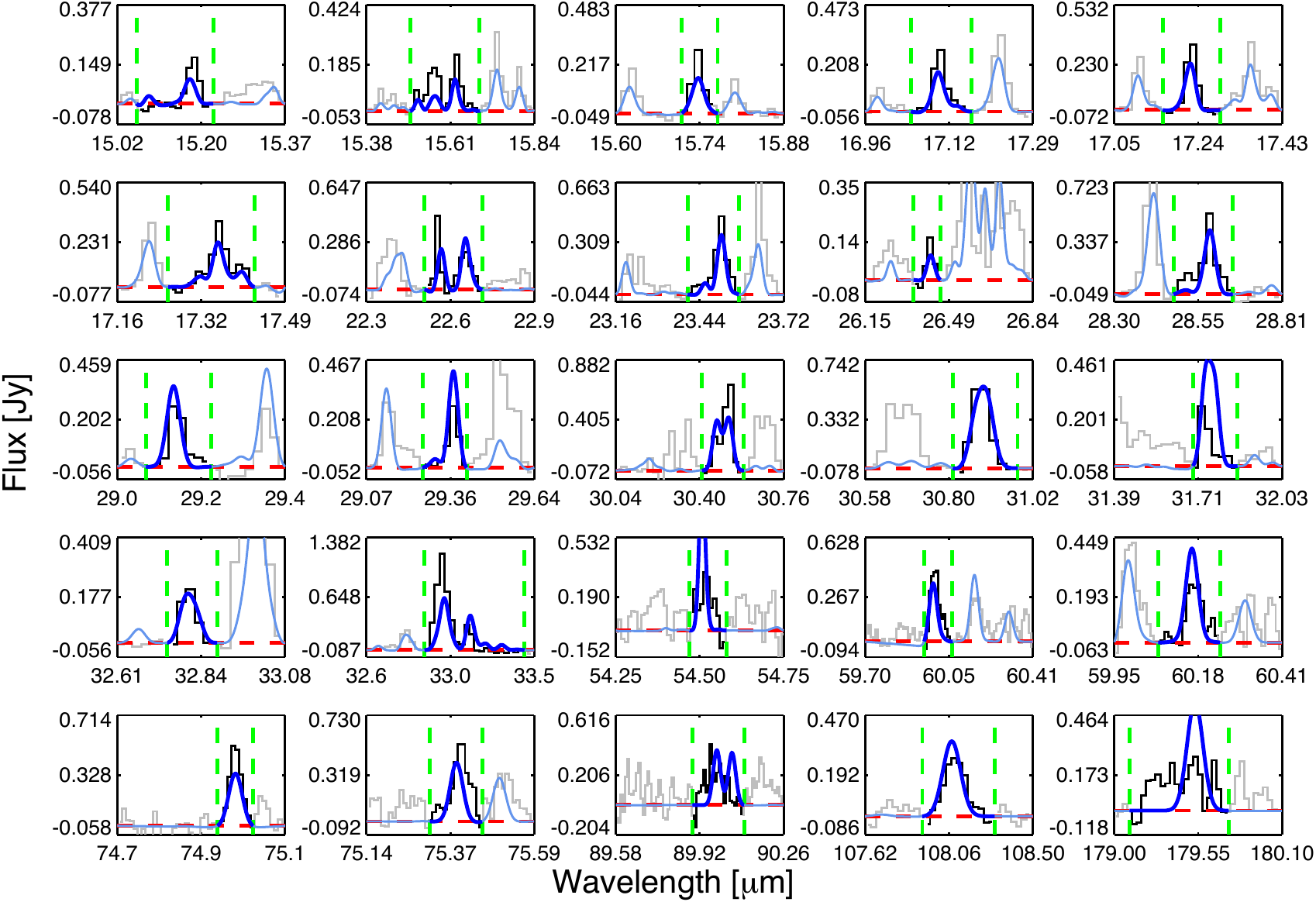}
\caption{Continuum-subtracted water line spectra for RNO 90. The data are drawn using a stepped line, while a smooth curve show the best-fitting synthetic Case II model spectrum. The horizontal dashed lines indicate zero line flux. The vertical dashed lines show the spectral bounds over which the line fluxes are integrated. }
\label{fig:rno90_lines}
\end{figure*}

Figure \ref{fig:rno90_lines} shows a gallery of the selected water lines and complexes for RNO 90 compared to the best-fit Case II models. Each line flux measurement was obtained by integrating the area under the data within a pre-selected wavelength range. The same range was used for both the model and data spectra. This procedure was performed for each disk in the sample and the integrated line fluxes are given in Tables \ref{tab:lineflux_spitzer} and \ref{tab:lineflux_herschel}.

\section{Marginalized confidence levels for the critical radii} 

\begin{figure*}[ht!]
\centering
\includegraphics[width=5.95cm]{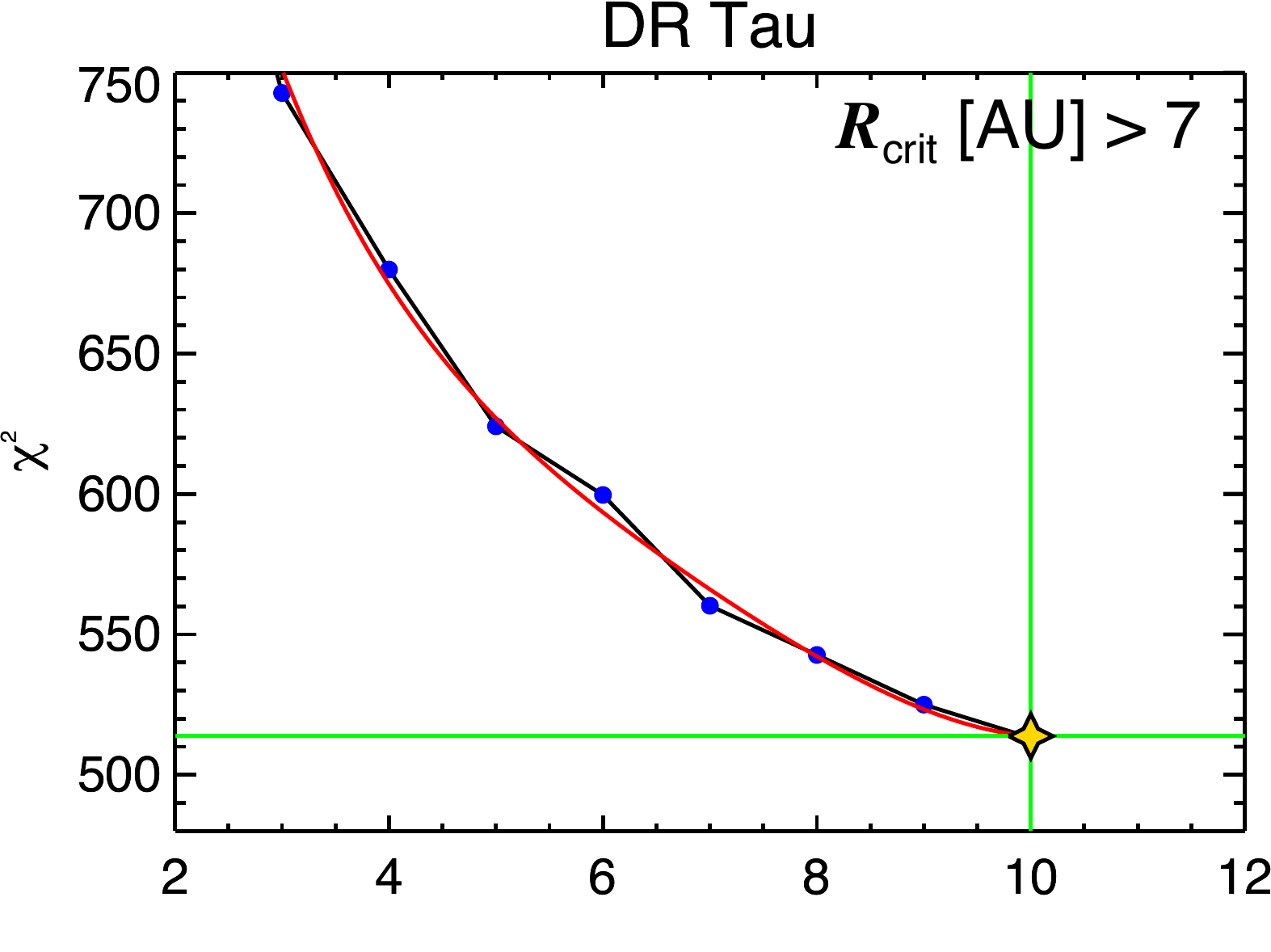}
\includegraphics[width=5.95cm]{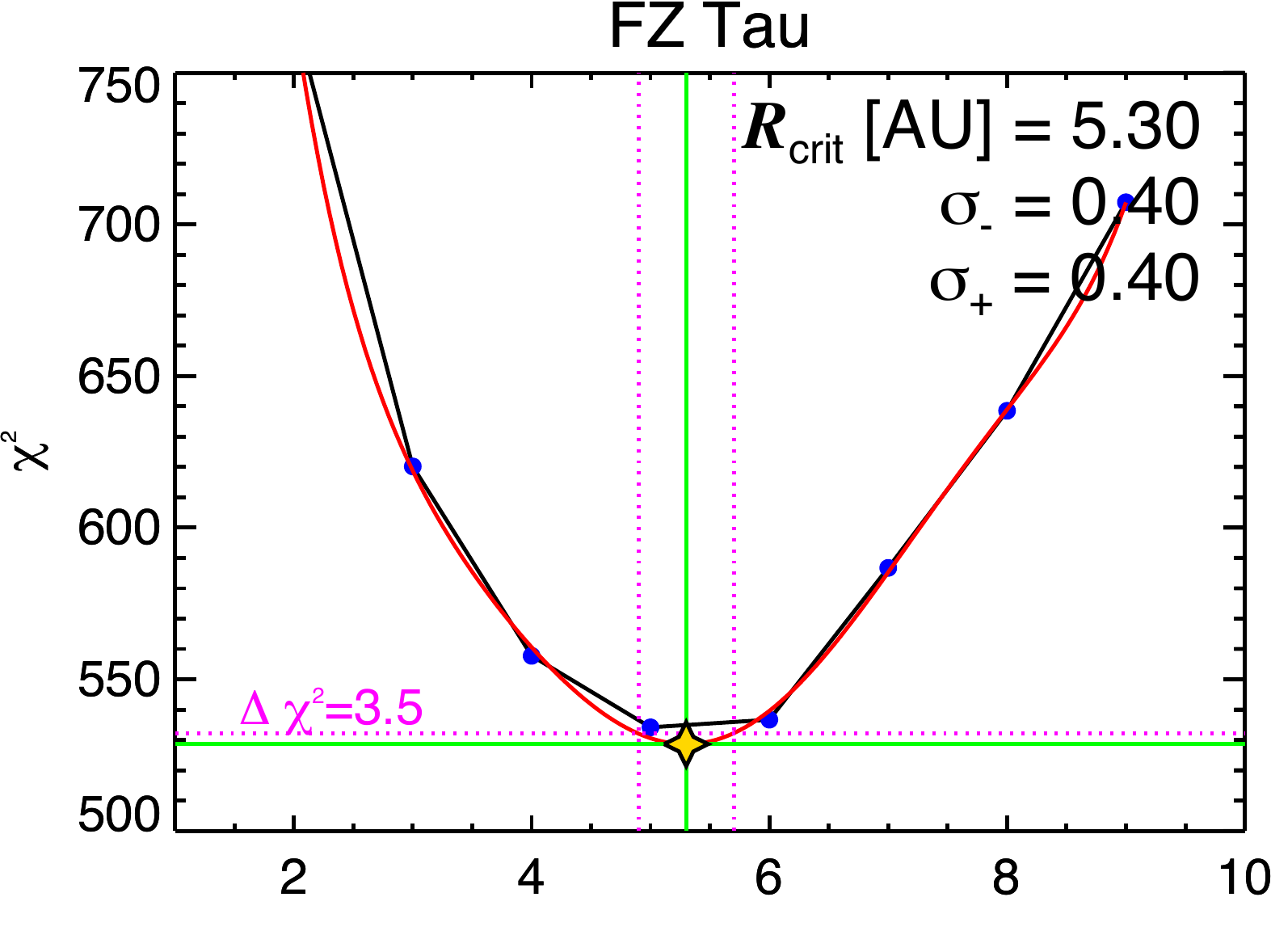}
\includegraphics[width=5.95cm]{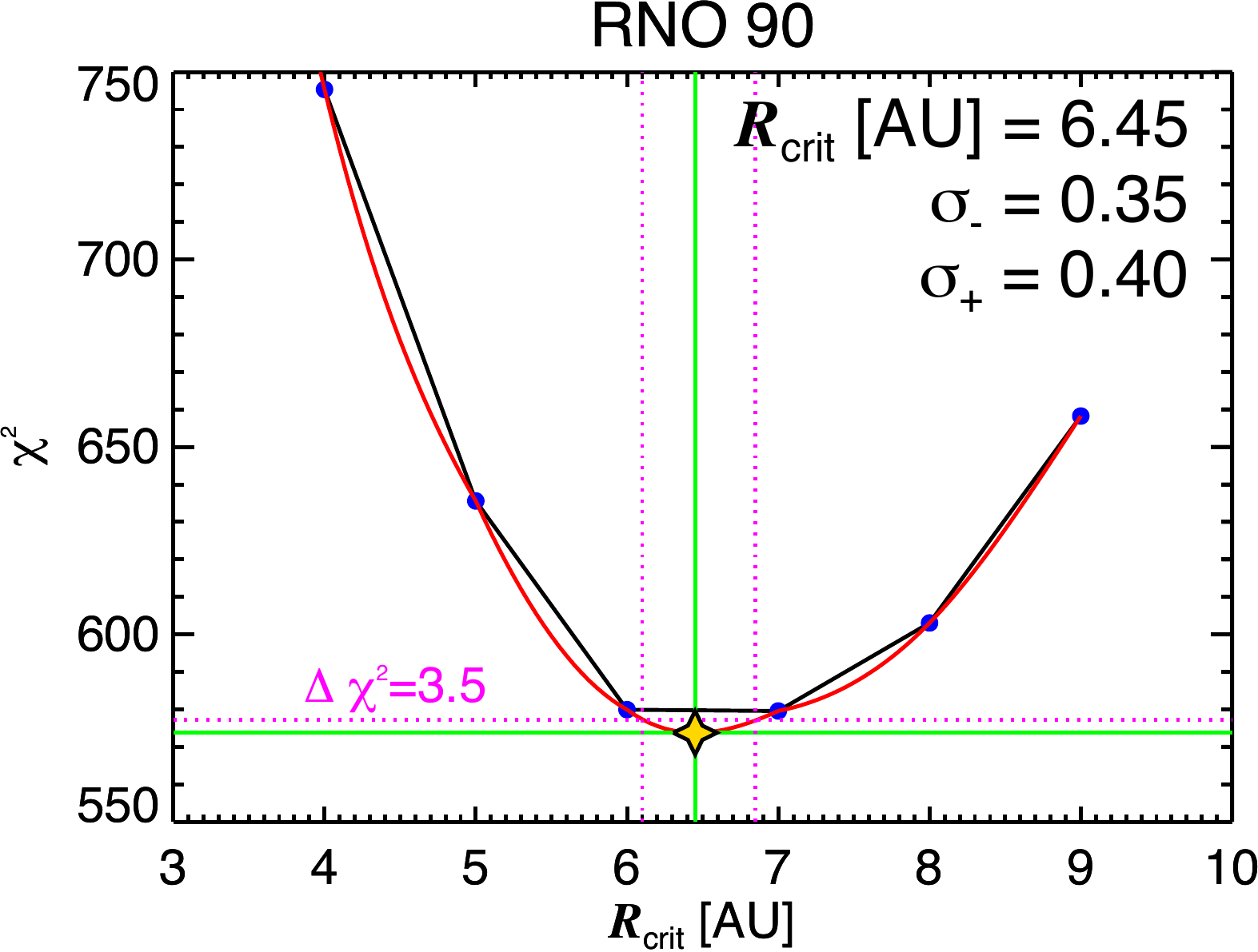}
\includegraphics[width=5.95cm]{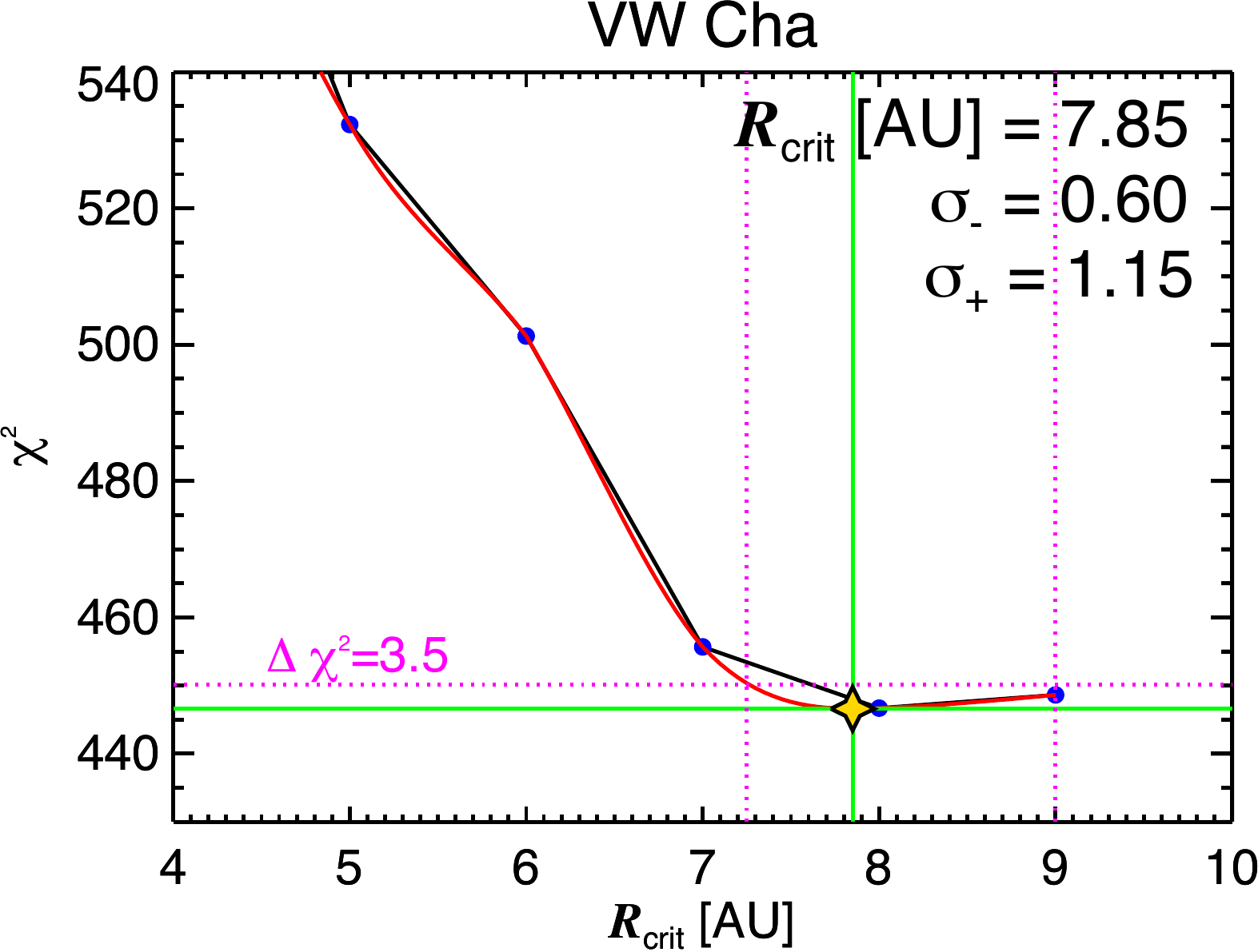}
\caption{Derivation of the confidence limits on the critical radii from Case I $\chi^{2}$ analysis in one-dimension. Our grid points are marked in blue, while the red curves are polynomial fits. The minimum found in each curve is indicated with a star symbol. The dotted lines on either side of the star represent the 68\% confidence level for three free parameters projected onto a one-dimensional space.}
\label{fig: redchi_1D_caseI}
\end{figure*}

\begin{figure*}[ht!]
\centering
\includegraphics[width=5.95cm]{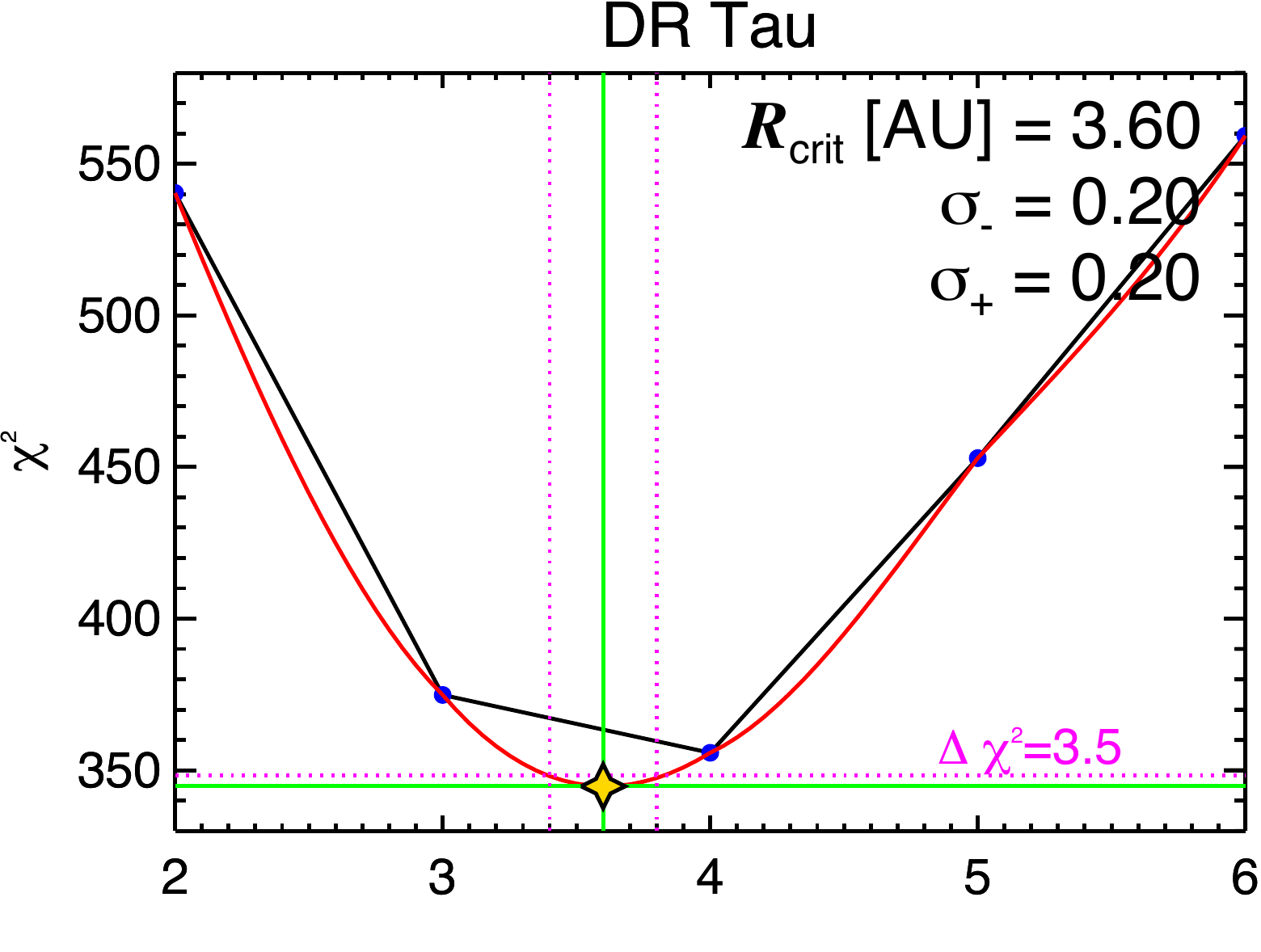}
\includegraphics[width=5.95cm]{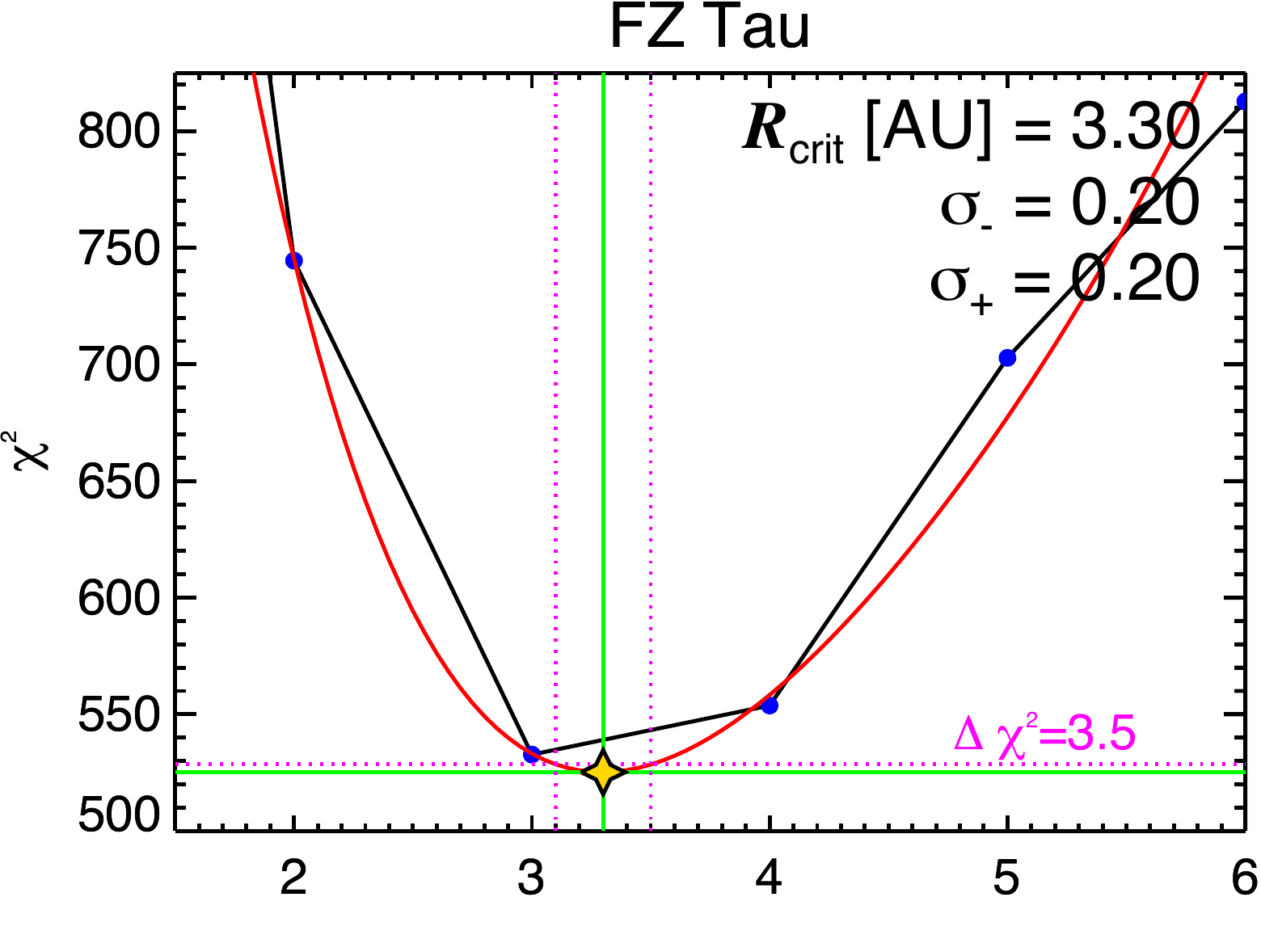}
\includegraphics[width=5.95cm]{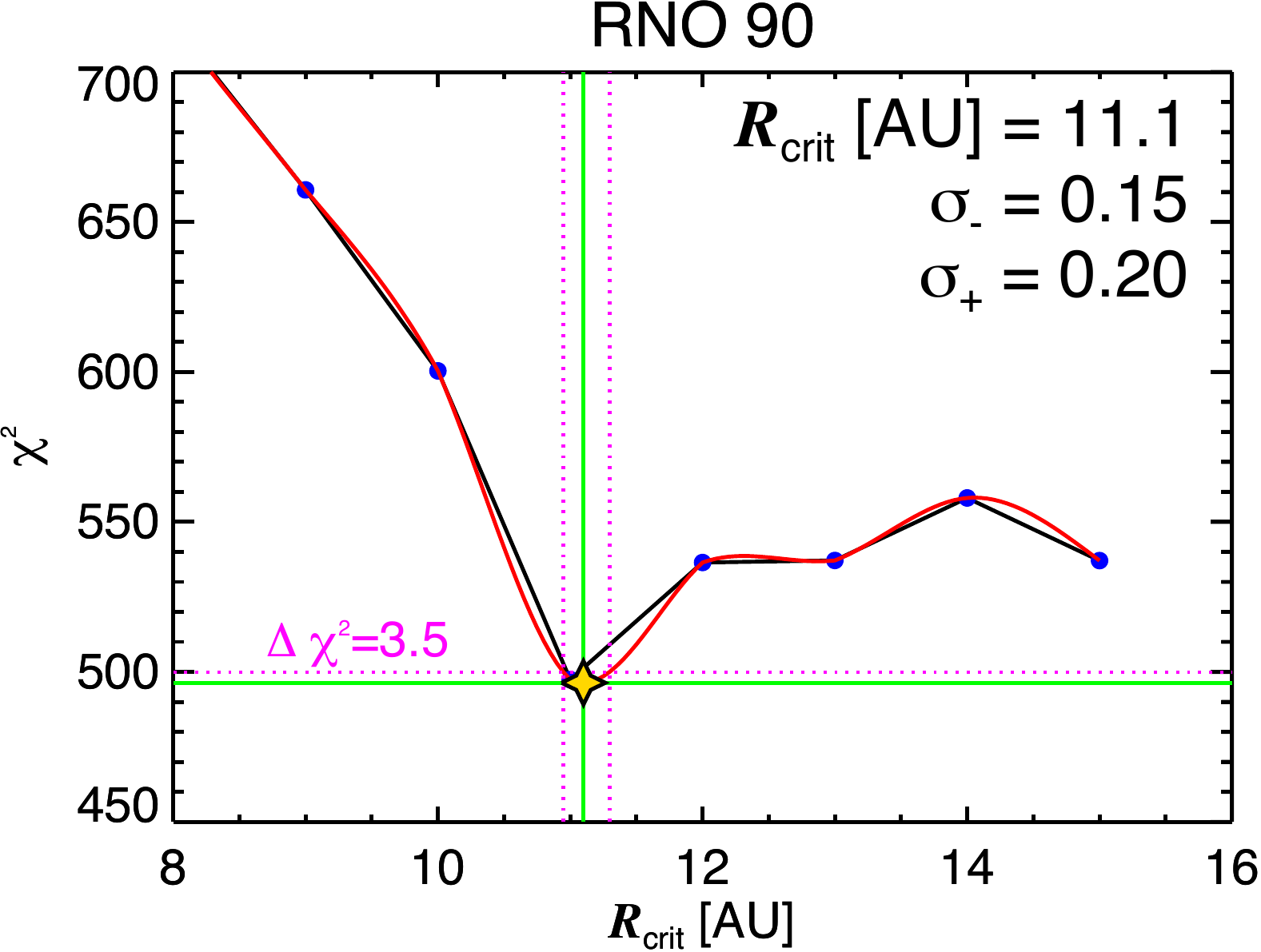}
\includegraphics[width=5.95cm]{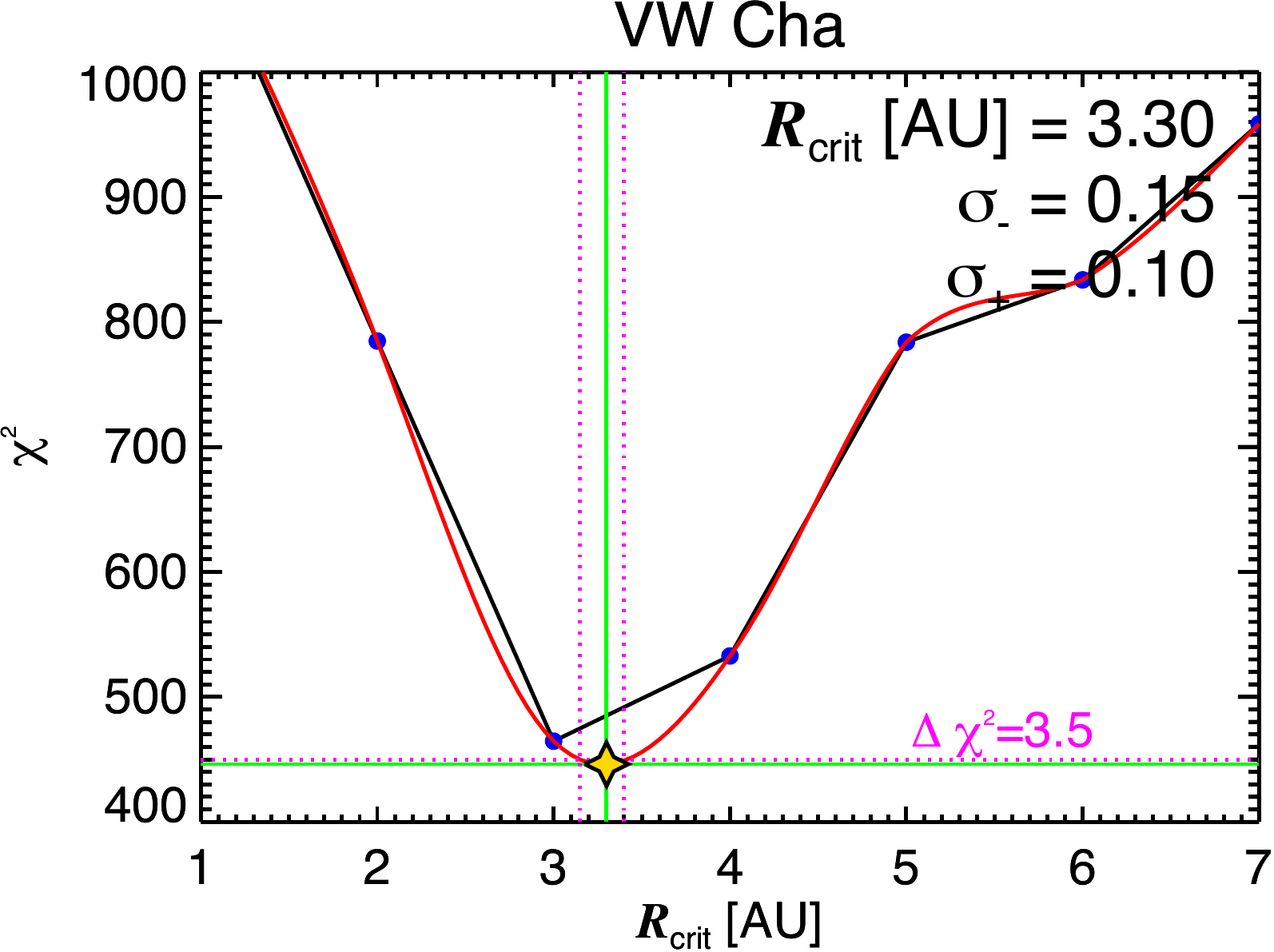}
\caption{Derivation of the confidence limits on the critical radii from Case II $\chi^{2}$ analysis in one-dimension. Our grid points are marked in blue, while the red curves are polynomial fits. The minimum found in each curve is indicated with a star symbol. The dotted lines on either side of the star represent the 68\% confidence level for three free parameters projected onto a one-dimensional space.}
\label{fig: redchi_1D_caseII}
\end{figure*}

The behavior of the goodness-of-fit of the model grids were investigated in detail in order to identify and quantify degeneracies of the gas-phase model. Following \cite{NR}, we marginalized (projected) the $\chi^2$ space to one dimension, as a function of $R_{\rm crit}$, and calculated the corresponding parameter errors as the marginalized 68\% confidence interval, as shown in Figs. \ref{fig: redchi_1D_caseI} and \ref{fig: redchi_1D_caseII} for the Case I and II models, respectively. Note that in the Case I models of DR Tau, there is no minimum in the $\chi^2$ space for $R_{\rm crit}$, indicating that the critical radius must be larger than the surface snow line to fit the data. The model cannot reproduce this scenario since the water vapor is always frozen out at such large radii. 

\section{Full spectral range data with best-fit models} 

\begin{figure*}[ht!]
\centering
\includegraphics[width=8.95cm]{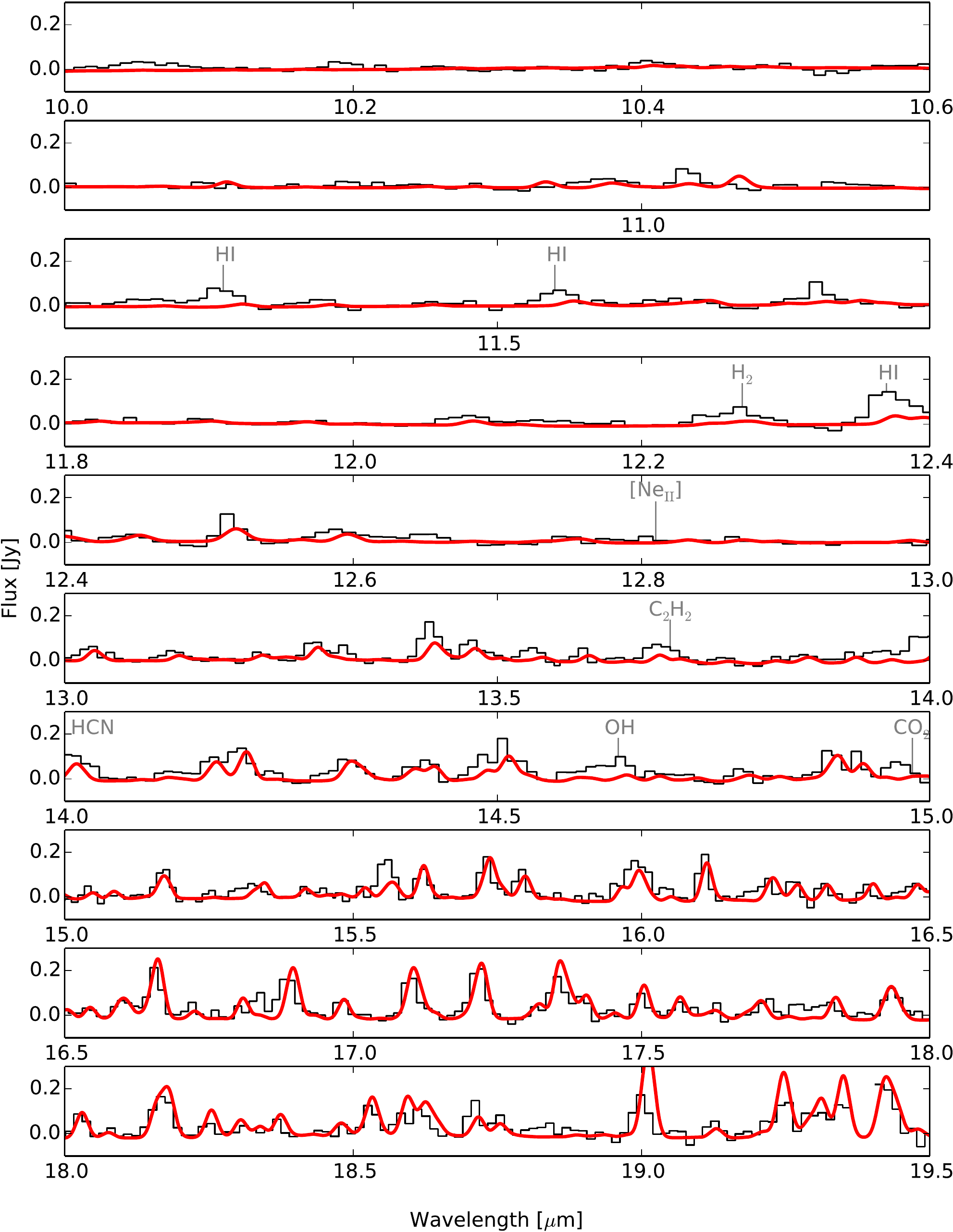}
\includegraphics[width=8.95cm]{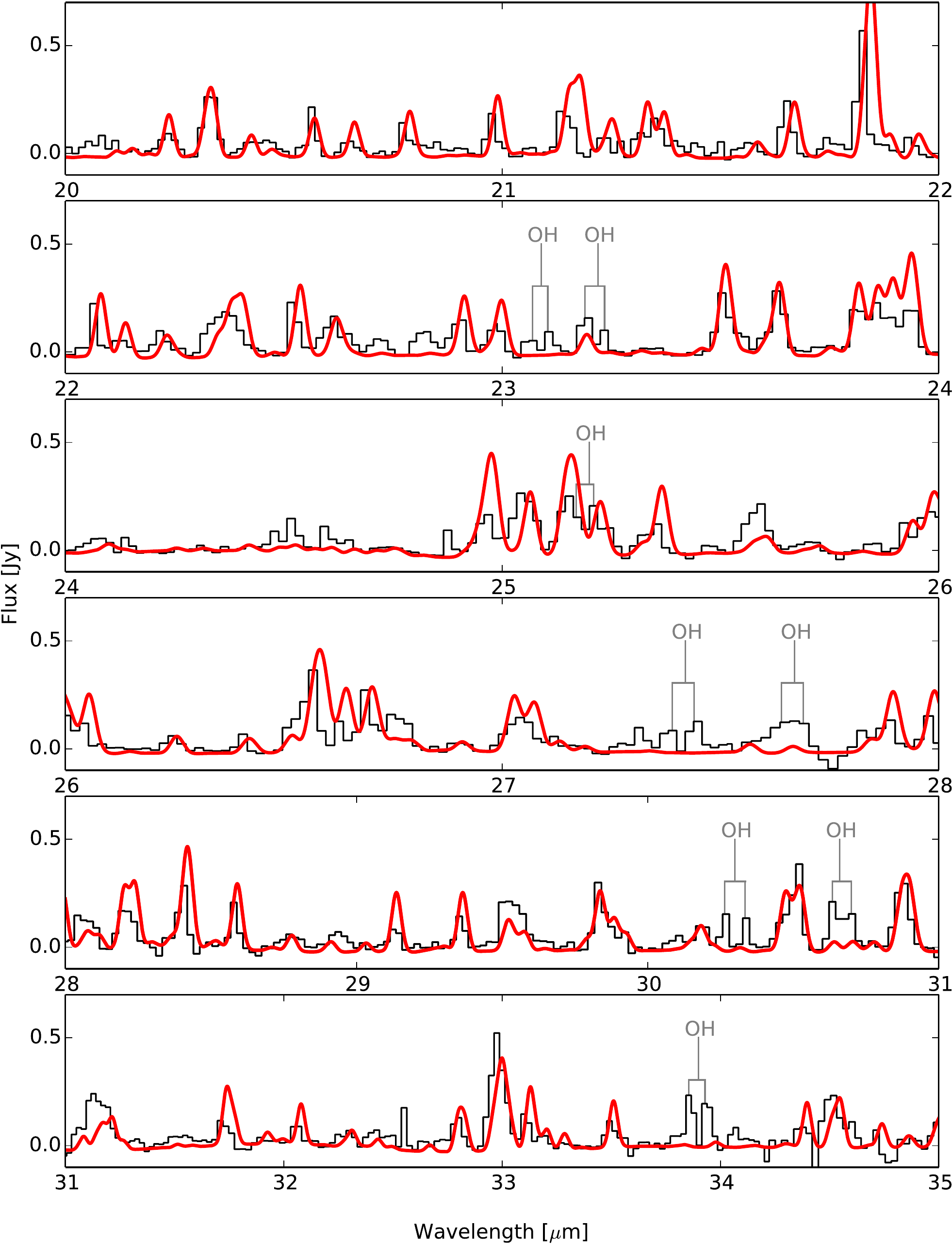}
\caption{\textbf{DR Tau:} Best-fit Case II model compared to the {\it Spitzer}-IRS spectral range. The model only fits the water emission features, but additional atomic and molecular emission features due to H I, H$_2$, [Ne II], C$_2$H$_2$, OH, and CO$_2$ are labeled.  }
\label{fig:drtau_irs}
\end{figure*}

\begin{figure*}[ht!]
\centering
\includegraphics[width=16.cm]{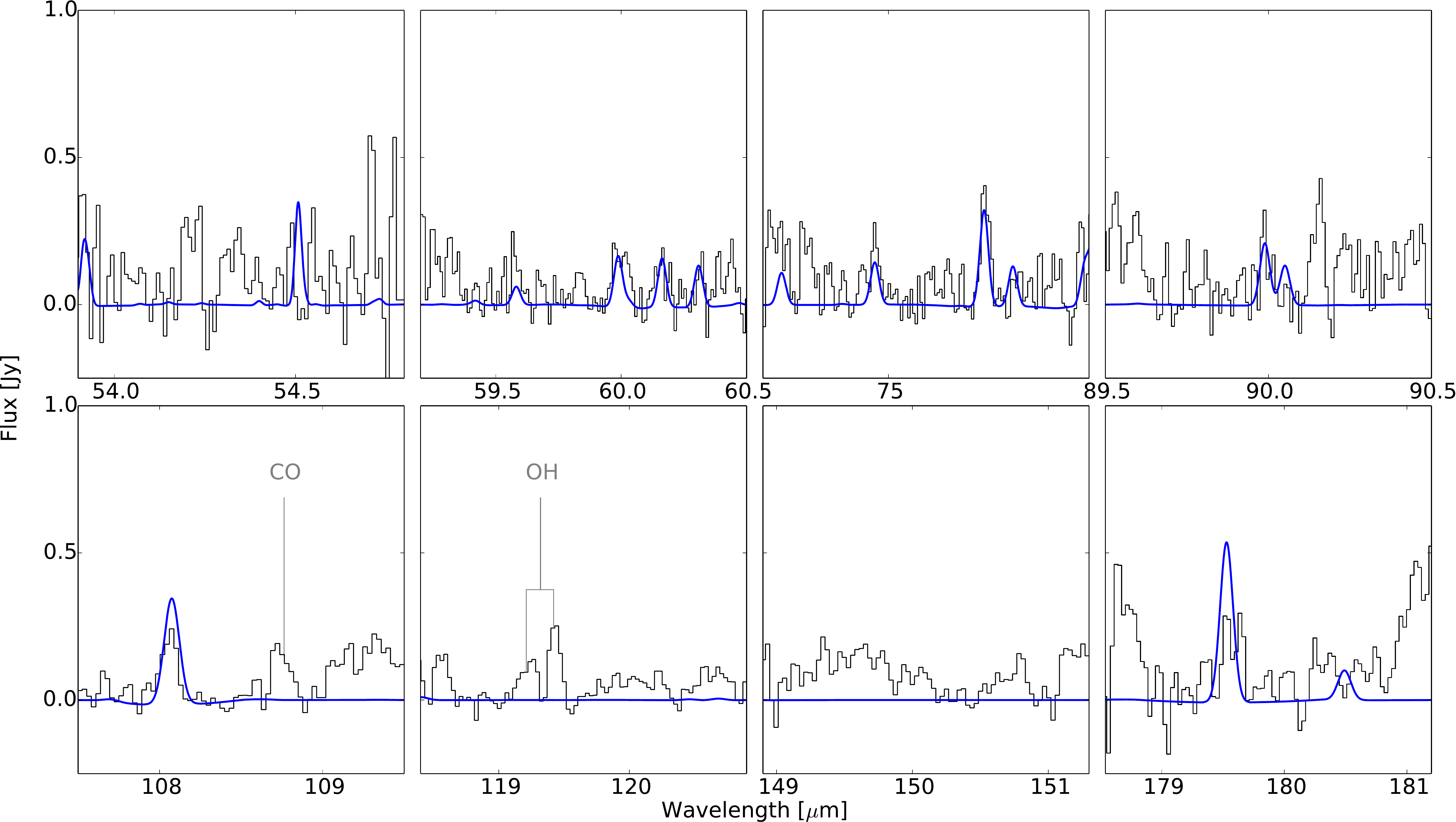}
\caption{\textbf{DR Tau:} Best-fit Case II model compared to the {\it Herschel}-PACS spectroscopy. Lines due to CO and OH, which are not included in the model, are labeled.}
\label{fig:drtau_pacs}
\end{figure*}

\begin{figure*}[ht!]
\centering
\includegraphics[width=8.95cm]{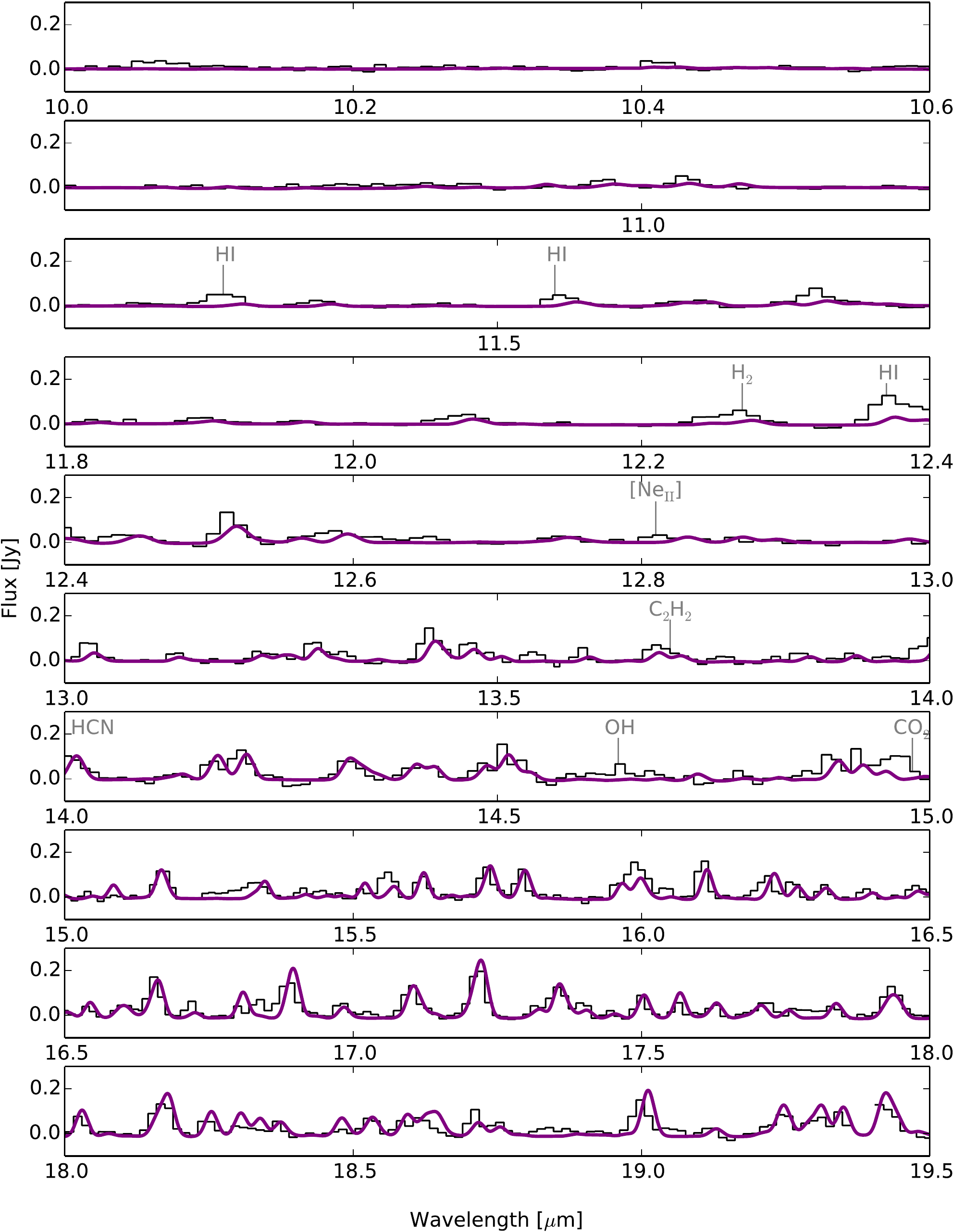}
\includegraphics[width=8.9cm]{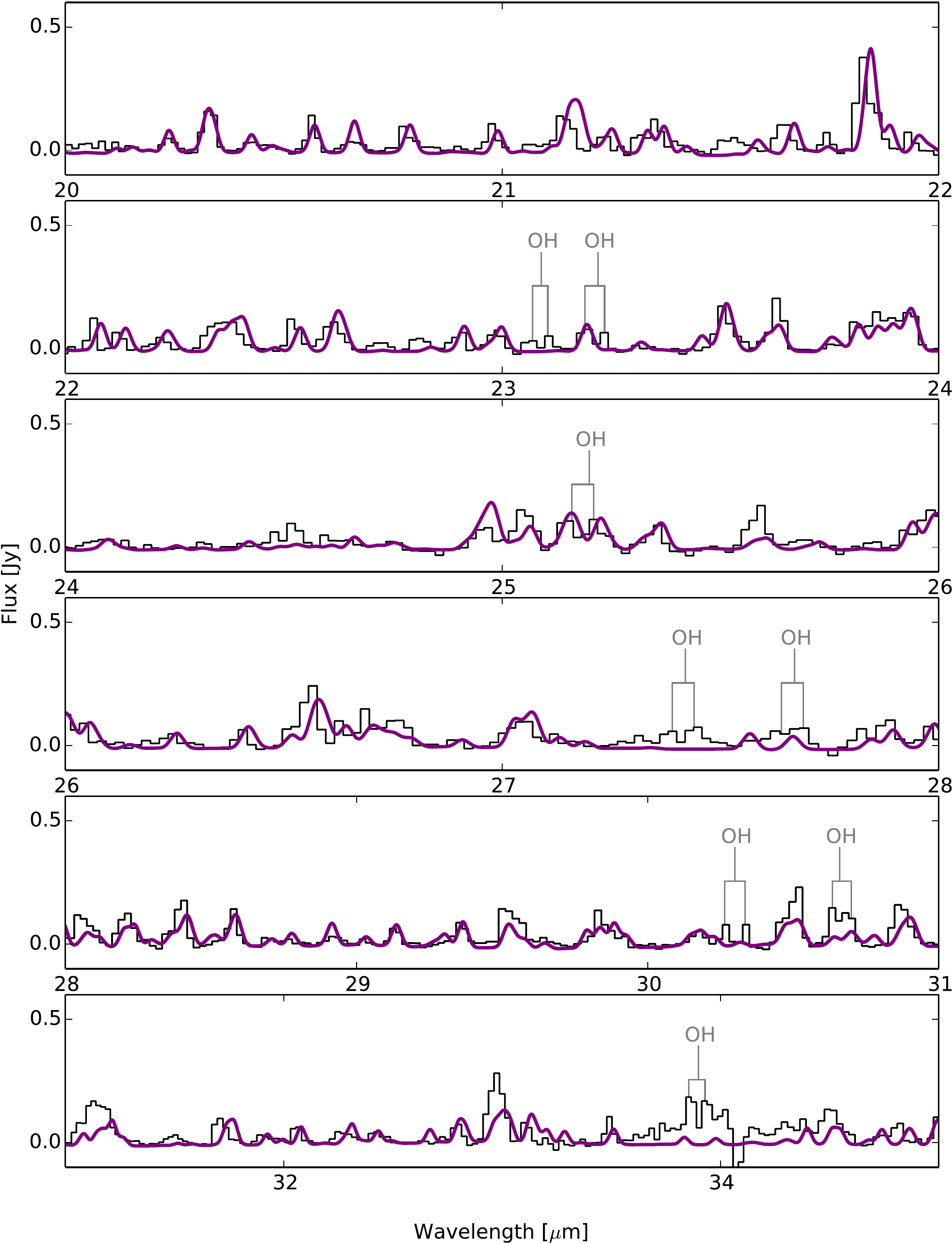}
\caption{\textbf{FZ Tau:} Best-fit Case II model compared to the {\it Spitzer}-IRS spectral range.}
\label{fig:fztau_irs}
\end{figure*}

\begin{figure*}[ht!]
\centering
\includegraphics[width=16.cm]{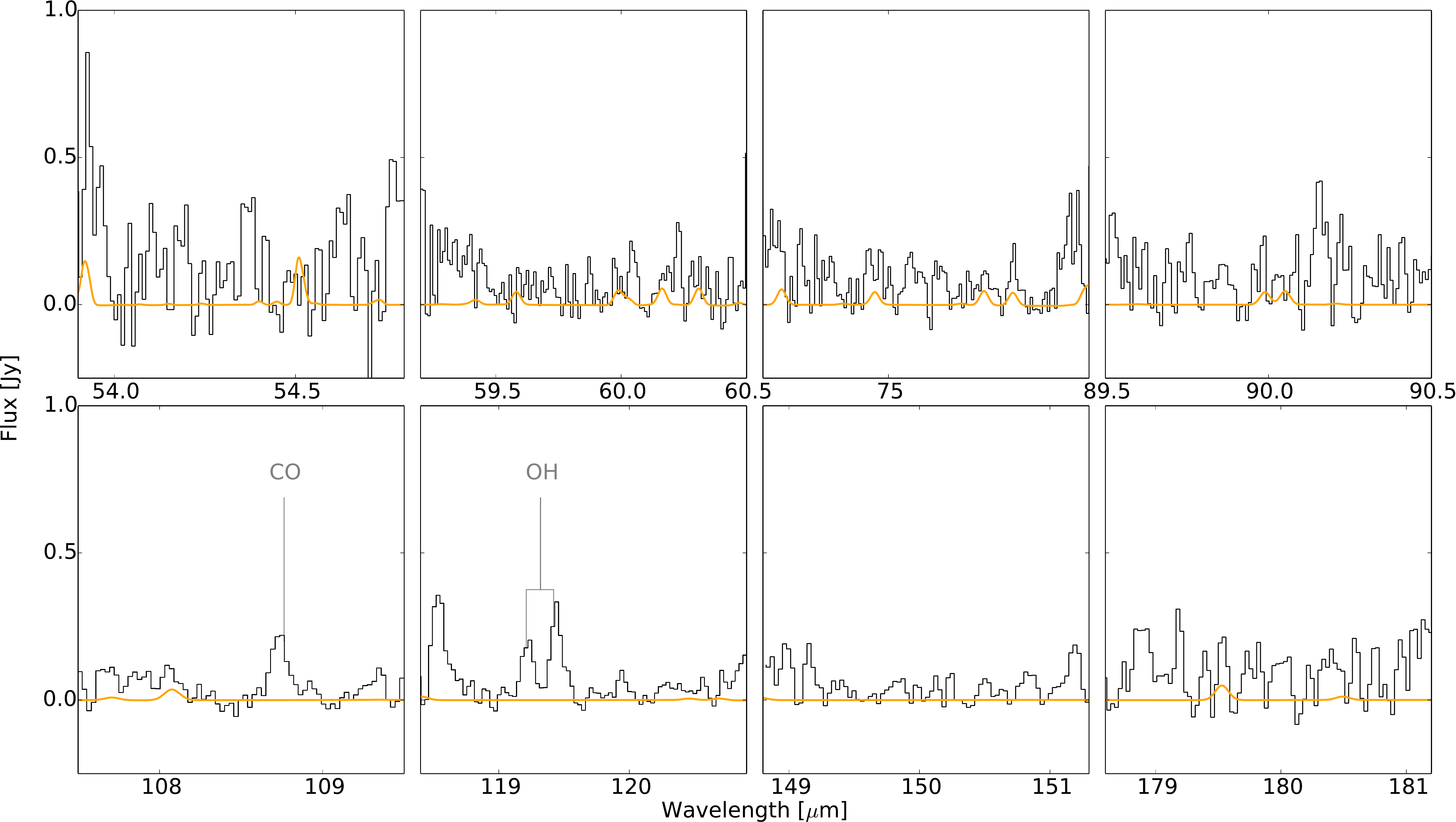}
\caption{\textbf{FZ Tau:} Best-fit Case II model compared to the {\it Herschel}-PACS spectroscopy.}
\label{fig:fztau_pacs}
\end{figure*}

\begin{figure*}[ht!]
\centering
\includegraphics[width=8.95cm]{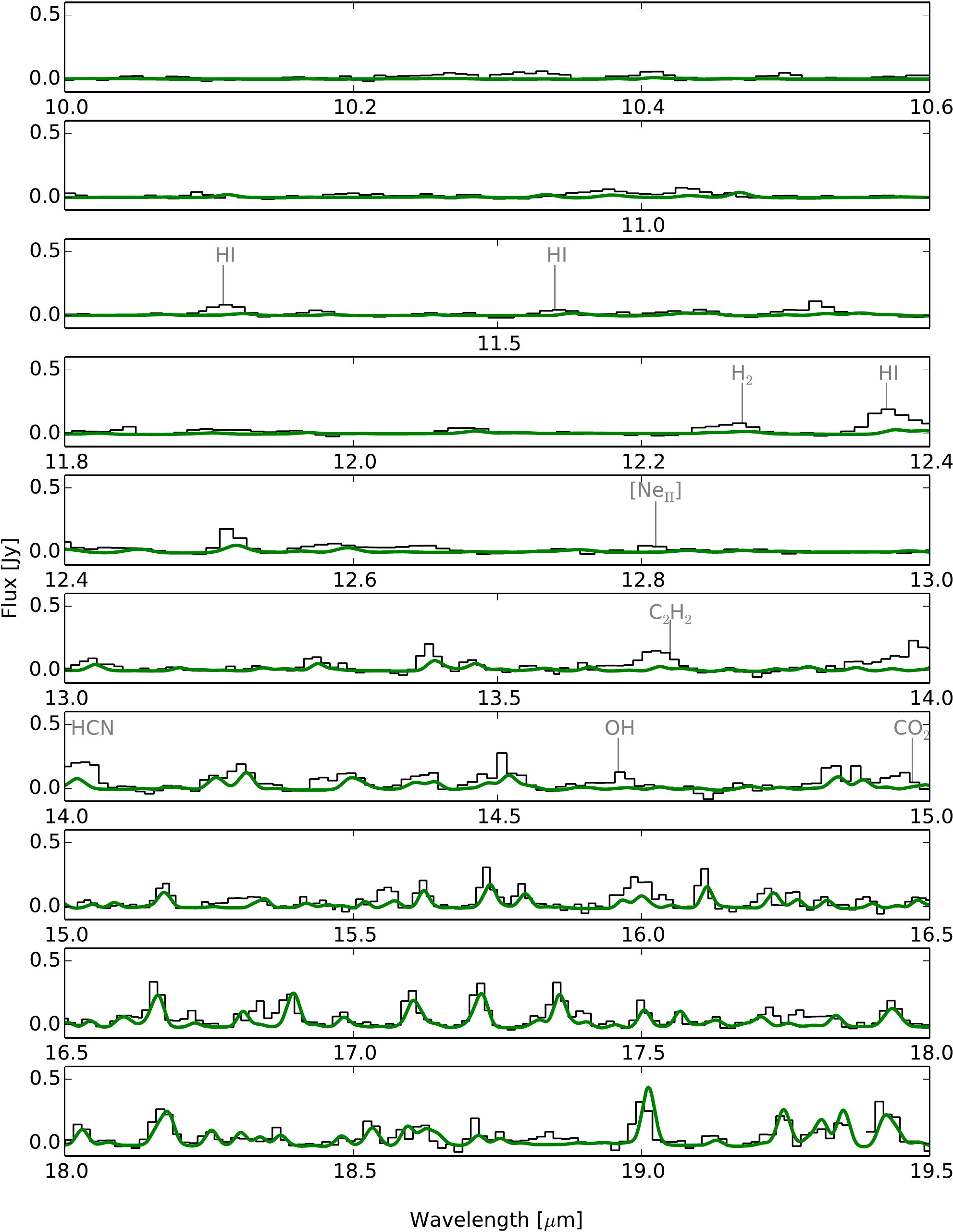}
\includegraphics[width=8.9cm]{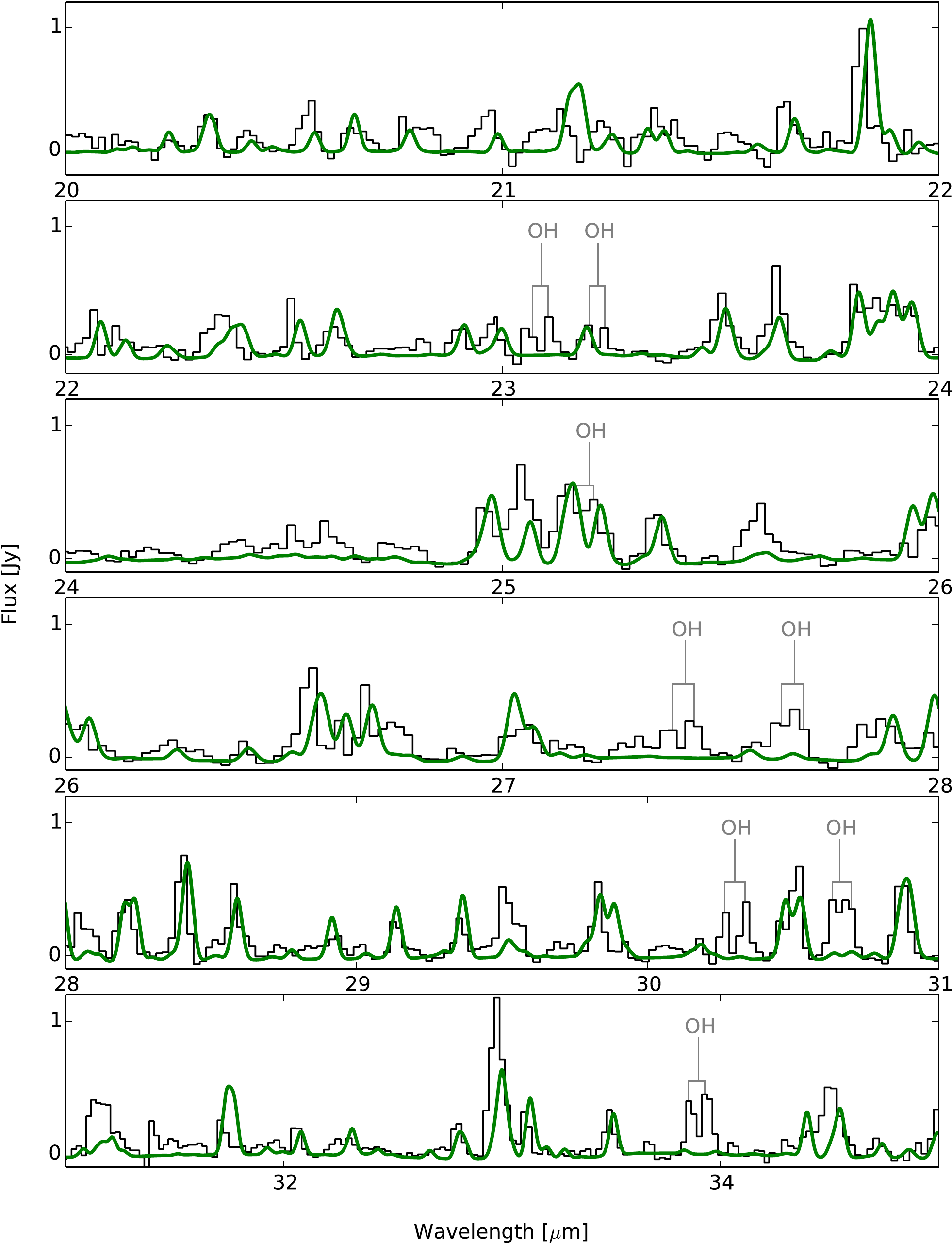}
\caption{\textbf{RNO 90:} Best-fit Case II model compared to the {\it Spitzer}-IRS spectral range.}
\label{fig:rno90_irs}
\end{figure*}

\begin{figure*}[ht!]
\centering
\includegraphics[width=16.cm]{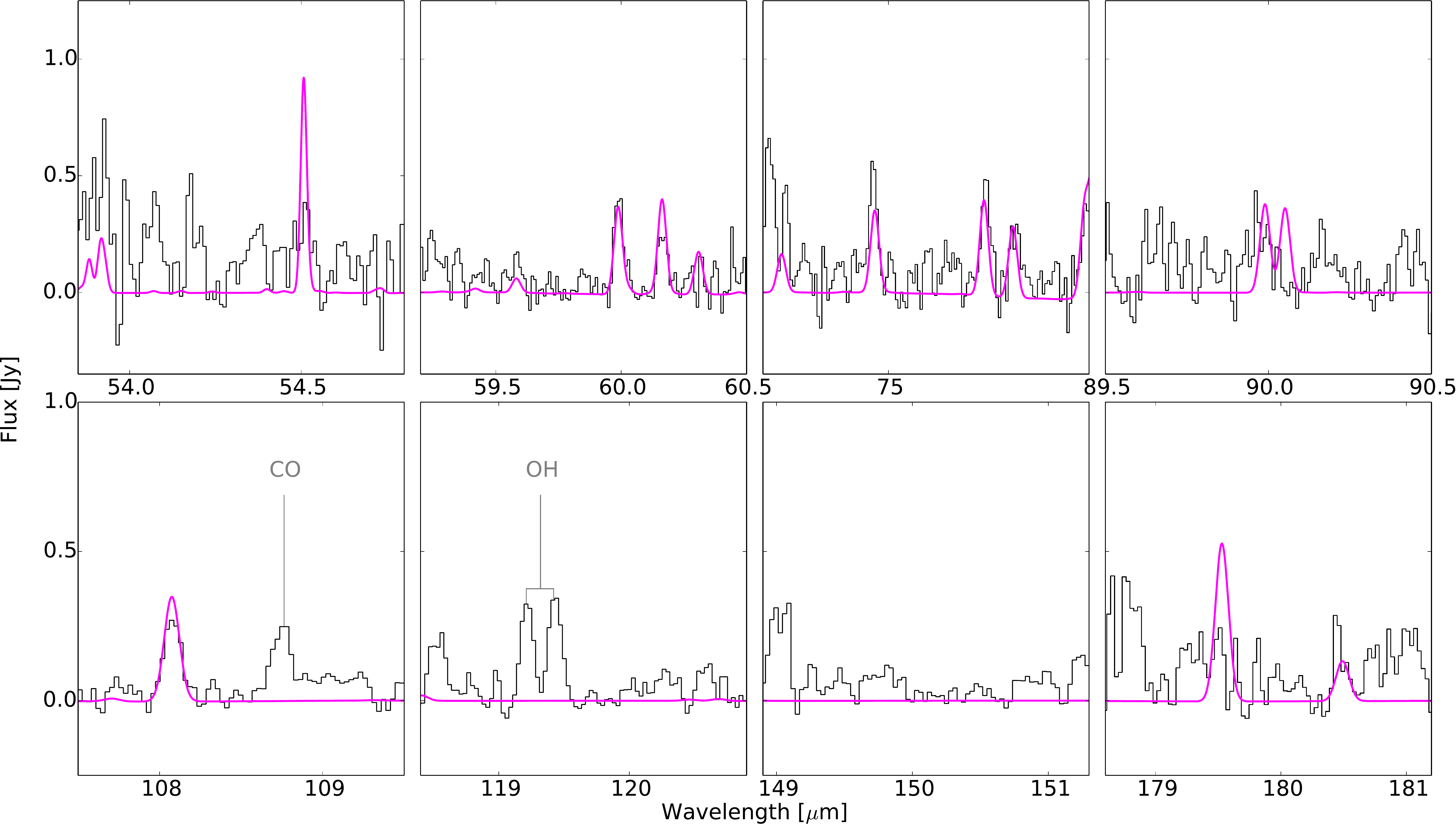}
\caption{\textbf{RNO 90:} Best-fit Case II model compared to the {\it Herschel}-PACS spectroscopy.}
\label{fig:rno90_pacs}
\end{figure*}

\begin{figure*}[ht!]
\centering
\includegraphics[width=8.95cm]{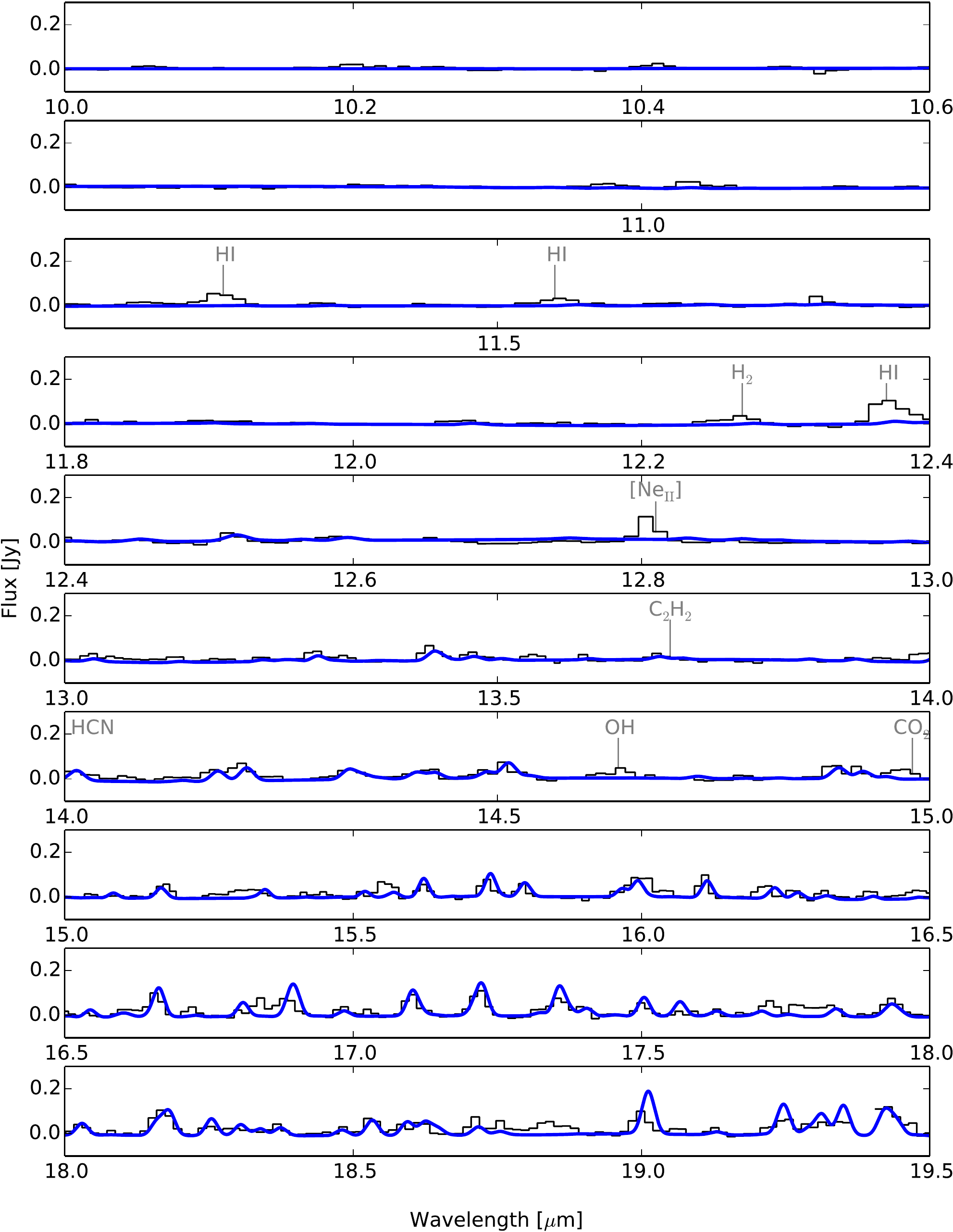}
\includegraphics[width=8.9cm]{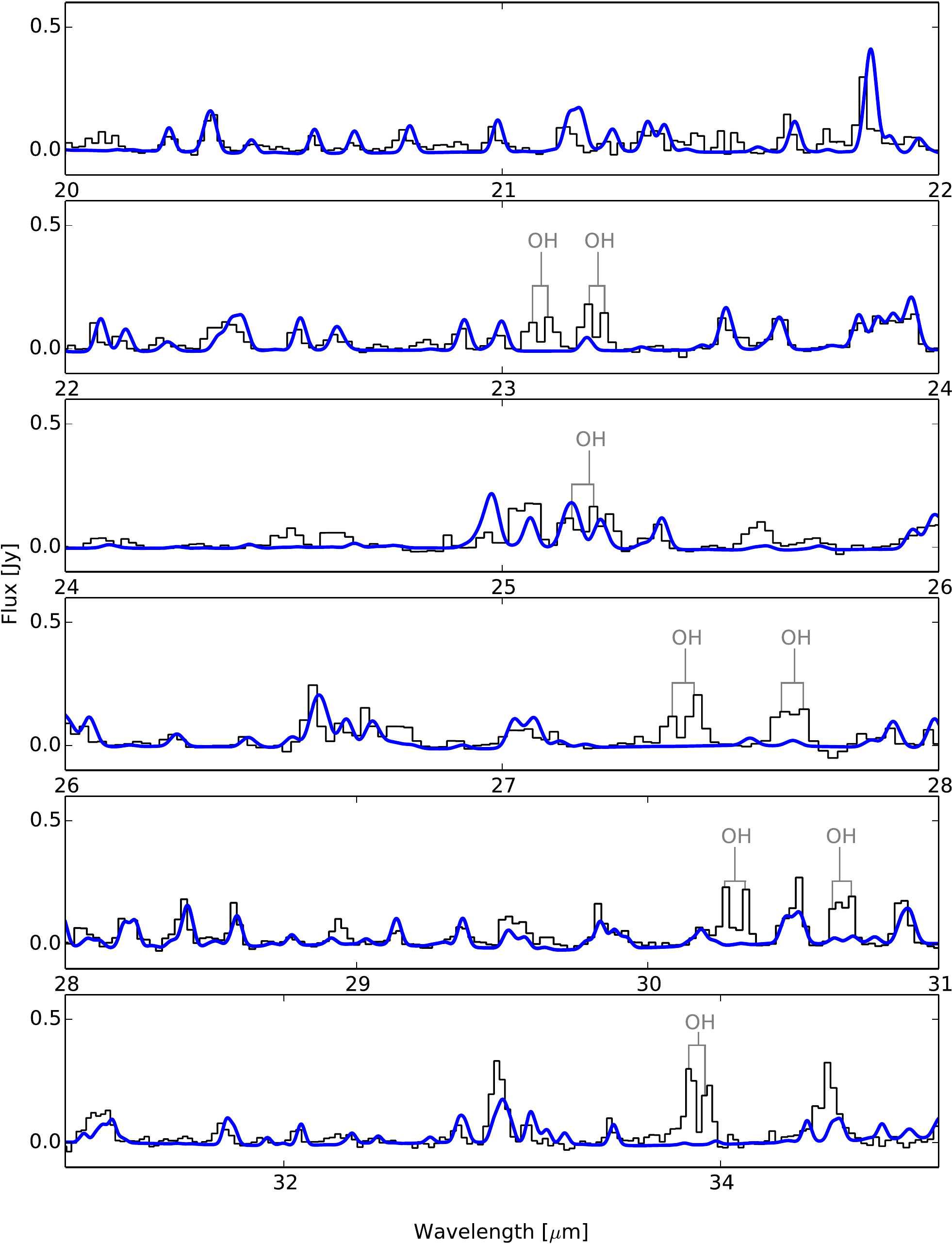}
\caption{\textbf{VW Cha:} Best-fit Case II model compared to the {\it Spitzer}-IRS spectral range.}
\label{fig:vwcha_irs}
\end{figure*}

\begin{figure*}[ht!]
\centering
\includegraphics[width=16.cm]{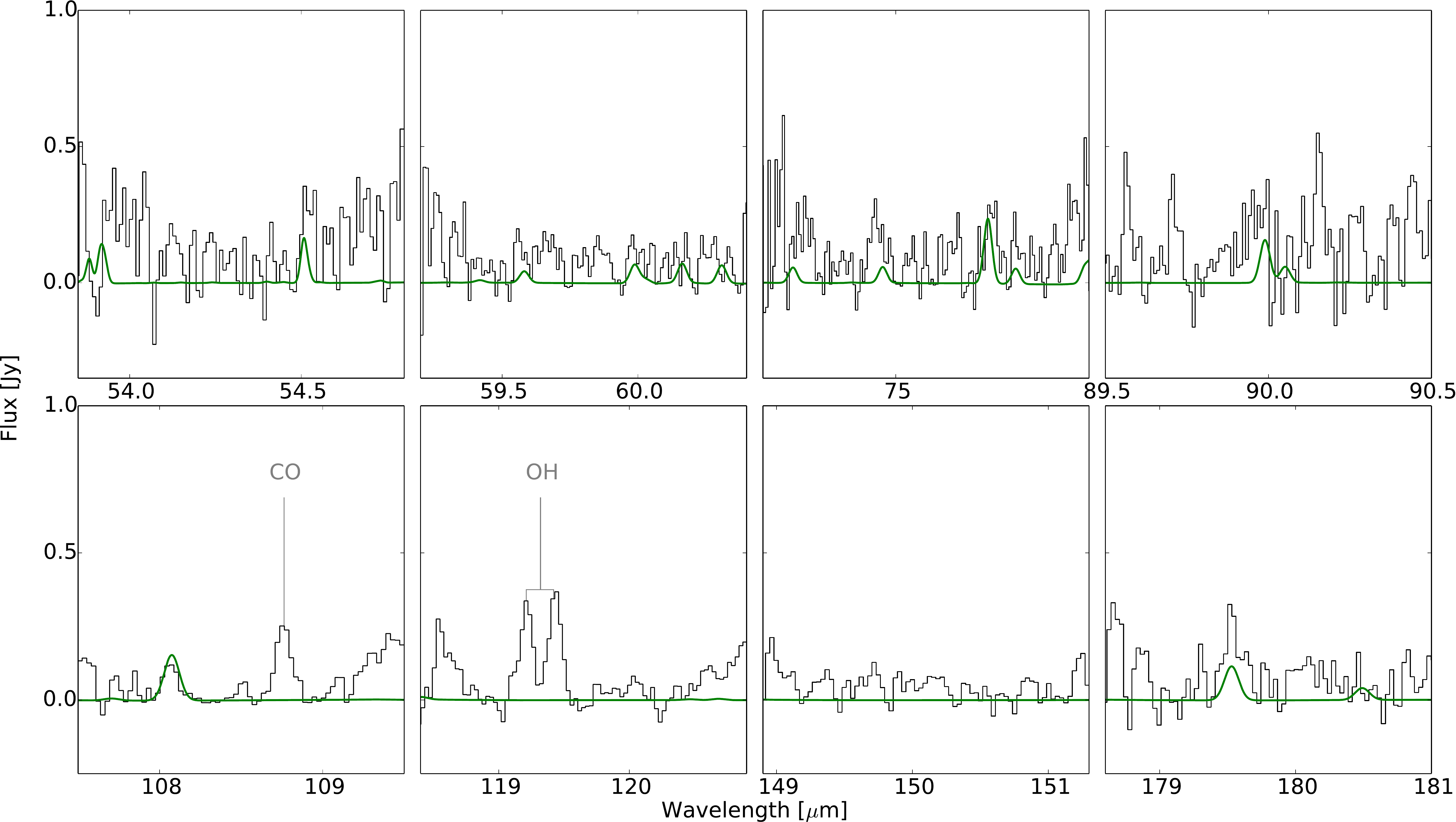}
\caption{\textbf{VW Cha:} Best-fit Case II model compared to the {\it Herschel}-PACS spectroscopy.}
\label{fig:vwcha_pacs}
\end{figure*}

Figures \ref{fig:drtau_irs}--\ref{fig:vwcha_pacs} show the full observed wavelength range, superimposed with the best-fitting Case II models for all four disks. There are two figures for each disk displaying the {\it Spitzer}-IRS SH ($\lambda = 10-19.5\,\mu$m) and LH ($\lambda = 20-36\,\mu$m), and {\it Herschel}-PACS ($\lambda = 53.9-181.2\,\mu$m) data and best-fit models. 

\section{Integrated water line fluxes} 

Tables \ref{tab:lineflux_spitzer} and \ref{tab:lineflux_herschel} list the integrated line fluxes used in the analysis and model fits for all four disks for the Spitzer and Herschel spectra, respectively. The tables also include molecular data for the dominant lines of blended water line complexes.

\clearpage
\LongTables
\begin{landscape}
\begin{deluxetable*}{lcccccccccccccccc}
\centering
\tablecolumns{17}
\tablewidth{0pt} 
\tablecaption{{\it Spitzer}-IRS \\ Integrated observed and model line fluxes}

\tablehead{
\colhead{ } & \colhead{ } & \colhead{ } & \colhead{ } & \colhead{ } & \multicolumn{3}{c}{DR Tau} & \multicolumn{3}{c}{FZ Tau} &   \multicolumn{3}{c}{RNO 90} &  \multicolumn{3}{c}{VW Cha} \\  
\colhead{Species} & \colhead{Transition} &    \colhead{Wavelength} & \colhead{$E_{\rm up}$} & \colhead{$A$} & $F_{\rm Obs}$ & $F_{\rm Case I}$ & $F_{\rm Case II}$ & $F_{\rm Obs}$ & $F_{\rm Case I}$ & $F_{\rm Case II}$ & $F_{\rm Obs}$& $F_{\rm Case I}$ & $F_{\rm Case II}$ &$F_{\rm Obs}$ & $F_{\rm Case I}$ & $F_{\rm Case II}$ \\
\colhead{ } & \colhead{ } & \colhead{($\mu$m)} & \colhead{(K)} & \colhead{($\rm s^{-1}$)} & \multicolumn{12}{c}{($10^{-14} \rm erg\,s^{-1}\,cm^{-2}$)} 
}

\startdata
%----------------------------------------------------------------------------------------------------------------------------------------------------------------------------------------------------------------------------------------------------------------------------------------------------------------------

o-H$_2$O   &   8$_{7\,2}$-7$_{4\,3}$     			&        15.16   &       2288.6   &          0.06     &  $<$ 4.9                            &  5.3            &   3.6  & 4.6 $\pm$ 0.3       & 5.3                 &              3.7                & $<$ 5.8 & 12.3	&  3.8   & $<$ 1.9                         & 2.1        &      1.3    \\
p-H$_2$O   &   10$_{6\,4}$-9$_{3\,7}$     		         &        15.17   &       2698.3   &          0.42     &                                            &                    &          &                                 &          	             &	                                &          	                   &     &                                       &                &            \\ 
\\ [-1.5ex]
\hline  \\ [-1.5ex]
%----------------------------------------------------------------------------------------------------------------------------------------------------------------------------------------------------------------------------------------------------------------------------------------------------------------------

p-H$_2$O   &   13$_{6\,8}$-12$_{3\,9}$     		&        15.57   &       3953.8   &          3.88     &   7.6 $\pm$ 0.7                  &          0.7     &   5.4  &  10.1 $\pm$ 0.5   &          5.9   &    7.0   &	18.2 $\pm$ 1.0    &          3.2	&  7.1   &     6.4 $\pm$ 0.4         &          1.4  &      3.1        \\
o-H$_2$O   &   13$_{3\,10}$-12$_{2\,11}$     		&        15.62   &       3474.2   &          2.92     &                                             &                     &           &	                          &                   &         &	                               &          	          &	&                                       &                  &              \\
\\ [-1.5ex]
\hline \\ [-1.5ex]
%----------------------------------------------------------------------------------------------------------------------------------------------------------------------------------------------------------------------------------------------------------------------------------------------------------------------

o-H$_2$O   &   12$_{3\,10}$-11$_{0\,11}$     		&        15.74   &       2823.6   &          1.09     &   $<$ 4.4                          &          2.2       &   4.5   &   	3.5 $\pm$ 0.3   &          2.3 &  3.7	&     7.3 $\pm$ 0.6   &          7.2	&  4.7	&     2.0 $\pm$ 0.2      &          1.6   &   2.7   \\
\\ [-1.5ex]
\hline \\ [-1.5ex]
%----------------------------------------------------------------------------------------------------------------------------------------------------------------------------------------------------------------------------------------------------------------------------------------------------------------------

o-H$_2$O   &   12$_{5\,8}$-11$_{2\,9}$     		&        17.10   &       3273.7   &          3.71     &   $<$ 6.9                          &          2.9        &   6.0  &   	4.5 $\pm$ 0.3   &          3.2 &  3.9	 &     9.0 $\pm$ 0.9   &          9.8	&  5.6	&     $<$ 2.7                 &          2.1    &    2.9  \\
\\ [-1.5ex]
\hline \\ [-1.5ex]
%----------------------------------------------------------------------------------------------------------------------------------------------------------------------------------------------------------------------------------------------------------------------------------------------------------------------

p-H$_2$O   &   11$_{3\,9}$-10$_{0\,10}$   		&        17.23   &       2438.8   &          0.96   	&      8.9 $\pm$ 0.8   &          6.0  & 	6.1      &       7.7 $\pm$ 0.3    &          5.6 &	 7.1   &   $<$ 9.4  	              &         0.9	&  6.9	  &      4.3 $\pm$ 0.3   &          2.8  &	4.2  \\
\\ [-1.5ex]
\hline \\ [-1.5ex]
%----------------------------------------------------------------------------------------------------------------------------------------------------------------------------------------------------------------------------------------------------------------------------------------------------------------------

o-H$_2$O   &   11$_{2\,9}$-10$_{1\,10}$   		&        17.36   &       2432.4   &          0.95   	&    $<$ 8.5             &          5.0   &	10.2      &         6.8 $\pm$ 0.4  &          4.8 &   6.2    &	14.6 $\pm$ 1.0       &         13.3	&  10.0	  &    4.4 $\pm$ 0.4   &          3.0     &   4.3    \\
\\ [-1.5ex]
\hline \\ [-1.5ex]
%----------------------------------------------------------------------------------------------------------------------------------------------------------------------------------------------------------------------------------------------------------------------------------------------------------------------

o-H$_2$O   &   10$_{8\,3}$-9$_{7\,2}$          		&        22.54   &       3243.4   &          32.72   &     8.4 $\pm$ 0.9 &	        9.1   &   	8.7       &	5.8 $\pm$	0.3 &         6.3  &    4.9   	& 12.5 $\pm$ 1.0   &        15.0	&	10.7    &   3.6 $\pm$ 0.4  &      	4.8	&       4.6    \\
o-H$_2$O   &   6$_{5\,2}$-5$_{2\,3}$          		&        22.62   &       1278.5   &          0.06   	 &  			      &		       &         	            &          		       &	       &                  &	 		      &		         &           &	   		        &            	&           \\
o-H$_2$O   &   5$_{5\,0}$-4$_{2\,3}$          		&        22.64   &       1067.6   &          0.01   	 &        		      &	                &		            &	          	       &	       &                  &			      &		         &            &		                 &		&  	 	\\
\\ [-1.5ex]
\hline \\ [-1.5ex]
%----------------------------------------------------------------------------------------------------------------------------------------------------------------------------------------------------------------------------------------------------------------------------------------------------------------------

p-H$_2$O   &   5$_{5\,1}$-4$_{2\,2}$     			&        23.46   &       1067.7   &         0.02   	&    6.2 $\pm$ 0.9   &        4.5   &	7.3        &       2.8 $\pm$ 0.3    &         3.9 &   4.6      &   12.8 $\pm$ 1.0 &          9.6	&	7.8      &  $<$ 2.1           &          2.3	  &       2.9        \\
o-H$_2$O   &   11$_{6\,5}$-10$_{5\,6}$     		&        23.51   &       3084.8   &         16.72   	&        			                &		        &				     &	               &              &		                   &		&             &		 	        &		  &          		\\
\\ [-1.5ex]
\hline \\ [-1.5ex]
%----------------------------------------------------------------------------------------------------------------------------------------------------------------------------------------------------------------------------------------------------------------------------------------------------------------------

o-H$_2$O   &   8$_{4\,5}$-8$_{1\,8}$         		&        26.42   &       1615.3   &          0.02   	&   $<$ 0.7   	     &          1.8  & 	0.9       &   $<$ 1.1   		&         1.2 	 &    1.3     & $<$ 2.0   	       &          2.5	&	1.4         & $<$ 0.3   	&          0.6   &     0.6	\\
\\ [-1.5ex]
\hline \\ [-1.5ex]
%----------------------------------------------------------------------------------------------------------------------------------------------------------------------------------------------------------------------------------------------------------------------------------------------------------------------

p-H$_2$O   &   7$_{7\,1}$-6$_{6\,0}$      		 &        28.59   &       2006.8   &          20.36   &  $<$ 3.2  	             &          4.3	 &	5.3         &  $<$ 2.4  		&          3.0	  &	2.4      & 11.5 $\pm$ 0.6   &          7.5    	&   6.7	 & $<$ 1.9          &          2.1  &      1.9	\\
\\ [-1.5ex]
\hline \\ [-1.5ex]
%----------------------------------------------------------------------------------------------------------------------------------------------------------------------------------------------------------------------------------------------------------------------------------------------------------------------

p-H$_2$O   &   9$_{5\,5}$-8$_{4\,4}$      		&        29.14   &       2122.1   &          8.87   	&            $<$ 1.0   	    &          2.1      &	3.8         &  $<$ 0.9   		 &          1.3  &	1.1       & $<$ 4.6         	  &          4.0	&  4.9	  & $<$ 1.0   	 &          1.4   &     1.4	\\
\\ [-1.5ex]
\hline \\ [-1.5ex]
%----------------------------------------------------------------------------------------------------------------------------------------------------------------------------------------------------------------------------------------------------------------------------------------------------------------------

p-H$_2$O   &   8$_{3\,5}$-7$_{2\,6}$      		 &        29.36   &       1510.9   &          1.78   	&                    $<$ 2.1     &	       5.0     &	 4.1        & $<$ 0.9   		  &          3.5  &	  2.4     & $<$ 0.8     	  &          7.8	&   6.7	   & $<$ 1.2   	 &          2.0    &    1.6	\\
\\ [-1.5ex]
\hline \\ [-1.5ex]
%----------------------------------------------------------------------------------------------------------------------------------------------------------------------------------------------------------------------------------------------------------------------------------------------------------------------

p-H$_2$O   &   8$_{5\,3}$-7$_{4\,4}$      		&        30.47   &       1806.9   &         8.92   	&   7.2 $\pm$ 0.7	    &          4.5     &       8.9          & 4.9 $\pm$ 0.2           &          3.2  &	2.6        & 12.2 $\pm$ 0.7 &         11.0	&	11.5     & 4.9 $\pm$ 0.3 &      4.2    &  3.3	\\
o-H$_2$O   &   7$_{6\,1}$-6$_{5\,2}$      		&        30.53   &       1749.8   &         13.55   	&                                       &		         &		      &                                     &          	&	              &		                 &         	&	     &		                &         	       & 	\\
\\ [-1.5ex]
\hline \\ [-1.5ex]
%----------------------------------------------------------------------------------------------------------------------------------------------------------------------------------------------------------------------------------------------------------------------------------------------------------------------

o-H$_2$O   &   8$_{5\,4}$-7$_{4\,3}$      		&        30.87   &       1805.9   &          8.69   	&   $<$ 5.5  		     &          4.8	 &	7.6          & 3.3 $\pm$ 0.3           &          3.7   &	2.1         & 10.5 $\pm$ 0.8   &         11.9	&	10.9      & 3.6 $\pm$ 0.4   &    4.8     &    2.6   \\
o-H$_2$O   &   6$_{3\,4}$-5$_{0\,5}$      		&        30.90   &        933.7    &          0.35   	&                                        &		          &		      &			                  &        	 &	         	     & 			           &        		&	      & 	        		&         		\\
\\ [-1.5ex]
\hline \\ [-1.5ex]
%----------------------------------------------------------------------------------------------------------------------------------------------------------------------------------------------------------------------------------------------------------------------------------------------------------------------

p-H$_2$O   &   8$_{4\,4}$-7$_{3\,5}$  		&        31.74   &       1628.3   &          4.70   	&  $<$ 1.9                          &          4.4	 &	 5.0          & $<$ 2.1   	            &          3.0    &   2.1         & $<$ 2.6      	 &          9.1	&	9.9       & $<$ 1.2   		&          3.7 & 	2.1 \\
o-H$_2$O   &   4$_{4\,1}$-3$_{1\,2}$  		&        31.77   &       702.3     &          0.02   	&                                          &		 &		       &          		   &		      &		        &         		 &		   	&	       &         		          &                 &     \\
\\ [-1.5ex]
\hline \\ [-1.5ex]
%----------------------------------------------------------------------------------------------------------------------------------------------------------------------------------------------------------------------------------------------------------------------------------------------------------------------

p-H$_2$O   &   11$_{4\,8}$-10$_{3\,7}$    	&        32.80   &       2651.7   &          5.25   	&  $<$ 1.8   		       &          1.0	&	3.2           & $<$ 0.9   		    &	1.0  	      &      1.6       & $<$ 3.2     	  &          2.5	&	3.2        & $<$ 0.7   	&          0.7   &	1.9 \\
o-H$_2$O   &   14$_{3\,12}$-13$_{2\,11}$    	&        32.83   &       3671.0   &          9.13   	&                                          &		&		       &		              &          	      &	                  &			  &          		&	         &    			 &         	   &	\\
\\ [-1.5ex]
\hline \\ [-1.5ex]
%----------------------------------------------------------------------------------------------------------------------------------------------------------------------------------------------------------------------------------------------------------------------------------------------------------------------

p-H$_2$O   &   5$_{5\,1}$-4$_{4\,0}$      		&        32.92   &       5827.7   &         11.36   	&  15.3 $\pm$ 5.3             &         9.0    &	 16.6            & 7.4 $\pm$ 1.9         &          6.2   &	  4.5   & 27.4 $\pm$ 4.9 &   23.7	&	18.9         & 7.9 $\pm$ 2.5  &          8.5   &   7.9 	 \\
o-H$_2$O   &   7$_{5\,2}$-6$_{4\,3}$      		&        32.99   &       1524.8   &         8.34   	&                                           &		 &		            &		                      &          	      &	           &			     &         	&	         &			  &		     &       	 \\
p-H$_2$O   &   6$_{6\,0}$-5$_{5\,1}$      		&        33.01   &       1503.6   &         12.79   	&  	  			        &         	 &	  	            &			             &         	      &	           &	 		     &         	&	         &		   	  &		     &          		

%----------------------------------------------------------------------------------------------------------------------------------------------------------------------------------------------------------------------------------------------------------------------------------------------------------------------			  

\enddata
\tablecomments{
Most {\it Spitzer}-IRS water line complexes include multiple transitions. Transitions that dominate each unresolved line complex are defined as contributing $>25\%$ relative to the total intensity of the complex. Upper limits are reported for integrated fluxes with $S/N<5$. The integrated line fluxes are listed as $F_{\rm Obs}$,  $F_{\rm Case I}$, and $F_{\rm Case II}$ for the observed data, the Case I model, and the Case II models, respectively. Lines that contribute to each water complex are grouped together and separated by solid lines. 
}
\label{tab:lineflux_spitzer}
\end{deluxetable*}

\clearpage
\end{landscape}

\clearpage
\LongTables
\begin{landscape}
\begin{deluxetable*}{lcccccccccccccccc}
\centering
\tablecolumns{17}
\tablewidth{0pt} 
\tablecaption{{\it Herschel}-PACS \\ Integrated observed and model line fluxes}

\tablehead{
\colhead{ } & \colhead{ } & \colhead{ } & \colhead{ } & \colhead{ } & \multicolumn{3}{c}{DR Tau} & \multicolumn{3}{c}{FZ Tau} &   \multicolumn{3}{c}{RNO 90} &  \multicolumn{3}{c}{VW Cha} \\  
\colhead{Species} & \colhead{Transition} &    \colhead{Wavelength} & \colhead{$E_{\rm up}$} & \colhead{$A$} & $F_{\rm Obs}$ & $F_{\rm Case I}$ & $F_{\rm Case II}$ & $F_{\rm Obs}$ & $F_{\rm Case I}$ & $F_{\rm Case II}$ & $F_{\rm Obs}$& $F_{\rm Case I}$ & $F_{\rm Case II}$ &$F_{\rm Obs}$ & $F_{\rm Case I}$ & $F_{\rm Case II}$ \\
\colhead{ } & \colhead{ } & \colhead{($\mu$m)} & \colhead{(K)} & \colhead{($\rm s^{-1}$)} & \multicolumn{12}{c}{($10^{-14} \rm erg\,s^{-1}\,cm^{-2}$)} 
}

\startdata

 o-H$_2$O   &   5$_{3\,2}$-5$_{0\,5}$      		&        54.51   &       732.1   &          0.04   	&       $<$ 2.6                	&    1.2       &	    0.9          & $<$ 4.1   	    &          0.8    &    0.5  & 3.8 $\pm$ 0.4	&   1.9        &        2.2        &  $<$ 1.9   	  &       0.7	     &      0.4	\\
\\ [-1.5ex]
\hline \\ [-1.5ex]
%----------------------------------------------------------------------------------------------------------------------------------------------------------------------------------------------------------------------------------------------------------------------------------------------------------------------

p-H$_2$O   &   7$_{2\,6}$-6$_{1\,5}$      		&        59.99   &       1021.0   &          1.33   	& 1.5 $\pm$ 0.1                  &    0.7      &          0.7           &	$<$ 1.1   	   &          0.5    &    0.2   & 1.9 $\pm$ 0.1   &   1.5       &	   2.2 	&	$<$ 1.2   	  &          0.6     &   0.3	\\
\\ [-1.5ex]
\hline \\ [-1.5ex]
%----------------------------------------------------------------------------------------------------------------------------------------------------------------------------------------------------------------------------------------------------------------------------------------------------------------------

p-H$_2$O   &   8$_{2\,6}$-7$_{3\,5}$      		&        60.16   &       1414.2   &          0.70   	& 0.6 $\pm$ 0.1                  &    0.6      &	   0.5           & $<$ 1.3        &          0.4    &     0.2    &	$<$ 0.7    	 &   0.1	&	  1.4         &       $<$ 0.9 	  &          0.4     &    0.2	\\
\\ [-1.5ex]
\hline \\ [-1.5ex]
%----------------------------------------------------------------------------------------------------------------------------------------------------------------------------------------------------------------------------------------------------------------------------------------------------------------------

o-H$_2$O   &   7$_{2\,5}$-6$_{3\,4}$      		&        74.94   &        1125.7    &          0.26   &  1.5 $\pm$ 0.1                &   0.3        &	   0.3           & $<$ 0.7        &          0.2	&	0.1    &       $<$ 0.7   &   0.7 	&	 0.8           & $<$ 0.9             &        0.3         &     0.1   \\
\\ [-1.5ex]
\hline \\ [-1.5ex]
%----------------------------------------------------------------------------------------------------------------------------------------------------------------------------------------------------------------------------------------------------------------------------------------------------------------------

o-H$_2$O   &   3$_{2\,1}$-2$_{1\,2}$      		&        75.38   &        305.2    &          0.33   	&  0.8 $\pm$ 0.1                 &  0.9	&	   0.7           & $<$ 0.4      &          0.3	&	0.1    & 1.1 $\pm$ 0.1 &  1.2 	&	 0.9            & $<$ 1.3  	      &          0.7	&      0.6  \\
\\ [-1.5ex]
\hline \\ [-1.5ex]
%----------------------------------------------------------------------------------------------------------------------------------------------------------------------------------------------------------------------------------------------------------------------------------------------------------------------

p-H$_2$O   &   3$_{2\,2}$-2$_{1\,1}$      		&        89.99   &        296.8    &          0.35   	&   $<$ 1.0   			&   0.7              &     0.5            & $<$ 0.1    &          0.3   &	0.1     & $<$ 0.9     	     &  1.1	&	  1.0          & $<$ 0.3   	      &          0.5	 &    0.3      \\
p-H$_2$O   &   7$_{4\,4}$-7$_{3\,5}$      		&        90.05   &        1334.8  &          0.35   	&   	  				&            	      &	  		    &	   	   &             		&	           &                 	     &          	&	   	        &		 	      &          		 &          \\
\\ [-1.5ex]
\hline \\ [-1.5ex]
%----------------------------------------------------------------------------------------------------------------------------------------------------------------------------------------------------------------------------------------------------------------------------------------------------------------------

o-H$_2$O   &   2$_{2\,1}$-1$_{1\,0}$      		&       108.07   &        194.1   &          0.26   	&   1.0 $\pm$ 0.1               &   1.2              &       1.1            &	$<$ 0.2 &          0.2	&	0.1       & 0.9 $\pm$ 0.1 & 1.0 &	     1.0          & $<$ 0.5            &          0.5 	  &      0.5   \\
\\ [-1.5ex]
\hline \\ [-1.5ex]
%----------------------------------------------------------------------------------------------------------------------------------------------------------------------------------------------------------------------------------------------------------------------------------------------------------------------

o-H$_2$O   &   2$_{1\,2}$-1$_{0\,1}$      		&       179.53   &        114.4   &          0.06   	&  $<$ 0.5   		        &  0.7               &	        0.7            & $<$ 0.2   &          0.1	&	  0.1     & $<$ 0.7   	       &  0.6 	&         0.6              &  $<$	0.4           &          0.3     &   	0.1			  

\enddata
\tablecomments{
The integrated line fluxes are listed as $F_{\rm Obs}$,  $F_{\rm Case I}$, and $F_{\rm Case II}$ for the observed data, the Case I model, and the Case II models, respectively. Lines that contribute to each water complex are grouped together and separated by solid lines. }
\label{tab:lineflux_herschel}
\end{deluxetable*}

\clearpage
\end{landscape}

\end{document}